\documentclass[10pt, final, conference, letterpaper, oneside, twocolumn]{IEEEtran}
\newif\ifISIT \ISITfalse                            %
\newif\ifStudentAward \StudentAwardtrue             %
\ifISIT
\newcommand{\ISIT}[1]{#1}
\newcommand{\full}[1]{}
\else
\newcommand{\ISIT}[1]{}
\newcommand{\full}[1]{#1}
\fi

\usepackage[hyperref]{Bamble}                       %
\usepackage[noadjust]{cite}
\usepackage{accents}
\usepackage{balance}
\usepackage{soul}
\usepackage[T1]{fontenc}
\pgfplotsset{compat=1.18}
\usepackage[mode=buildnew]{standalone}
\usepackage{xfrac}
\IEEEoverridecommandlockouts
\usepackage{colortbl}
\usepackage{bm}
\newcommand{\bigO}{\mathcal{O}}
\newcommand{\ourell}{\ell}
\newcommand{\reals}{\mathbb{R}}

\newcommand{\fminsum}{\tilde{f}}
\newcommand{\ourStar}{\star}
\newcommand{\ourRu}{R_{\mathrm{U}}}
\newcommand{\ourRl}{R_{\mathrm{L}}}
\newcommand{\ourpos}[1]{\mathrm{pos}\left\langle#1\right\rangle}
\newcommand{\ourneg}[1]{\mathrm{neg}\left\langle#1\right\rangle}
\newcommand{\labeler}{\lambda}
\newcommand{\ourZ}{Z}
\newcommand{\ourZopt}{\ourZ^{\ourStar}}
\newcommand{\ourzeta}{\zeta}
\newcommand{\ourfloor}[1]{\lfloor #1 \rfloor}
\newcommand{\ourPe}{P_{\mathrm{e}}}
\newcommand{\ind}[1]{[#1]}
\newcommand{\Gdeep}{\mathcal{G}^*}
\newcommand{\Edeep}{\mathcal{E}^*}
\definecolor{colorSetA}{rgb}{0.52, 0.73, 0.4}
\definecolor{colorSetB}{rgb}{0.89, 0.66, 0.34}
\newcommand{\highlight}[2]{\colorbox{#1}{$\displaystyle #2$}}

\newcommand{\arikan}{Ar{\i}kan\xspace}
\newcommand{\MSA}{MSA\xspace}

\SetKwFunction{algDecode}{Decode}
\SetKwFunction{algMakeDecision}{MakeDecision}
\SetKwFunction{algGenieCorrect}{GenieCorrect}
\SetKwData{alga}{a}
\SetKwData{algb}{b}
\SetKwData{algc}{c}
\SetKwData{algv}{v}
\newcommand{\algLambdaf}{\Lambda_f}
\newcommand{\algLambdag}{\Lambda_g}

\makeatletter
\def\IEEElabelanchoreqn#1{\bgroup
\def\@currentlabel{\p@equation\theequation}\relax
\def\@currentHref{\@IEEEtheHrefequation}\label{#1}\relax
\Hy@raisedlink{\hyper@anchorstart{\@currentHref}}\relax
\Hy@raisedlink{\hyper@anchorend}\egroup}
\makeatother
\newcommand{\subnumberinglabel}[1]{\IEEEyesnumber
\IEEEyessubnumber*\IEEElabelanchoreqn{#1}}

\newcommand{\twobibs}[2]{#2} %

\begin{document}
\IEEEoverridecommandlockouts %
\title{An Analytical Study of the Min-Sum Approximation for Polar Codes}
\date{}
\author{
\IEEEauthorblockN{Nir Chisnevski, Ido~Tal, Shlomo~Shamai~(Shitz)\\
The Andrew and Erna Viterbi Faculty of Electrical and Computer Engineering,\\
Technion, Haifa 32000, Israel.\\
Email: \{\texttt{nir.ch@campus}, \texttt{idotal@ee}, \texttt{sshlomo@ee}\}\texttt{.technion.ac.il}}
}
\maketitle

\begin{abstract}
\label{sec: abstract}
\ISIT{\ifStudentAward THIS PAPER IS ELIGIBLE FOR THE STUDENT PAPER AWARD. \fi}%
The min-sum approximation is widely used in the decoding of polar codes.
Although it is a numerical approximation, hardly any penalties are incurred in practice.
We give a theoretical justification for this.
We consider the common case of a binary-input, memoryless, and symmetric channel, decoded using successive cancellation and the min-sum approximation.
Under mild assumptions, we show the following.
For the finite length case, we show how to exactly calculate the error probabilities of all synthetic (bit) channels in time $\bigO(N^{1.585})$, where $N$ is the codeword length.
This implies a code construction algorithm with the above complexity. 
For the asymptotic case, we develop two rate thresholds, denoted $\ourRl = \ourRl(\labeler)$ and $\ourRu =\ourRu(\labeler)$, where $\labeler(\cdot)$ is the labeler of the channel outputs (essentially, a quantizer).
For any $0 < \beta < \frac{1}{2}$ and any code rate $R < \ourRl$, there exists a family of polar codes with growing lengths such that their rates are at least $R$ and their error probabilities are at most $2^{-N^\beta}$.
That is, strong polarization continues to hold under the min-sum approximation.
Conversely, for code rates exceeding $\ourRu$, the error probability approaches $1$ as the code-length increases, irrespective of which bits are frozen.
We show that $0 < \ourRl \leq \ourRu \leq C$, where $C$ is the channel capacity.
The last inequality is often strict, in which case the ramification of using the min-sum approximation is that we can no longer achieve capacity.
\end{abstract}

\section{Introduction} \label{sec: introduction}
Polar codes are a family of capacity-achieving error correcting codes with efficient encoding and decoding algorithms, introduced by \arikan \cite{Arikan:09p}.
In this paper, we study the setting of a binary-input, memoryless and symmetric channel.
Although many generalizations to this case exist \cite{HondaYamamoto:12p, STA:09p,  Sasoglu:12c, SasogluTal:19p, ShuvalTal:19.2p, Wang+:15c, Wang+:16c, TPFV:22p, AravaTal:23c, PfisterTal:21c, Arikan:10c, KoradaUrbanke:10c, HofShamai:10a, MahdavifarVardy:11p, ShuvalTal:24c}, 
it is arguably the most basic and common one. Moreover, it affords a very efficient hardware implementation using the numerical min-sum approximation (\MSA) in the decoder.

The seminal decoding algorithm of polar codes is called successive-cancellation (SC) decoding. It is a recursive algorithm that makes repeated use of the following two functions:
\begin{IEEEeqnarray}{rCl}
    f(L_a,L_b) &=& 2\tanh^{-1}
    \left(
    \tanh\left(\frac{L_a}{2}\right)\cdot
    \tanh\left(\frac{L_b}{2}\right)
    \right) \; , 
    \label{eq: definition of f fucntion} \\
    g_{u}(L_a,L_b) &=& {(-1)^{u}\cdot L_a} + L_b \; .
    \label{eq: definition of g fucntion}
\end{IEEEeqnarray}
The functions $g_0$ and $g_1$ are simple to implement, since addition and subtraction are hardware-friendly operations.
However, the $f$ function is somewhat complicated, since hyperbolic functions are expensive in terms of calculation time and power consumption.
Therefore, in many practical implementations the \MSA is used \cite{LTVG:10c}.
That is, similar to what is done in LDPC decoder implementation \cite{768759}, the $f$ function is replaced with a simpler function $\fminsum$ given by
\begin{IEEEeqnarray}{rCl}
    \fminsum(L_a,L_b) &=& \sign(L_a) \cdot \sign(L_b) \cdot \min{\{|L_a|,|L_b|\}} \; ,
    \label{eq: definition of min-sum f function}
\end{IEEEeqnarray}
where \(\sign(\cdot)\) is the sign function defined as
\begin{IEEEeqnarray*}{rCl}
    \sign(x) \triangleq
    \begin{cases}
	\phantom{-}1 & \text{if } x>0 \; , \\
        -1 & \text{if } x<0 \; , \\
	\phantom{-}0 & \text{if } x=0 \; .
    \end{cases}     
\end{IEEEeqnarray*}
\begin{figure}
    \centering
    \includegraphics[scale=1]{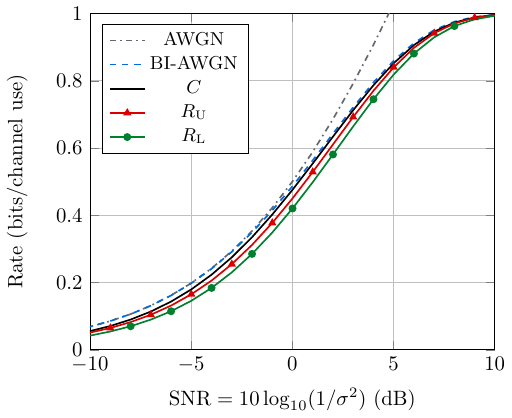}
    \caption{The capacity $C$ and the thresholds $\ourRu$ and $\ourRl$ of a BI-AWGN with 3-bit quantized output\full{, using the labeling function $\labeler$ given in \eqref{eq: awgn labeler}}.
    For reference, the capacities of the corresponding non-quantized BI-AWGN and AWGN are also given.
}
    \label{fig: awgnRateComparison}
\end{figure}%
For the non-approximated setting, $L_a$, $L_b$, and the outputs of $f$ and $g$ are log-likelihood ratios (LLRs) corresponding to certain channel outputs.
For the approximated setting, we use the generalized term `labels' for the corresponding quantities.
At the base of the recursion the labels $L_a$ and $L_b$ are obtained by applying a labeling function $\labeler(\cdot)$ on the channel outputs.
The full definition of $\labeler(\cdot)$ is given in \cref{sec: notation}.
Informally, $\labeler(y)$ is a quantized version of the LLR corresponding to the channel output $y$, up to a positive scaling constant.

The \MSA is also used in decoders that are derivatives of the SC decoder, such as the SC list decoder \cite{TalVardy:15p} and the SC stack decoder \cite{NiuChen:12p}. %
Often, the \MSA incurs only a small penalty in error rate \cite[Figure~7]{Leroux:13p}, \cite[Figures~7-8]{MSN:19c}, and \cite[Figure~4]{Balatsoukas:15p}.  In this paper, we analyze this phenomenon.

\begin{figure}
    \centering
    \includegraphics[scale=1]{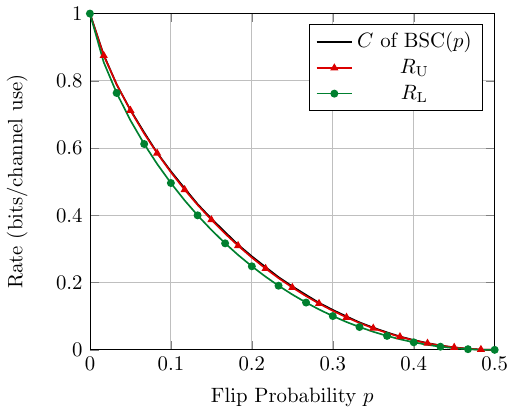}
    \caption{The capacity $C$ and the thresholds $\ourRu$ and $\ourRl$ for the $\mathrm{BSC}(p)$.}
    \label{fig: bscRateComparison}
\end{figure}

The following theorem is our main result for the asymptotic case.
The theorem promises two rate thresholds, $\ourRl$ and $\ourRu$, when employing the \MSA in SC decoding.
Below $\ourRl$, strong polarization is guaranteed, while above $\ourRu$ the error probability approaches $1$.
Figures \ref{fig: awgnRateComparison} and \ref{fig: bscRateComparison} plot these thresholds and the channel capacity $C$ for the binary-input additive white Gaussian noise (BI-AWGN) channel with quantized output, and for the binary symmetric channel (BSC), respectively.
As can be seen in these figures, $\ourRl$, $\ourRu$, and $C$ are all rather close.
However, note that $\ourRu$ is strictly smaller than $C$.
That is, in these cases using the \MSA means that we can no longer achieve capacity.
The theorem assumes that a ``fair labeler'' is used, as defined in \Cref{def: fair labeler} below. 

\begin{theorem}
Let $W$ be a binary-input, memoryless and symmetric channel.
Fix $0 < \beta < \frac{1}{2}$.
Let $\labeler(\cdot)$ be a fair labeler. Then, there exist thresholds $\ourRl = \ourRl(\labeler)$ and $\ourRu = \ourRu(\labeler)$, such that $0 < \ourRl \leq \ourRu$.
When using SC decoding and the \MSA, the following holds.
For any code rate $R <\ourRl$ there exists a family of polar codes with growing \ISIT{length $N$}\full{lengths} such that their rates are at least $R$ and their word error probabilities are at most $2^{-N^\beta}$\full{, where $N$ is the codeword length}.
Conversely, for code rates exceeding $\ourRu$, the word error probability approaches $1$ as the code-length increases, irrespective of which bits are frozen.
\label{thm: threshold rates for the asymptotic case}
\end{theorem}

If we only assume a fair labeler, $\ourRl$ is weak but still positive, and $\ourRu$ is trivial.
For a significant subclass of fair labelers, ``good labelers'' (\Cref{def: good labeler}), both bounds can be significantly strengthened. A good labeler is often the case in practice.

For the finite-length case and the good labeler setting, we develop an algorithm for calculating the exact error probability of each min-sum synthetic channel, defined in \eqref{eq: definition of Q-tilde joint-distribution}.
The running time of our algorithm is $\bigO(N^{1.585})$.
Note that in the non-approximated setting, no such algorithm exists, only a method to calculate bounds on the error probabilities \cite{TalVardy:13p}. 

\section{Notation}
\label{sec: notation}
Denote by $W: \mathcal{X} \rightarrow \mathcal{Y}$  a general binary-input, memoryless, and symmetric channel with input alphabet $\mathcal{X}=\{0,1\}$ and output alphabet $\mathcal{Y}$.
For each pair $x \in \mathcal{X}$ and $y \in \mathcal{Y}$ the input probability is $p(x)$, and the transition probability is $W(y|x)$.
Hence, the joint probability is given by $W(y;x) = p(x) \cdot W(y|x)$.
We will assume that \(p(x)\) is symmetric, i.e.\ $p(0) = p(1) = 1/2$.

For $n \in \mathbb{N}$ denote $N = 2^n$ and let $(X_i,Y_i)_{i=0}^{N-1}$ be $N$ i.i.d.\ pairs, each distributed according to $W(y;x)$. Denote by $U_0^{N-1}$ the polar transform of $X_0^{N-1}$.
For $0 \leq i <N$, define the following synthetic joint distribution\footnote{We find it notationally easier to track joint distributions instead of channels.
The latter is simply obtained from the former by multiplying by $2$.}:
\begin{multline}
    W_N^{(i)}(y_0^{N-1}, u_0^{i-1}; u_i) = \\ \Pr( Y_0^{N-1} = y_0^{N-1}, U_0^{i-1} = u_0^{i-1}, U_i = u_i ) \; .
    \label{eq: definition of W joint-distribution}
\end{multline}
By \cite[Proposition 3]{Arikan:09p},
\begin{multline}
    W_N^{(2j)} \left( y_0^{N-1},u_0^{2j-1};u_{2j} \right) = \\
    \smash{\sum_{u_{2j+1}}}
    W_{N/2}^{(j)} \left( y_0^{\frac{N}{2}-1},u_{0,e}^{2j-1} \oplus u_{0,o}^{2j-1};u_{2j} \oplus u_{2j+1} \right) \\
    \cdot W_{N/2}^{(j)} \left( y_{\frac{N}{2}}^{N-1},u_{0,o}^{2j-1};u_{2j+1}\right),
   \label{eq: minus transform of W joint-distribution} 
\end{multline}
and
\begin{multline}
    W_N^{(2j+1)} \left( y_0^{N-1},u_0^{2j};u_{2j+1} \right) = \\
    W_{N/2}^{(j)} \left( y_0^{\frac{N}{2}-1},u_{0,e}^{2j-1} \oplus u_{0,o}^{2j-1};u_{2j} \oplus u_{2j+1} \right) \\
    \cdot W_{N/2}^{(j)} \left( y_{\frac{N}{2}}^{N-1},u_{0,o}^{2j-1};u_{2j+1} \right),
    \label{eq: plus transform of W joint-distribution}
\end{multline}
where $W_{1}^{(0)}(y;x) = W(y;x)$ and ``$\oplus$'' is addition over $\mathrm{GF}(2)$.
In the above, $u_{0,e}^{2j-1}$ and $u_{0,o}^{2j-1}$ are the even and odd entries of $u_{0}^{2j-1}$, respectively.
As shown in \cite{Arikan:09p}, $W_N^{(2j)}$ and $W_N^{(2j+1)}$ are the result of applying the ``$-$'' and ``+'' transforms, respectively, on $W_{N/2}^{(j)}$, up to a relabeling of the output.
For each joint distribution $W_N^{(i)}$ we define the LLR $L_N^{(i)}$ as
\begin{equation}
    L_N^{(i)}( y_0^{N-1},u_0^{i-1}) %
    \triangleq \log_{2} {\left( \frac
    {W_N^{(i)}( y_0^{N-1},u_0^{i-1};u_i=0 )}
    {W_N^{(i)}(y_0^{N-1},u_0^{i-1};u_i=1 )} \right)} \; .  
    \label{eq: LLR defintion}
\end{equation}
\ISIT{Using the relations described in \eqref{eq: minus transform of W joint-distribution} and \eqref{eq: plus transform of W joint-distribution} we obtain recursive transforms for the LLRs.
Namely, these are \eqref{eq: minus transform of labels} and \eqref{eq: plus transform of labels} below, once we remove all the tildes.}
\full{Using the relations described in \eqref{eq: minus transform of W joint-distribution} and \eqref{eq: plus transform of W joint-distribution} we obtain the following recursive transforms for the LLRs:
\begin{IEEEeqnarray}{l}
    \subnumberinglabel{eq: transforms of llrs}
    L_N^{(2j)}\left( y_0^{N-1},u_0^{2j-1}\right ) =
    \IEEEeqnarraynumspace \label{eq: minus transform of llrs} \\
    \quad
    f \! \left(\!
    L_{N/2}^{(j)} \! \left( \! y_0^{N/2-1},u_{0,e}^{2j-1} \oplus u_{0,o}^{2j-1}\!\right)\!,\!
    L_{N/2}^{(j)}\! \left( \!  y_{N/2}^{N-1},u_{0,o}^{2j-1}\right) \!
    \right) , \IEEEnonumber
    \\
    L_N^{(2j+1)}\left( y_0^{N-1},u_0^{2j}\right ) =
    \IEEEeqnarraynumspace\label{eq: plus transform of llrs} \\ 
    \quad
    g_{u_{2j}} \! \left( \!
    L_{N/2}^{(j)} \! \left( \! y_0^{N/2-1},u_{0,e}^{2j-1} \oplus u_{0,o}^{2j-1}\!\right)\!,\!
    L_{N/2}^{(j)} \! \left( \! y_{N/2}^{N-1},u_{0,o}^{2j-1}\right) \!
    \right) , \IEEEnonumber
\end{IEEEeqnarray}
where $f$ and $g$ are defined in \eqref{eq: definition of f fucntion} and \eqref{eq: definition of g fucntion}, respectively.}
The starting condition for this recursion is
\full{\begin{IEEEeqnarray}{c}
		L_1^{(0)}(y) = \mathrm{LLR}(y) \triangleq \log_2 \left( W(y;0)/W(y;1) \right) \; .
    \label{eq: LLRs  basis condition} 
\end{IEEEeqnarray}}\ISIT{$L_1^{(0)}(y) = \log_2 (W(y;0)/W(y;1) )$.}

The SC decoder uses $f$ and $g$ to recursively calculate the LLRs of all synthetic joint distributions, yielding a decoding algorithm with running time $\mathcal{O}(N \log N)$.

The min-sum SC decoder is a simplified version of the original SC decoder, as it uses $\fminsum$ (see \eqref{eq: definition of min-sum f function}) instead of the computationally heavier $f$ during the recursion.
\full{A graphical comparison between these two functions if given in \Cref{fig: fvsfminsum}.
\begin{figure}
    \centering
    \includegraphics[scale=1]{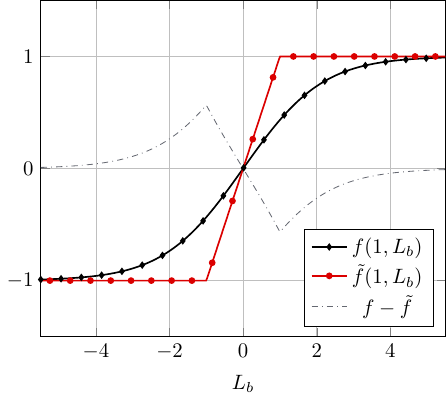}
    \caption{A comparison between the non-approximated function $f(L_a,L_b)$ and the approximated function $\fminsum(L_a,L_b)$ for $L_a=1$.}
    \label{fig: fvsfminsum}
\end{figure}}Unlike $f$, both $\fminsum$ and $g$ are positive homogeneous (i.e. multiplying both inputs by a positive constant multiplies the output by the same constant).
This implies that the min-sum decoder is not affected by scaling.
Therefore, we further extend the approximation and allow the initial labels at the base of the recursion not to be LLRs, but some values obtained by applying a labeling function $\labeler$ on the channel outputs.
We now list 3 properties required of a labeler $\labeler$ to be called a ``good labeler''.
\begin{definition}[Good labeler] A labeler $\labeler: \mathcal{Y} \to \reals$ is a good labeler with respect to a binary-input memoryless symmetric channel $W : \mathcal{X} \to \mathcal{Y}$ if the following holds:
	\begin{enumerate}
		\item Symmetry preservation: since $W$ is symmetric, there exists a permutation $\pi : \mathcal{Y} \to \mathcal{Y}$ such that for all $y \in \mathcal{Y}$, $W(y|1) = W(\pi(y) | 0)$ and $\pi(\pi(y)) = y$.
        We require that $\labeler(\pi(y)) = -\labeler(y)$ for all $y \in \mathcal{Y}$.
		\item \label{itm: sign consistency} Sign consistency: for all positive $t$ we have $\alpha_t \geq \alpha_{-t}$, where $\alpha_t = \sum_{y : \labeler(y) = t} W(y|0)$, and the inequality is strict for at least one $t$.
		\item Finite integer range: the range of $\labeler$ is contained in $\{-\gamma, -\gamma+1, \ldots, \gamma-1, \gamma\}$, for some positive integer $\gamma$.
	\end{enumerate}
	\label{def: good labeler}
\end{definition}
\full{Note that the requirement of a strict inequality in the second property rules out channels with capacity zero. }%
We also define a ``fair labeler'' as follows.
\begin{definition}[Fair labeler] A labeler $\labeler$ is a fair labeler if the first two requirements of a good labeler are met.
\label{def: fair labeler}
\end{definition}

Note that if we were to take $\labeler(y) = \mathrm{LLR}(y)$, we would have a fair labeler, for any channel with positive capacity.
The last property of the good labeler is required only for computational reasons, and is often the case due to quantization. The justification for it is by the homogeneous property of $\fminsum$ and $g$ and its implications, as described above. 

Under the \MSA, labels are calculated recursively by
\begin{IEEEeqnarray}{l}
    \subnumberinglabel{eq: transforms of labels}
    \tilde{L}_N^{(2j)}\left( y_0^{N-1},u_0^{2j-1}\right ) =
    \IEEEeqnarraynumspace \label{eq: minus transform of labels} \\
    \quad
    \fminsum \! \left(\!
    \tilde{L}_{N/2}^{(j)} \! \left( \! y_0^{N/2-1},u_{0,e}^{2j-1} \oplus u_{0,o}^{2j-1}\!\right)\!,\!
    \tilde{L}_{N/2}^{(j)}\! \left( \!  y_{N/2}^{N-1},u_{0,o}^{2j-1}\right) \!
    \right) , \IEEEnonumber
    \\
    \tilde{L}_N^{(2j+1)}\left( y_0^{N-1},u_0^{2j}\right ) =
    \IEEEeqnarraynumspace\label{eq: plus transform of labels} \\ 
    \quad
    g_{u_{2j}} \! \left( \!
    \tilde{L}_{N/2}^{(j)} \! \left( \! y_0^{N/2-1},u_{0,e}^{2j-1} \oplus u_{0,o}^{2j-1}\!\right)\!,\!
    \tilde{L}_{N/2}^{(j)} \! \left( \! y_{N/2}^{N-1},u_{0,o}^{2j-1}\right) \!
    \right) , \IEEEnonumber
\end{IEEEeqnarray}
\full{where $\fminsum$ and $g$ are defined in \eqref{eq: definition of min-sum f function} and \eqref{eq: definition of g fucntion}, respectively. Note the similarity between \eqref{eq: transforms of llrs} and \eqref{eq: transforms of labels}. As opposed to the starting condition \eqref{eq: LLRs basis condition} for the non-approximated setting, the starting condition under the \MSA is
\begin{IEEEeqnarray}{c}
    \tilde{L}_1^{(0)}(y) = \labeler(y) \; .
    \label{eq: labels basis condition} 
\end{IEEEeqnarray}
} 
\ISIT{with starting condition $\tilde{L}_1^{(0)}(y) = \labeler(y)$.}
\full{\par The use of a labeling function $\labeler$ is beneficial in practice, since it allows us to avoid the estimation of unknown channel parameters.
For example, consider the case of an AWGN channel with unknown noise level $\sigma^2$.
Thus, the LLR of output symbol $y$ is given by $2y/\sigma^2$. However, using $\lambda(y) = \mathrm{LLR}(y) = 2y/\sigma^2$ will give exactly the same results as using $\lambda(y) = \alpha \cdot y$, where $\alpha > 0$ is some fixed positive constant. The utility of the latter fair labeling function is that $\sigma^2$ need not be estimated. In practice, we use the following good labeler, which is a quantized version of the previous fair labeler,
\begin{IEEEeqnarray*}{rCl}
\labeler(y)=
    \begin{cases}
        \sign(y) \cdot \ourfloor{\alpha \cdot |y|} & \text{if } |y| < \gamma/\alpha \; , \\
         \sign(y) \cdot\gamma & \text{if } |y| \geq \gamma/\alpha \; .
    \end{cases}     
\end{IEEEeqnarray*}
We optimize $\alpha$ and $\gamma$ to work well over the range $\sigma^2$ is likely to belong to.}
\section{Posynomial representation}
\label{sec: posynomial representation}
For a fair labeler, we now define the synthetic joint distributions (on the label $t$ and input $u_i$) at stage $i$ of the SC decoder and min-sum SC decoder. These are, respectively,
\begin{IEEEeqnarray}{rCl}
    Q_N^{(i)}(t;u_{i}) &\triangleq& \!\!\!\! \sum_{\substack{y_0^{N-1},u_0^{i-1}:\\ L_N^{(i)}(y_0^{N-1},u_0^{i-1}) = t}} W_N^{(i)}(y_0^{N-1},u_0^{i-1};u_{i}) \; ,
    \label{eq: definition of Q joint-distribution}
    \\[5pt]
    \tilde{Q}_N^{(i)}(t;u_{i}) &\triangleq& \!\!\!\! \sum_{\substack{y_0^{N-1},u_0^{i-1}:\\ \tilde{L}_N^{(i)}(y_0^{N-1},u_0^{i-1}) = t}} W_N^{(i)}(y_0^{N-1},u_0^{i-1};u_{i}) \; .
    \label{eq: definition of Q-tilde joint-distribution}
\end{IEEEeqnarray}%
\ifISIT
Denote by $\tilde{\mathcal{T}}_N^{(i)}$ the support of $\tilde{Q}_N^{(i)}(t;u_i)$ with respect to $t$. 
In the setting of a good labeler, by definition, $\mathcal{\tilde{T}}_1^{(0)} \subseteq \{-\gamma, \ldots, \gamma\}$.
As will soon become apparent, under this setting,
\begin{equation}
	\tilde{\mathcal{T}}_N^{(i)} \subseteq \{-2^{\mathrm{wt}(i)} \cdot \gamma, \ldots, 2^{\mathrm{wt}(i)} \cdot \gamma\} \;  ,
	\label{eq: tilde TNi}
\end{equation}
\else
Denote by $\mathcal{T}_N^{(i)}$ and $\tilde{\mathcal{T}}_N^{(i)}$ the support of $Q_N^{(i)}(t;u_i)$ and $\tilde{Q}_N^{(i)}(t;u_i)$, respectively, with respect to $t$. 
\fi
\ifISIT
and $\mathrm{wt}(i)$ is the Hamming weight of the vector whose entries are the binary representation of $i$. Indeed, this follows by \eqref{eq: minus transform of Q-tilde joint-distribution} and \eqref{eq: plus transform of Q-tilde joint-distribution} below, and the definitions of $\fminsum$ and $g$ in \eqref{eq: definition of min-sum f function} and \eqref{eq: definition of g fucntion}.
\fi

Using the relations in \ISIT{\eqref{eq: minus transform of W joint-distribution}--\eqref{eq: transforms of labels}}\full{\eqref{eq: minus transform of W joint-distribution}--\eqref{eq: labels basis condition}}, we obtain the following minus and plus transforms of synthetic \ISIT{min-sum }joint distributions.
\begin{lemma}[Transforms of synthetic joint distributions]
    \label{lemm: recursive relation on Q-tilde joint-distribution}
    \begin{IEEEeqnarray}{l}
        \subnumberinglabel{eq: Transforms of Q-tilde joint-distributions}
        \tilde{Q}_N^{(2j)}(t;u_{2j}) = 
        \label{eq: minus transform of Q-tilde joint-distribution} \\
        \quad\quad\quad\; \sum_{\substack{t_a,t_b,u_{2j+1}:\\ \fminsum(t_a,t_b) = t}} \!\!\! \tilde{Q}_{N/2}^{(j)}(t_a; u_{2j} \oplus u_{2j+1}) \cdot \tilde{Q}_{N/2}^{(j)}(t_b; u_{2j+1}) \; , \IEEEnonumber \\
        \tilde{Q}_N^{(2j+1)}(t;u_{2j+1}) =
        \label{eq: plus transform of Q-tilde joint-distribution} \\
        \quad\quad\quad\; \sum_{\substack{t_a,t_b,u_{2j}:\\ g_{u_{2j}}(t_a,t_b) = t}} \!\!\! \tilde{Q}_{N/2}^{(j)}(t_a; u_{2j} \oplus u_{2j+1}) \cdot \tilde{Q}_{N/2}^{(j)}(t_b; u_{2j+1}) \; . \IEEEnonumber
    \end{IEEEeqnarray}
    \full{The above continues to hold if we remove all tildes.}
\end{lemma}
The following lemma ensures that symmetry holds for the min-sum synthetic distributions.
\begin{lemma}[Symmetry of synthetic joint distribution]
    \label{lemm: symmetry of Q-tilde joint-distribution}
    \begin{IEEEeqnarray}{c}
        \tilde{Q}_N^{(i)}(t;u_{i}) = \tilde{Q}_N^{(i)}(-t;u_{i} \oplus 1) \; .
        \label{eq: symmetry of Q-tilde joint-distribution}
    \end{IEEEeqnarray}
    \full{The above continues to hold if we remove all tildes.}
\end{lemma}
\ifISIT
\else
	We now give intuition as to why a setting in which the \MSA and a good labeler are used is much easier in terms of exactly calculating quantities of interest, as opposed to the non-approximated setting.

	In the non-approximated setting, for the minus transform we have that
	\begin{equation}
		\label{eq:T minus size bound}
		|\mathcal{T}_N^{(2j)}| \leq |\mathcal{T}_{N/2}^{(j)}|^2 \; .
	\end{equation}
	Indeed, this follows since in \eqref{eq: minus transform of Q-tilde joint-distribution} with the tildes removed we sum over all $(t_a, t_b) \in \mathcal{T}_{N/2}^{(j)} \times \mathcal{T}_{N/2}^{(j)}$ which determine $t = f(t_a,t_b)$, and also over $u_{2j+1} \in \{0,1\}$, which does not appear in $f(t_a,t_b)$. For the plus transform, we have by inspection of \eqref{eq: plus transform of Q-tilde joint-distribution} with the tildes removed that $|\mathcal{T}_N^{(2j+1)}| \leq 2 \cdot |\mathcal{T}_{N/2}^{(j)}|^2$. In fact, for a symmetric channel, this can be strengthened to 
	\begin{equation}
		\label{eq:T plus size bound}
		|\mathcal{T}_N^{(2j+1)}| \leq |\mathcal{T}_{N/2}^{(j)}|^2 \; .
	\end{equation}
Indeed, by \cref{lemm: symmetry of Q-tilde joint-distribution}, $t \in \mathcal{T}_{N/2}^{(j)}$ iff $-t \in \mathcal{T}_{N/2}^{(j)}$.
	Thus, \eqref{eq:T plus size bound} follows by inspection of $g$ in \eqref{eq: definition of g fucntion}. Hence, by \eqref{eq:T minus size bound}, \eqref{eq:T plus size bound}, and a straightforward induction,
\begin{equation}
	|\mathcal{T}_N^{(i)}| \leq |\mathcal{T}_1^{(0)}|^{2^{\mathrm{wt}(i)} } \; ,
	\label{eq: TNi size bound}
\end{equation}
where $\mathrm{wt}(i)$ is the Hamming weight of the vector whose entries are the binary representation of $i$.

	In contrast, consider the case of the \MSA and a good labeler. By definition, $\mathcal{\tilde{T}}_1^{(0)} \subseteq \{-\gamma, \ldots, \gamma\}$. By inspection of \eqref{eq: definition of g fucntion}, \eqref{eq: definition of min-sum f function}, \eqref{eq: Transforms of Q-tilde joint-distributions}, and a straightforward induction, it follows that
\begin{equation}
	\tilde{\mathcal{T}}_N^{(i)} \subseteq \{-2^{\mathrm{wt}(i)} \cdot \gamma, \ldots, 2^{\mathrm{wt}(i)} \cdot \gamma\} \;  .
	\label{eq: tilde TNi}
\end{equation}
This is the reason we can carry out our calculations efficiently in this case: as opposed to \eqref{eq: TNi size bound}, the size  of $\tilde{\mathcal{T}}_N^{(i)}$ grows linearly with $N$, since $\mathrm{wt}(i)$ is at most $n$ and $N = 2^n$.

\fi
\ISIT{The above symmetry implies that the probability of error at the $i$-th stage of the min-sum SC decoder, when aided by a genie that reveals the correct values of $u_0^{i-1}$, is given by}%
\full{A further consequence of the symmetry is \cref{lemm: symmetry of Q-tilde joint-distribution} is the following lemma. It gives a simple expression for the probability of error at the $i$-th stage of the min-sum SC decoder when aided by a genie that reveals the correct values of $u_0^{i-1}$.}
\ifISIT
\begin{IEEEeqnarray}{c}
    \ourPe\left(\tilde{Q}_N^{(i)}\right) =  \tilde{Q}_N^{(i)}(0;0) + 2 \cdot \sum_{t < 0}\tilde{Q}_N^{(i)}(t;0) \; .
    \label{eq: pe using Q-tilde joint-distribution}
\end{IEEEeqnarray}
\else
\begin{lemma}
	\label{lemm: pe using Q-tilde joint-distribution}
\begin{IEEEeqnarray}{c}
    \ourPe\left(\tilde{Q}_N^{(i)}\right) = \tilde{Q}_N^{(i)}(0;0) + 2 \cdot \sum_{t < 0}\tilde{Q}_N^{(i)}(t;0) \; .
    \label{eq: pe using Q-tilde joint-distribution}
\end{IEEEeqnarray}
\end{lemma}
\fi
To derive a Bhattacharyya-like upper bound on $\ourPe$, and to aid in notation in general, we abuse notation and define the following posynomial, in the indeterminate $\xi$:
\begin{IEEEeqnarray}{c}
    \tilde{Q}_N^{(i)}(\xi) \triangleq \sum_t \tilde{Q}_N^{(i)}(t;0) \cdot \xi^t \; .
    \label{eq: posynomial definition}
\end{IEEEeqnarray}
The above is indeed a posynomial: all the coefficients are non-negative as they are probabilities, while $t$ is not restricted to non-negative numbers.
\full{\par
	The following corollary justifies why in \eqref{eq: posynomial definition} we define the posynomial $\tilde{Q}_N^{(i)}(\xi)$ without taking into account terms of the form $\tilde{Q}_N^{(i)}(t;1)$.
\begin{corollary}
    \label{cor: symmetry of posynomial}
    $\tilde{Q}_N^{(i)}(t;1)$ equals the coefficient of $\xi^t$ in $\tilde{Q}_N^{(i)}(1/\xi)$.
\end{corollary}
}
We further define the following:
\begin{IEEEeqnarray}{c}
	\ourZ\left(\tilde{Q}_N^{(i)},\xi\right) \triangleq 2\cdot \tilde{Q}_N^{(i)}(\xi) \; .
    \label{eq: z-ish definition}
\end{IEEEeqnarray}
Our upper bound on $\ourPe$ is presented in the following lemma.
\begin{lemma}[Bhattacharyya-like bound]    \label{lemm: Bhattacharyya-like bound}
    For $0 < \xi_0 \leq 1$, 
    \begin{IEEEeqnarray}{c}	    
        \ourPe\left(\tilde{Q}_N^{(i)}\right) \leq \ourZ\left(\tilde{Q}_N^{(i)},\xi_0\right) \; .
        \label{eq: Bhattacharyya-like bound}
    \end{IEEEeqnarray}
\end{lemma}
We remark that setting $\xi_0 = 1/\sqrt{2}$ and removing the tildes yields the Bhattacharyya bound on the error probability at the $i$-th stage of the non-approximated genie-aided SC decoder.
Also, we may optimize over $\xi_0$ to yield the tightest upper bound, denoted
\begin{IEEEeqnarray}{c}
    \ourZopt\left(\tilde{Q}_N^{(i)}\right) \triangleq \min_{0 < \xi_0 \leq 1} \ourZ \left(\tilde{Q}_N^{(i)},\xi_0\right) \; .
    \label{eq: Z-ish-opt definition}
\end{IEEEeqnarray}
The above optimization is an instance of geometric programming, and can thus be efficiently computed \cite[Section 4.5]{BoydVandenberghe:04b}. 

The following shows that the evolution of $\ourZ$ and $\ourZopt$ is similar to the evolution of the Bhattacharyya parameter in the non-approximated setting.
\begin{lemma}[Bhattacharyya-like evolutions]
\label{lemm: Bhattacharyya-like evolutions}
For $0 < \xi_0 \leq 1$ and $0 \leq j < N/2$ we have
	\begin{IEEEeqnarray}{rCl} 
        \subnumberinglabel{eq: Z-ish evolutions}
            \ourZ\left(\tilde{Q}_{N}^{(2j)},\xi_0\right) & \leq & 2 \cdot \ourZ\left(\tilde{Q}_{N/2}^{(j)},\xi_0\right) \; ,
            \label{eq: Z-ish minus evolution} \\
            \ourZ\left(\tilde{Q}_{N}^{(2j+1)},\xi_0\right) & = & \left(\ourZ\left(\tilde{Q}_{N/2}^{(j)},\xi_0\right) \right)^2 \; .
            \label{eq: Z-ish plus evolution}
	\end{IEEEeqnarray}
	Furthermore,
	\begin{IEEEeqnarray}{rCl}
        \subnumberinglabel{eq: Z-ish-opt evolutions}
		\ourZopt\left(\tilde{Q}_{N}^{(2j)}\right) & \leq & 2 \cdot \ourZopt\left(\tilde{Q}_{N/2}^{(j)}\right) \; ,
            \label{eq: Z-ish-opt minus evolution} \\
		\ourZopt\left(\tilde{Q}_{N}^{(2j+1)}\right) & = & \left(\ourZopt\left(\tilde{Q}_{N/2}^{(j)}\right) \right)^2 \; .
            \label{eq: Z-ish-opt plus evolution}
	\end{IEEEeqnarray}
\end{lemma}

To prove the above, we state the following two lemmas.
\begin{lemma}[Bound on posynomial minus transform]
\label{lemm: posynomial minus transform bound}
For all $0 < \xi_0 \leq 1$ we have
\begin{IEEEeqnarray}{c}
    \tilde{Q}_N^{(2j)}(\xi_0) \leq 2 \cdot \tilde{Q}_{N/2}^{(j)}(\xi_0) \; .
    \label{eq: posynomial minus transform bound}
\end{IEEEeqnarray}
\end{lemma}

\begin{lemma}[Posynomial plus transform]
\label{lemm: posynomial plus transform}
\begin{IEEEeqnarray}{c}
    \tilde{Q}_N^{(2j+1)}(\xi) = 2\cdot\left(\tilde{Q}_{N/2}^{(j)}(\xi) \right)^2 \; .
    \label{eq: posynomial plus transform}
\end{IEEEeqnarray}
\end{lemma}

The previous lemma implies that the coefficients of $\tilde{Q}_N^{(2j+1)}(\xi)$ can be calculated efficiently from those of $\tilde{Q}_{N/2}^{(j)}(\xi)$. We now show an analogous result for $\tilde{Q}_N^{(2j)}(\xi)$.
In aid of this, we define the ``above'' and ``below'' posynomials:
\begin{IEEEeqnarray}{rCl}
	\tilde{A}_N^{(i)}(\xi) & \triangleq & \sum_{t \in \tilde{\mathcal{T}}_N^{(i)}} \left( \sum_{t'>t} \tilde{Q}_N^{(i)}(t';0) \right) \cdot \xi^t \; , 
        \label{eq: definition of A posynomial}\\
	\tilde{B}_N^{(i)}(\xi) & \triangleq & \sum_{t \in \tilde{\mathcal{T}}_N^{(i)}} \left( \sum_{t'<t} \tilde{Q}_N^{(i)}(t';0) \right) \cdot \xi^t \; .
        \label{eq: definition of B posynomial}
\end{IEEEeqnarray}
Namely, if we write out $\tilde{Q}_{N}^{(i)}(\xi)$ in ascending order of powers of $\xi$, then the coefficient of $\xi^t$ in $\tilde{A}_N^{(i)}(\xi)$ (resp.\ $\tilde{B}_N^{(i)}(\xi)$) is the sum of the coefficients strictly above (resp. below) the monomial $\tilde{Q}_N^{(i)}(t;0) \xi^t$.

Let $\Gamma(\xi)$ and $\Lambda(\xi)$ be two posynomials. Denote by $[\xi^t] \; \Gamma(\xi)$ the coefficient of $\xi^t$ in $\Gamma(\xi)$. Define the ``positive'' and ``negative'' operators, and Hadamard (element-wise) product: these operators return posynomials, where for all $t$,
\begin{IEEEeqnarray}{rCl}
	{[\xi^t]} \; \ourpos{\Gamma(\xi)} &=&
	\begin{cases}
		[\xi^t] \; \Gamma(\xi) & t \geq 0 \; , \\
		[\xi^{-t}] \; \Gamma(\xi) & t < 0 \; .
	\end{cases}
        \label{eq: definition of pos operator} \\
	{[\xi^t]} \; \ourneg{\Gamma(\xi)} &=&
	\begin{cases}
		[\xi^t] \; \Gamma(\xi) & t \leq 0 \; , \\
		[\xi^{-t}] \; \Gamma(\xi) & t > 0 \; .
	\end{cases}
        \label{eq: definition of neg operator} \\
	{[\xi^t]} \; \big( \Gamma(\xi) \odot \Lambda(\xi) \big) &=& \left( [\xi^t] \; \Gamma(\xi) \right) \cdot  \left( [\xi^t] \; \Lambda(\xi) \right) \; .
        \label{eq: definition of Hadamard product}
\end{IEEEeqnarray}

\begin{lemma}[Posynomial minus transform]
\label{lemm: posynomial minus transform}
\begin{IEEEeqnarray}{rCl}
    \IEEEeqnarraymulticol{3}{l}{\tilde{Q}_N^{(2j)}(\xi) = }   \IEEEnonumber \\
\quad &&\! \phantom{{} + {} }  2 \left(\tilde{Q}_{N/2}^{(j)}(\xi) \odot \left( 2\cdot \ourpos{\tilde{A}_{N/2}^{(j)}(\xi)} + \tilde{Q}_{N/2}^{(j)}(\xi) \right) \right) \IEEEnonumber \\
      && \! {} + 2\left(\tilde{Q}_{N/2}^{(j)}(1/\xi) \odot \left(2 \cdot \ourneg{\tilde{B}_{N/2}^{(j)}(\xi)} + \tilde{Q}_{N/2}^{(j)}(1/\xi) \right) \right) \IEEEnonumber \\
          && \! {} - 2 \left([\xi^0] \; \tilde{Q}_{N/2}^{(j)}(\xi) \right)^2 \; .
          \label{eq: posynomial minus transform}
\end{IEEEeqnarray}
\end{lemma}

\section{Finite-Length Case}
\label{sec: finite length case}
In this section, we assume a good labeler. For the finite length case, our aim is to calculate $\ourPe\left(\tilde{Q}_N^{(i)}\right)$ for all $0 \leq i < N$, where the codeword length is $N=2^n$.
The expression for this is given in \eqref{eq: pe using Q-tilde joint-distribution}, which we can recast using \eqref{eq: definition of neg operator} as
\begin{equation}
	\ourPe\left(\tilde{Q}_N^{(i)}\right) = \left.\ourneg{\tilde{Q}_N^{(i)}(\xi)}\right|_{\xi=1} \; .
	\label{eq: pe using posynomial}
\end{equation}
We use \eqref{eq: posynomial plus transform} and \eqref{eq: posynomial minus transform} to calculate $\tilde{Q}_N^{(i)}(\xi)$ for all $i$, and then apply \eqref{eq: pe using posynomial} to yield the error probability.
The following two lemmas specify the complexity of calculating $\tilde{Q}_N^{(2j)}(\xi)$ and $\tilde{Q}_N^{(2j+1)}(\xi)$ from $\tilde{Q}_{N/2}^{(j)}(\xi)$.
Namely, the complexity of calculating all the coefficients of the former, given all the coefficients of the latter.
Recall that $\mathcal{\tilde{T}}_N^{(i)}$ is defined in \eqref{eq: tilde TNi}. 
\full{\begin{lemma}[Complexity of posynomial minus transform]}
	\ISIT{\begin{lemma}}
    \label{lemm: complexity of calc minus posynomial}
    The complexity of calculating $\tilde{Q}_N^{(2j)}(\xi)$ from $\tilde{Q}_{N/2}^{(j)}(\xi)$ is $\bigO\left(|\mathcal{\tilde{T}}_{N/2}^{(j)}|\right)$. 
\end{lemma}
\full{\begin{lemma}[Complexity of posynomial plus transform]}
\ISIT{\begin{lemma}}
    \label{lemm: complexity of calc plus posynomial}
    The complexity of calculating $\tilde{Q}_N^{(2j+1)}(\xi)$ from $\tilde{Q}_{N/2}^{(j)}(\xi)$ is $\bigO\left(|\mathcal{\tilde{T}}_{N/2}^{(j)}|\cdot \log(|\mathcal{\tilde{T}}_{N/2}^{(j)}|)\right)$. 
\end{lemma}

The following theorem is our main result for this section.
It shows that the complexity of calculating all the probabilities of error $\ourPe\left(\tilde{Q}_N^{(i)}\right)$ is polynomial in the codeword length $N$ and in $\gamma$ (recall \Cref{def: good labeler}).
\begin{theorem}[Total complexity of evaluating $\ourPe$]
    \label{thm: complexity of finite length case}
    When using a good labeler $\labeler$, the complexity of calculating $\ourPe\left(\tilde{Q}_N^{(i)}\right)$ for all $0 \leq i < N$ is $\bigO( N^{\log_2 3} \log N \cdot \gamma \log \gamma )$. We simplify this to $\bigO( N^{1.585} \cdot \gamma \log \gamma )$.
\end{theorem}

\section{Asymptotic Case}
\label{sec: asymptotic case}
In this section we prove \Cref{thm: threshold rates for the asymptotic case}.
We first do so assuming a fair labeler, and then show how to significantly improve the thresholds $\ourRl$ and $\ourRu$ for the case of a good labeler.
The following three results are required for deriving $\ourRl$. 
\begin{proposition}
 	\label{prop: fromUniversal}
	Let $B_1, B_2, \ldots$ be i.i.d.\ random variables such that $\Pr(B_i=0)=\Pr(B_i=1)=1/2$. Let $S_0, S_1, \ldots$ be a $[0,1]$-valued random process that satisfies
\begin{IEEEeqnarray}{c}
	S_{n+1} \leq \kappa \cdot
    	\begin{cases}
		S_n, & B_{n+1}=0 \; , \\
       		S^2_n, &  B_{n+1}=1 \; ,
   	 \end{cases}
	 \quad n \geq 0 \; .
	\label{eq: evolution contidion on z}
\end{IEEEeqnarray}
Then, for every $\epsilon'>0$ and $\delta'>0$ there exist $n' = n'(\epsilon',\delta',\kappa)$ and $\eta = \eta(\epsilon',\delta',\kappa) > 0$ such that if $S_0 \leq \eta$ then
\begin{IEEEeqnarray}{c}
	\label{eq: fromUniversal}
	\Pr \left( S_n \leq \epsilon' \textup{ for all } n \geq n' \right) \geq 1 - \delta' \; .
\end{IEEEeqnarray}
\end{proposition}
This is \cite[Equation 171]{ShuvalTal:25aa}, and is the crux of proving \cite[Proposition 49]{ShuvalTal:25aa}.
\full{\par}%
The expression for $\eta$ is given in the penultimate displayed equation in \cite[Appendix A]{ShuvalTal:25aa}, where $r$ is defined slightly before as the largest positive solution of $\kappa^r +(2\kappa)^{-r} = 2$.
In our setting, $S_n$ will be related to $\ourZopt$.
Thus, by \eqref{eq: Z-ish-opt evolutions}, we specialize to $\kappa = 2$.
Plugging $x=2^r$ into $2^r +4^{-r} = 2$ yields $x+ 1/x^2  = 2$.
The three roots of this equation are $1$, $\varphi$, and $-\varphi^{-1}$, where $\varphi =\frac{1}{2}\cdot \left(1 +\sqrt{5}\right)$ is the golden ratio.
Thus, $r = \log_2(\varphi)$ and %
\ISIT{$\eta(\delta')=\frac{1}{8} \cdot (\delta'/2)^{1/\log_2\varphi}$.}%
\full{\begin{equation}
 	\eta(\delta')=\frac{1}{8} \cdot (\delta'/2)^{1/\log_2\varphi} \; .
	\label{eq: eta of delta}
\end{equation}}

The following result is an immediate corollary.
\begin{corollary}
	\label{corollary: of universal}
	Let $S_0,S_1,\ldots$ be as in \Cref{prop: fromUniversal}, with $\kappa =2$. Fix $\epsilon'>0$ and $\eta > 0$. Then there exists $n' = n'(\epsilon',\eta)$ such that if $S_0 \leq \eta$ then
	\begin{equation}
		\label{eq: of universal}
	\Pr \left( S_n \leq \epsilon' \textup{ for all } n \geq n' \right) \geq 1 - \delta'(\eta) \; ,
\end{equation}
\ISIT{\vspace{-1pt} where \vspace{-4pt}}\full{where}
\begin{equation}
	\delta'(\eta) \triangleq 2 \cdot (8 \eta)^{\log_2 \varphi} \quad \mbox{and} \quad \varphi =(1 + \sqrt{5})/2 \; .
\label{eq: delta' definition}
\end{equation}
\end{corollary}

The following result is of primary importance and will be used directly to prove \Cref{thm: threshold rates for the asymptotic case}.
\begin{proposition}
\label{prop: augmented proposition}
Let $S_0, S_1, \ldots$ be as in \Cref{prop: fromUniversal}, with $\kappa = 2$.
Fix $0<\beta<1/2$, $\eta>0$, and $\delta > \delta'(\eta)$, where $\delta'(\eta)$ is given in \eqref{eq: delta' definition}. Then, there exists $n_0 = n_0\left(\beta, \delta - \delta'(\eta)\right)$ such that if $S_0 \leq \eta$ then
\begin{equation}
    \label{eq: augmented proposition}
    \Pr \left( S_n \leq 2^{-2^{n \beta}} \textup{ for all } n \geq n_0 \right) \geq 1 - \delta \; . 
\end{equation}
\end{proposition}

\subsection{Fair Labeler}
\label{subsec: fair labeler}
\begin{IEEEproof}[Proof of \Cref{thm: threshold rates for the asymptotic case}]
For $\ourRu$, we first recall that any decoder operates on the output of $W$, after it has been labeled by $\labeler$.
Thus, it effectively sees the channel $\tilde{Q}(t|x) = 2 \cdot \tilde{Q}_1^{(0)}(t;x)$, as defined in \eqref{eq: definition of Q-tilde joint-distribution}.
We take $\ourRu$ as the capacity of this channel, which is valid by the strong converse to the coding theorem, see \cite[Theorem 5.8.5]{Gallager:68b}.

We now work towards deriving $\ourRl$.
Consider a polar code of length $N = 2^n$ with non-frozen index set $\mathcal{A} = \{ 0\leq i <N: \ourZopt(\tilde{Q}_N^{(i)}) < 2^{-N^{\beta'}} \}$, where $\beta' = \frac{\beta + 1/2}{2}$.
By the ``genie-aided decoder'' argument in \cite{Arikan:09p}, the union bound, and \Cref{lemm: Bhattacharyya-like bound}, the error probability of such a code is at most
\(
    |\mathcal{A}|\cdot 2^{-N^{\beta'}} \leq N\cdot 2^{-N^{\beta'}} < 2^{-N^\beta} \; ,
\)
where the last inequality holds for $N$ large enough.
Thus, we must find an $\ourRl$ such that for $R < \ourRl$ fixed and all $N$ large enough, $|\mathcal{A}| \geq N \cdot R$.
Consider the set $\mathcal{A}' = \{0 \leq i < N : \ourzeta_N^{(i)} < 2^{-N^{\beta'}} \}$, where $\ourzeta_1^{(0)} = \ourZopt(\tilde{Q}_1^{(0)})$ and
\begin{IEEEeqnarray}{c}
    \ourzeta_N^{(i)} =
    \begin{cases}
        2 \cdot \ourzeta_{N/2}^{(i/2)} & i\;  \mbox{is even}, \\
        \left(\ourzeta_{N/2}^{((i-1)/2)}\right)^2 & i\; \mbox{is odd}.
    \end{cases}
    \label{eq: zeta recursive}
\end{IEEEeqnarray}

By \eqref{eq: Z-ish-opt evolutions}, we have for all $i$ that $\ourzeta_N^{(i)} \geq \ourZopt(\tilde{Q}_N^{(i)})$.
Namely, $\mathcal{A}' \subseteq \mathcal{A}$.
Thus, it suffices to find an $\ourRl$ such that for $R < \ourRl$ fixed and all $N$ large enough, $|\mathcal{A}'| \geq N \cdot R$.
For any $M = 2^m$, we use the definition of $\delta'(\cdot)$ in \eqref{eq: delta' definition} and define the following:
\begin{IEEEeqnarray}{c}
    \label{eq: fair definition of RL}
    \ourRl(M) \triangleq \frac{1}{M} \sum_{j=0}^{M-1} \max \left\{1 - \delta'(\ourzeta_M^{(j)}), 0\right\} \; .
\end{IEEEeqnarray}

Proving the following two items will complete the proof:
\begin{enumerate}
\item For a given $M$ and $R<\ourRl(M)$ there exists $n_0$ such that for all $n \geq n_0$ we have $|\mathcal{A}'| \geq N \cdot R$.
\item There exists an $M$ such that $\ourRl(M) > 0$.
\end{enumerate}

To prove the first item, assume that $R$, and therefore $\ourRl(M)$, are positive, otherwise the claim is trivial.
For each one of the $M$ indices $0 \leq j < M$, we invoke \Cref{prop: augmented proposition} with $\delta = \delta'(\zeta_M^{(j)}) + (\ourRl(M) - R)$, $\eta = S_0 = \zeta_M^{(j)}$, and $\beta'' = \frac{\beta' + 1/2}{2}$ in place of $\beta$. 
Denote the $n_0$ promised by the proposition as $n_0^{(j)}$.
Now define $n_0^{\mathrm{max}}=\max_{j}n_0^{(j)}$.
By \eqref{eq: augmented proposition}, for $n \geq m + n_0^{\mathrm{max}}$ the fraction of indices $ 0 \leq i < N$ such that $\zeta_N^{(i)} \leq 2^{-2^{(n-m)\beta''}}$ is at least
\begin{IEEEeqnarray*}{rCl}
	\IEEEeqnarraymulticol{3}{l}{\frac{1}{M} \sum_{j=0}^M \max\left\{ 1- \left( \delta'(\zeta_M^{(j)}) + \ourRl(M) - R \right), 0 \right\}} \\
	\full{& = &\frac{1}{M} \sum_{j=0}^M \max\left\{ 1- \delta'(\zeta_M^{(j)}) + R - \ourRl(M) , 0 \right\} \\}
	& \geq &\frac{1}{M} \sum_{j=0}^M \max\left\{ 1- \delta'(\zeta_M^{(j)}) + R - \ourRl(M), R - \ourRl(M) \right\} \\
	      & = &\frac{1}{M} \sum_{j=0}^M \max\left\{ 1- \delta'(\zeta_M^{(j)}),0  \right\} + R - \ourRl(M) \\
	& = & R \; .
\end{IEEEeqnarray*}
For the first item to hold, we take $n_0 \geq m + n_0^{\mathrm{max}}$ large enough such that for all $n \geq n_0$ we have $2^{-2^{(n-m)\beta''}} \leq 2^{-2^{n\beta'}} = 2^{-N^{\beta'}}$ (ensuring $|\mathcal{A}'| \geq N\cdot R$) and $N \cdot 2^{-N^{\beta'}} < 2^{-N^\beta}$.

We now prove the second item.
That is, it is always possible to find an $M$ such that $\ourRl(M) > 0$.
We first show that $\ourZopt(\tilde{Q}_1^{(0)}) < 1$. Indeed, 
\[
    \left. \ourZ(\tilde{Q}_1^{(0)},\xi)\right|_{\xi = 1} = 1 \quad \mbox{and} \quad \left. \frac{d}{d \xi}\ourZ(\tilde{Q}_1^{(0)},\xi) \right|_{\xi = 1} > 1 \; ,
\]
where the inequality follows by item \ref{itm: sign consistency} in \Cref{def: good labeler}.
Hence, for $\xi_0 < 1$ sufficiently close to $1$ it must hold that $\ourZ(\tilde{Q}_1^{(0)},\xi_0) < 1$.
Thus, $\ourzeta_1^{(0)}=\ourZopt(\tilde{Q}_1^{(0)}) < 1$.
Next, note that $\ourzeta_M^{(M-1)} = \left(\ourzeta_1^{(0)}\right)^M$.
Take $M$ as the smallest power of $2$ that is at least $\log_a b$ where $a  = \ourzeta_1^{(0)}$ and $b = \eta(1/2) \approx 0.327254$.
For this choice, $\ourRl(M) \geq \frac{1}{2M} > 0$, by considering the last term in \eqref{eq: fair definition of RL}.
\end{IEEEproof}

\subsection{Good Labeler}
\label{subsec: good labeler}
We now show how both thresholds $\ourRl$ and $\ourRu$ can be strengthened in the case of a good labeler.
We give a simplified description here. We give a full and more nuanced description in the expanded version \ISIT{\cite{ChisnevskiTalShamai:25a}}\full{\cref{sec: improved thresholds}}.
For $\ourRl$, we observe the following regarding the proof of \Cref{thm: threshold rates for the asymptotic case}.
Any definition of $\ourzeta_N^{(i)}$ that satisfies $\ourzeta_N^{(i)} \geq \ourZopt(\tilde{Q}_N^{(i)})$ for all $0 \leq i < N$ is valid.
Thus, for a parameter $V = 2^v \leq M$, and all indices $0 \leq k < V$,  define $\ourzeta_V^{(k)} = \ourZopt(\tilde{Q}_V^{(k)})$.
For $N > V$, define $\zeta_N^{(i)}$ recursively according to \eqref{eq: zeta recursive}.
This improves $\ourRl(M)$, which we now denote as $\ourRl(V,M)$, since for polarization stage $v$ we are calculating the exact values of $\ourZopt(\tilde{Q}_V^{(k)})$, as opposed to bounds on them.
By \Cref{lemm: complexity of calc plus posynomial,lemm: complexity of calc minus posynomial}, we can indeed calculate $\tilde{Q}_V^{(k)}$ efficiently.

To strengthen $\ourRu$, we now define $\ourRu(V)$ as the average capacity of the channels corresponding to $\tilde{Q}_V^{(k)}$ over $0 \leq k < V$.
The proof of this threshold being valid is given in \ISIT{\cite{ChisnevskiTalShamai:25a}}\full{\cref{sec: improved thresholds}}.
In essence, we employ a so called ``block-genie'' that corrects us after $N/V$ decisions have been made.
Each block of size $N/V$ corresponds to $N/V$ uses of one of the above channels, and hence we cannot code for this block at a rate exceeding the capacity of that channel.

\ISIT{The two figures in this paper were derived with $V = 12$ and $M = 36$, and pruning as described in \cite{ChisnevskiTalShamai:25a}.}

\ifISIT
\clearpage
\bibliographystyle{IEEEtran}
\IEEEtriggeratref{15}
\bibliography{mybib.bib}
\fi
\section{Improved Thresholds}
\label{sec: improved thresholds}
In this section we give a full description of how $\ourRl$ and $\ourRu$ were calculated in \Cref{fig: awgnRateComparison,fig: bscRateComparison}. These methods can be applied to any setting in which a good labeler is used. 

\subsection{Definition of $\ourRl(\mathcal{G}, \mathcal{E})$ and $\ourRu(\mathcal{G})$}
\label{subsec: definition of Rl and Ru}
We start by defining two sets: $\mathcal{G}$ and $\mathcal{E}$. Both sets contain depth-index pairs $(d,j)$, where $d \geq 0$ and $0 \leq j < 2^d$.
We think of each pair in these sets as a vertex of a full binary tree\footnote{A binary tree in which each node has either two children or no children.}.
\begin{figure}
    \centering
    \includegraphics[scale=0.58]{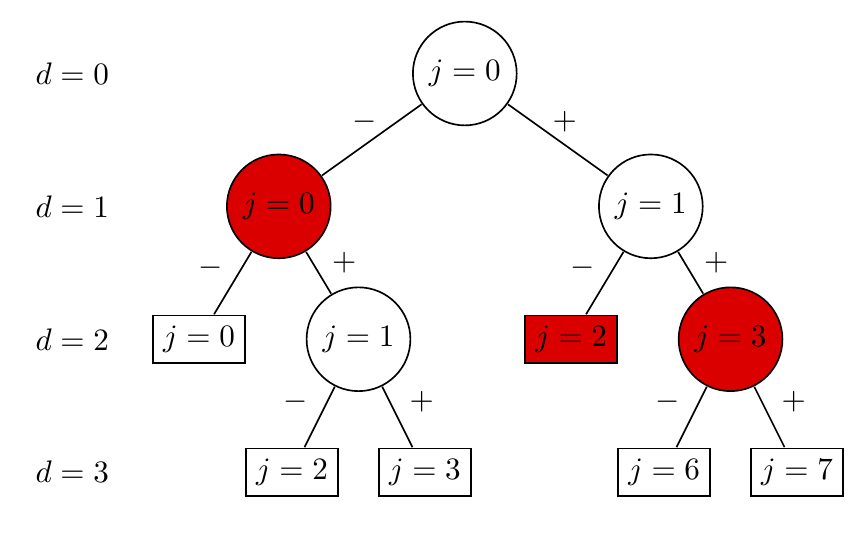}
    \caption{A full binary tree, where the nodes in $\mathcal{G}$ are red and the nodes in $\mathcal{E}$ are rectangular (leaves). The node $(d=2,j=2)$ is both in $\mathcal{G}$ and $\mathcal{E}$.}
    \label{fig: tree}
\end{figure}
An example of such a tree is presented in \cref{fig: tree}.
The root of the tree is $(d=0,j=0)$.
A vertex $(d,j)$ has either no children, in which case it is a leaf, or two children: $(d+1,2j)$ as the left child and $(d+1,2j+1)$ as the right child.
Hence, we can view $j$ as representing a path of $d$ edges labelled ``$-$'' and ``$+$'' starting at the root and ending at $(d,j)$. The binary representation of $j = \sum_{i=0}^{d-1} b_i 2^i$ dictates the corresponding path. Namely, $b_i = 0$ means that the $(d-i)$-th edge is a ``$-$'' (left) edge, while $b_i=1$ means that the $(d-i)$-th edge is a ``$+$'' (right) edge.

Given a full binary tree, the sets $\mathcal{E}$ and $\mathcal{G}$ are defined as follows.
The set $\mathcal{E}$ contains the leaves of the tree. We call such a set valid. The set $\mathcal{G}$ is defined such that any path from the root to a leaf contains exactly one vertex in $\mathcal{G}$. Such a pair $(\mathcal{G},\mathcal{E})$ is termed valid.
Note that if we were to delete all descendants of vertices in $\mathcal{G}$ from the tree, we would again have a full binary tree whose leaves are $\mathcal{G}$. Hence, if $(\mathcal{G},\mathcal{E})$ is a valid pair, then both $\mathcal{G}$ and $\mathcal{E}$ are valid.

For now, we assume that $\mathcal{G}$ and $\mathcal{E}$ are given (we will latter describe how to choose them). Our thresholds are now denoted $\ourRl(\mathcal{G}, \mathcal{E})$ and $\ourRu(\mathcal{G})$. For $\ourRl(\mathcal{G}, \mathcal{E})$, we generalize \eqref{eq: fair definition of RL} to
\begin{IEEEeqnarray}{c}
	\label{eq: general Rl}
	\ourRl(\mathcal{G}, \mathcal{E}) = \sum_{(d,j) \in \mathcal{E}} \frac{1}{2^d} \cdot \max \left\{1 - \delta'(\ourzeta_{2^d}^{(j)}), 0\right\} \; ,
\end{IEEEeqnarray}
where $\delta'$ is defined in \eqref{eq: delta' definition} and $\ourzeta_{2^d}^{(j)}$ is defined recursively in \eqref{eq: zeta recursive}, with the following starting conditions: 
\begin{equation}
	\label{eq: ourzeta starting condition}
	\ourzeta_{2^d}^{(j)} = \ourZopt(\tilde{Q}_{2^d}^{(j)}) \; , \quad \mbox{for all} \; (d,j) \in \mathcal{G} .
\end{equation}
Note that with respect to the simplified description in \cref{subsec: good labeler}, if  we define (for $v \leq m$):
\begin{IEEEeqnarray}{rCl}
	\mathcal{G}(V) &=& \left\{ (d,j) : d = v \quad \mbox{and} \quad 0 \leq j < V = 2^v \right\} \; ,\label{eq: GV} \\ 
	\mathcal{E}(M) &=& \left\{ (d,j) : d = m \quad \mbox{and} \quad 0 \leq j < M = 2^m \right\} \; , \IEEEnonumber
\end{IEEEeqnarray}
then $\ourRl(V,M) = \ourRl\left(\mathcal{G}(V),\mathcal{E}(M)\right)$. 
For $\ourRu(\mathcal{G})$, we have
\begin{IEEEeqnarray}{c}
	\label{eq: general Ru}
        \ourRu(\mathcal{G}) = \sum_{(d,j) \in \mathcal{G}} \frac{1}{2^d} \cdot I\left(\tilde{Q}_{2^d}^{(j)}\right) \; ,
\end{IEEEeqnarray}
where $I\left(\tilde{Q}_{2^d}^{(j)}\right)$ is the mutual information corresponding to the joint distribution $\tilde{Q}_{2^d}^{(j)}$. That is, the capacity of the channel corresponding to $\tilde{Q}_{2^d}^{(j)}$.
Note that with respect to the simplified description in \cref{subsec: good labeler}, $\ourRu(V) = \ourRu(\mathcal{G}(V))$. 

Computationally, given $\mathcal{G}$ and $\mathcal{E}$, the calculation of $\ourRl(\mathcal{G}, \mathcal{E})$ and $\ourRu(\mathcal{G})$ is implemented as follows. We carry out a pre-order scan of the tree starting from the root. That is, we scan the root, scan the subtree rooted at its left child recursively, and then scan the subtree rooted at its right child recursively. The first node scanned is thus the root, for which $\tilde{Q}_1^{(0)}(\xi)$ is given. Assume we are currently scanning a node $(d,j)$ which is not the root. Hence, this node has a parent, $(d',j')=(d-1,\lfloor j/2 \rfloor)$.
\begin{itemize}
	\item If the path from the root to $(d,j)$ has not yet traversed a vertex in $\mathcal{G}$, then by induction we have already calculated $\tilde{Q}_{2^{d'}}^{(j')}(\xi)$, and now calculate $\tilde{Q}_{2^{d}}^{(j)}(\xi)$ according to either \eqref{eq: posynomial plus transform} or \eqref{eq: posynomial minus transform}, depending on the parity of $j$.
		\begin{itemize}
			\item If $(d,j) \in \mathcal{G}$, then we also calculate $\ourZopt(\tilde{Q}_{2^{d}}^{(j)}(\xi))$ and set $\ourzeta_{2^d}^{(j)} = \ourZopt(\tilde{Q}_{2^d}^{(j)})$, in accordance with the starting condition \eqref{eq: ourzeta starting condition}.
		\end{itemize}
	\item If the path from the root to $(d,j)$ has already traversed a vertex in $\mathcal{G}$, then by induction we have already calculated $\ourzeta_{2^{d'}}^{(j')}$ and now calculate $\ourzeta_{2^{d}}^{(j)}$ according to \eqref{eq: zeta recursive}.
		\begin{itemize}
			\item If $(d,j) \in \mathcal{E}$, we do not recursively continue the scan, since we have reached a leaf.
		\end{itemize}
\end{itemize}

We now describe how we chose $\mathcal{G}$ and $\mathcal{E}$ and calculated $\ourRl$ and $\ourRu$ in \cref{fig: awgnRateComparison,fig: bscRateComparison}. We set parameters $d_\mathcal{G} = 12$ and  $d_\mathcal{E} = 36$ as the maximal depth of a vertex in $\mathcal{G}$ and $\mathcal{E}$, respectively. We further set a numeric threshold $\epsilon = 10^{-3}$ that allows us to add vertices to $\mathcal{G}$ and $\mathcal{E}$ at a depth shallower than $d_\mathcal{G}$ and $d_\mathcal{E}$, respectively, in case sufficient polarization has already occurred. Conceptually, we carry out a pre-order scan of a perfect binary tree\footnote{A full binary tree in which all the leaves are at the same depth.} of height $d_\mathcal{E}$, trimming it as we go along. That is, $\mathcal{G}$ and $\mathcal{E}$ are generated dynamically as the scan progresses. We initialize variables $\ourRl=\ourRu=0$. Each time a vertex is added to $\mathcal{G}$, $\ourRu$ is incremented according to \eqref{eq: general Ru}. Each time a vertex is added to $\mathcal{E}$, $\ourRl$ is incremented according to \eqref{eq: general Rl}.

During the scan of vertex $(d,j)$ as described in the itemed list above:
\begin{itemize}
	\item If the path from the root to $(d,j)$ has not yet traveresed a vertex in $\mathcal{G}$, we add $(d,j)$ to both $\mathcal{G}$ and $\mathcal{E}$ and increment $\ourRu$ and $\ourRl$ if
		\begin{itemize}
			\item $I(\tilde{Q}_{2^{d}}^{(j)}(\xi)) < \epsilon$, or
			\item $1 - \delta'(\ourZopt(\tilde{Q}_{2^{d}}^{(j)}(\xi)) ) > 1- \epsilon$.
		\end{itemize}
		Otherwise, we add $(d,j)$ to $\mathcal{G}$ and increment $\ourRu$ if $d = d_\mathcal{G}$.
	\item If the path from the root to $(d,j)$ has already traversed a vertex in $\mathcal{G}$, we add $(d,j)$ to $\mathcal{E}$ and increment $\ourRl$ if
		\begin{itemize}
			\item $1-\delta'(\ourzeta_{2^{d}}^{(j)}(\xi)) > 1-\epsilon$, or
			\item $\ourzeta_{2^{d}}^{(j)}(\xi) > 1$, or
			\item $d = d_\mathcal{E}$.
		\end{itemize}
\end{itemize}

The curves for $\ourRl, \ourRu$ and $C$ in \cref{fig: awgnRateComparison} are plotted with respect to a BI-AWGN channel quantized by a labeler $\labeler$ to have eight possible outputs.
At the input we assume a normalized BPSK mapping from $\mathcal{X}= \{0,1\}$ to $\mathcal{X}'=\{1,-1\}$ such that $x'=1-2x$.
At the output we assume that the labeler maps $\mathcal{Y}=\reals$ to $\{-4,-3,-2,-1,1,2,3,4\}$.
The channel is defined by the above two mappings and by the relation $y=x'+ \nu$, where $\nu$ is the realization of a Gaussian random variable with zero mean and variance $\sigma^2$.
The labeler is
\begin{IEEEeqnarray}{rCl}
    \labeler(y) =
    \left\{ \,
    \begin{IEEEeqnarraybox}[][c]{r?rCcCl}
	    \IEEEstrut
	    4 & q_3 & \leq & y\makebox[0cm]{$\;\;\; ,$}         \\
	    3 & q_2 &\leq& y & <  & q_3  \; ,  \\
	    2 & q_1 &\leq& y & < & q_2  \; ,  \\
	    1 & 0 &\leq& y & < & q_1  \; ,  \\
	    -1 & -q_1 &\leq& y & < & 0  \; ,  \\
	    -2 & -q_2 &\leq& y & < & -q_1  \; ,  \\
	    -3 & -q_3 &\leq& y & < & -q_2  \; ,  \\
	    -4 &      &    & y & < & -q_3  \; ,  
	    \IEEEstrut
	    \end{IEEEeqnarraybox}
\right.
    \label{eq: awgn labeler}
\end{IEEEeqnarray}
where we used $q_1=0.2$, $q_2=0.6$, and $q_3=1.2$ to define the labeler regions.
For \cref{fig: bscRateComparison} we have a BSC with $\mathcal{Y}=\{0,1\}$ and the labeler is  $\labeler(y)=1-2y$.
Note that both labelers above are good labelers.

We now state and prove two propositions that justify $\ourRl(\mathcal{G}, \mathcal{E})$ and $\ourRu(\mathcal{G})$ as valid thresholds. These are generalizations of claims and proofs made in \Cref{sec: asymptotic case} for simpler choices of $\ourRl$ and $\ourRu$.

\subsection{Justification of $\ourRl(\mathcal{G}, \mathcal{E})$}
\begin{proposition}
\label{prop: RlGE good}
	Setting $\ourRl = \ourRl(\mathcal{G}, \mathcal{E})$ in \Cref{thm: threshold rates for the asymptotic case} is valid.
\end{proposition}

\begin{IEEEproof}
Recall that in \Cref{thm: threshold rates for the asymptotic case} we assume that $R < \ourRl(\mathcal{G}, \mathcal{E})$, and our aim is to prove the existence of a family of polar codes with growing lengths such that their rates are at least $R$ and their word error probabilities at most $2^{-N^\beta}$, where $N$ is the codeword length.

As in \cref{subsec: fair labeler}, we use the recursive relation \eqref{eq: zeta recursive} to define $\ourzeta_N^{(i)}$, where now the starting conditions are $\ourzeta_{2^d}^{(j)} = \ourZopt(\tilde{Q}_{2^d}^{(j)})$ for $(d,j) \in \mathcal{G}$. Note that by the definition of $\mathcal{E}$ and $\mathcal{G}$ and our description of the steps carried out when a node is scanned, the value of $\ourzeta_{2^d}^{(j)}$ calculated during the scan of $(d,j) \in \mathcal{E}$ is the same $\ourzeta_{2^d}^{(j)}$ defined by the above recursion.

Again by \eqref{eq: Z-ish-opt evolutions}, we have for all $N$ large enough and $0\leq i < N$ that $\ourzeta_N^{(i)} \geq \ourZopt(\tilde{Q}_N^{(i)})$.
Namely, $\mathcal{A}' \subseteq \mathcal{A}$, where $\mathcal{A}$ and $\mathcal{A}'$ are defined in \cref{subsec: fair labeler}.
Thus, it suffices to show that for $R < \ourRl(\mathcal{G}, \mathcal{E})$ fixed and all $N$ large enough, $|\mathcal{A}'| \geq N \cdot R$.

Assume that $R$, and therefore $\ourRl(\mathcal{G}, \mathcal{E})$ are positive, otherwise the claim is trivial.
As before, denote $\beta' = \frac{\beta + 1/2}{2}$.
For each one of the pairs $(d,j) \in \mathcal{E}$, we invoke \Cref{prop: augmented proposition} with $\delta = \delta'(\zeta_{2^d}^{(j)}) + (\ourRl(\mathcal{G}, \mathcal{E})) - R)$, $\eta = S_0 = \zeta_{2^d}^{(j)}$, and $\beta'' = \frac{\beta' + 1/2}{2}$ in place of $\beta$. 
Denote the $n_0$ promised by the proposition as $n_0^{(d,j)}$.
Now define $n_0^{\mathrm{max}}=\max_{(d,j) \in \mathcal{E}}n_0^{(d,j)}$ and $d^{\mathrm{max}}_\mathcal{E}=\max_{(d,j) \in \mathcal{E}}d$. Thus, for any $(d,j) \in \mathcal{E}$, $2^{-2^{(n-d)\beta''}} \leq 2^{-2^{(n-d^{\mathrm{max}}_\mathcal{E})\beta''}}$. Hence, by \eqref{eq: augmented proposition}, for $n \geq d^{\mathrm{max}}_\mathcal{E} + n_0^{\mathrm{max}}$ the fraction of indices $ 0 \leq i < N$ such that $\zeta_N^{(i)} \leq 2^{-2^{(n-d^{\mathrm{max}}_\mathcal{E})\beta''}}$ is at least
\begin{IEEEeqnarray*}{rCl}
\IEEEeqnarraymulticol{3}{l}{\sum_{(d,j) \in \mathcal{E}} \frac{1}{2^d} \max\left\{ 1- \left( \delta'(\zeta_{2^d}^{(j)}) + \ourRl(\mathcal{G}, \mathcal{E})- R \right), 0 \right\}} \\
	& = & \sum_{(d,j) \in \mathcal{E}} \frac{1}{2^d}     \max\left\{ 1- \delta'(\zeta_{2^d}^{(j)}) + R -  \ourRl(\mathcal{G}, \mathcal{E})  , 0 \right\} \\
	&\geq& \sum_{(d,j) \in \mathcal{E}} \frac{1}{2^d}  \max \left\{ 1 {-} \delta'(\zeta_{2^d}^{(j)}) {+} R{-}  \ourRl(\mathcal{G}, \mathcal{E}),R {-} \ourRl(\mathcal{G}, \mathcal{E}) \right\} \\
	      & = &\sum_{(d,j) \in \mathcal{E}} \frac{1}{2^d}  \left(  \max\left\{ 1- \delta'(\zeta_{2^d}^{(j)}),0  \right\}  + R -  \ourRl(\mathcal{G}, \mathcal{E})  \right) \\
	      & \eqann{a} &\sum_{(d,j) \in \mathcal{E}} \frac{1}{2^d}  \left(  \max\left\{ 1- \delta'(\zeta_{2^d}^{(j)}),0  \right\} \right) + R -  \ourRl(\mathcal{G}, \mathcal{E})   \\
	& = & R \; ,
\end{IEEEeqnarray*}
where \eqannref{a} follows since the Kraft inequality \cite[Equation 5.8]{CoverThomas:91b} is tight on full binary trees, as can easily be proven by induction.

We take $n_0$ in \cref{thm: threshold rates for the asymptotic case} such that  $n_0 \geq d^{\mathrm{max}}_\mathcal{E} + n_0^{\mathrm{max}}$.  We further require that $n_0$ is large enough so that for all $n \geq n_0$ we have $2^{-2^{(n-d^{\mathrm{max}}_\mathcal{E})\beta''}} \leq 2^{-2^{n\beta'}} = 2^{-N^{\beta'}}$. By the above, this ensures that $|\mathcal{A}'| \geq N \cdot R$. Lastly, we require that $n_0$ is large enough such that for all $n \geq n_0$ we have $N \cdot 2^{-N^{\beta'}} < 2^{-N^\beta}$. This ensures that the word error rate is at most $2^{-N^\beta}$.
\end{IEEEproof}

\subsection{Definition of block-genie and justification of $\ourRu(\mathcal{G})$}
Our aim now is to prove an analogous claim to \cref{prop: RlGE good} for $\ourRu$. This is \cref{prop: RuG good} below. In the proof of \cref{prop: RuG good} we use a ``block-genie'', a concept we now define. Recall that in the seminal paper \cite{Arikan:09p}, a genie-aided decoder is used. That is, a variant of SC decoding, in which at stage $i$ the genie reveals $u_1^{i-1}$. Thus, at stage $i$, the relevant distribution is $W_N^{(i)}$, given in \eqref{eq: definition of W joint-distribution}. The genie-aided decoder is used since it is easier to analyze than SC decoding, but still has exactly the same word error rate as the SC decoder. Our block-genie will have this property as well.

The block-genie-aided SC decoder is defined in \Cref{alg: makeDecision,alg: genieCorrect,alg: Decode}.
\begin{algorithm}
    \caption{Make Decision}
    \DontPrintSemicolon
    $\algMakeDecision(\lambda,i)$ \;
    \eIf{$i \in \mathcal{A}$}
    {$\hat{u}_i =
    \begin{cases}
        0 & \lambda \geq 0 \\
        1 & \lambda < 0 
    \end{cases}
    $}{$\hat{u}_i = 0$\;}
    \Return{ $\hat{u}_i$}
    \label{alg: makeDecision}
\end{algorithm}

\begin{algorithm}
    \caption{Genie Correct}
    \DontPrintSemicolon
    $\algGenieCorrect(i,T)$ \;
    \Return{ $(u_i, u_{i+1},\ldots,u_{i+T-1}) \cdot B_T \cdot  F^{\otimes t} $ } \;
    \tcc{$B_T$ is the bit reversal matrix, $t = \log_2 T$, $F = \left( \begin{array}{cc} 1 & 0 \\ 1 & 1 \end{array} \right)$, and $\otimes$ is the Kronecker product}
    \label{alg: genieCorrect}
\end{algorithm}

\begin{algorithm}
    \caption{Decode}
    \DontPrintSemicolon
    $\algDecode(\lambda_0,\lambda_1,\ldots,\lambda_{T-1};d,j)$ \\
    \eIf{$T=1$}{$\algc = \algMakeDecision(\lambda_0,j)$ \tcp*[r]{$\algc = (c_0)$} }
    {
    $\algLambdaf = \big(f(\lambda_0,\lambda_1), \ldots,f(\lambda_{T-2},\lambda_{T-1})\big)$\;
    \tcp{In the \MSA, $f$ is replaced by $\fminsum$}
    $\alga = \algDecode(\algLambdaf;d+1,2j)$ \tcp*[r]{$\alga = a_0^{T/2-1}$}  
    $\algLambdag = \big(g_{a_0}(\lambda_0,\lambda_1), \ldots,g_{a_{T/2-1}}(\lambda_{T-2},\lambda_{T-1})\big)$\;
    $\algb= \algDecode(\algLambdag;d+1,2j+1)$ \tcp*[r]{$\algb = b_0^{T/2-1}$}  
    $\algc = \big(a_0 \oplus b_0,b_0,\ldots,a_{T/2-1} \oplus b_{T/2-1},b_{T/2-1}\big)$\;
    \tcp*[r]{$\algc = c_0^{T-1}$}
    }
    \If{$(d,j) \in \mathcal{G}$}
    {
	    $i=j\cdot T$ \\
	    \tcc{Genie corrects decisions on $\hat{u}_i,\hat{u}_{i+1}, \ldots,\hat{u}_{i+T-1}$, after \underline{all} these are made}
        $\algc = \algGenieCorrect(i,T)$ \tcp*[r]{$\algc = c_0^{T-1}$}
    }
    \Return{$\algc$}
    \label{alg: Decode}
\end{algorithm}
For a code of length $N$ and a received word $y_0^{N-1}$, decoding is preformed by calling \cref{alg: Decode} with $(\labeler(y_0), \labeler(y_1), \ldots, \labeler(y_{N-1}))$ and $d=j=0$.
Note that the set $\mathcal{G}$ is used in \Cref{alg: Decode}. Conceptually, we break the task of decoding $u_0,u_1,\ldots,u_{N-1}$ into the decoding of $|\mathcal{G}|$ blocks. We assume a code of length $N = 2^n$, where $n \geq d^{\mathrm{max}}_\mathcal{G}=\max_{(d,j) \in \mathcal{G}}d$. For $(d,j) \in \mathcal{G}$, the corresponding block is $u_i,u_{i+1},\ldots,u_{i+T-1}$, where $i = j \cdot T$ and $T = 2^{n-d}$. When decoding this block, the genie has already revealed $u_1^{i-1}$ and thus the relevant distribution under the \MSA is $\tilde{Q}_{2^d}^{(j)}$ (applying $\fminsum$ in place of $f$ in \cref{alg: Decode}). Specifically, after this block has been decoded, the genie corrects any errors the decoder may have introduced. This is done by invoking the $\algGenieCorrect$ function defined in \cref{alg: genieCorrect} and used at the bottom of \cref{alg: Decode}.

Note that for $\mathcal{G}=\emptyset$, \Cref{alg: makeDecision,alg: genieCorrect,alg: Decode} simply describe SC decoding, without any help from a genie. Moreover, for $\mathcal{G} = \mathcal{G}(N)$ as defined in \eqref{eq: GV}, \Cref{alg: makeDecision,alg: genieCorrect,alg: Decode} describe \arikan's genie-aided SC decoding.
For the above two choices of $\mathcal{G}$, as well as for any other valid choice, the word error probability is the same, since correction are made only after decisions on $\hat{u}_i$ have been made in \cref{alg: makeDecision}. 
\begin{proposition}
	\label{prop: RuG good}
	Setting $\ourRu = \ourRu(\mathcal{G})$ in \Cref{thm: threshold rates for the asymptotic case} is valid.
\end{proposition}

\begin{IEEEproof}
	Fix $\Delta > 0$ and consider a code with rate $R \geq \ourRu(\mathcal{G}) + \Delta$. Denote the information set of this code as $\mathcal{A}$ and its length as $N=2^n$, Thus $|\mathcal{A}|=N\cdot R$. Assume that $n \geq d_{\mathcal{G}}^{\mathrm{max}}=\max_{(d,j)\in\mathcal{G}}d$.

	Consider the block corresponding to $(d,j) \in \mathcal{G}$. The number of indices in this block is $T = 2^{n-d}$. Of these, denote the indices in $\mathcal{A}$ by
	\[
		\mathcal{A}_{(d,j)} = \{j\cdot T \leq  i < j \cdot (T+1): i \in \mathcal{A}\} \; .
	\]
	Thus, the rate at which this block is coded for is
	\[
		R_{(d,j)} \triangleq |\mathcal{A}_{(d,j)}|/2^{n-d} \; .
	\]

	Since every index $0 \leq i < N$ is contained in exactly one block,
	\[
		R = \sum_{(d,j) \in \mathcal{G}} \frac{1}{2^d} R_{(d,j)} \; .
	\]
	Thus, by the above and \eqref{eq: general Ru},
	\[
		\Delta \leq R - \ourRu(\mathcal{G}) =  \sum_{(d,j) \in \mathcal{G}} \frac{1}{2^d} \left(R_{(d,j)} - I\left(\tilde{Q}_{2^d}^{(j)}\right)\right) \; .
	\]
	By the pigeon-hole principle and the Kraft inequality being tight for a full binary tree, there exists at least one $(d,j) \in \mathcal{G}$ such that  
\[
	R_{(d,j)} - I\left(\tilde{Q}_{2^d}^{(j)}\right) \geq \Delta \; . 
\]
By the strong converse to the coding theorem \cite[Theorem 5.8.5]{Gallager:68b}, the probability of misdecoding such a block converges to $1$, as the block size tends to infinity. Thus, the word error rate must converge to $1$ as $N$ tends to infinity, since all blocks have lengths that tend to infinity with $N$.
\end{IEEEproof}

We end this section by stating the following proposition. As will become apparent in the proof, this is a special case of \cref{cor: monotonicty of rate thresholds}, given in the following subsection.
\begin{proposition}
\label{prop: Ru leq C}
    $\ourRu(\mathcal{G}) \leq C$, where $C$ is the capacity of $W$.
\end{proposition}

\subsection{Monotonic Properties of $\ourRl(\mathcal{G}, \mathcal{E})$ and $\ourRu(\mathcal{G})$}
In this section, we show that the deeper we carry out our calculations, the tighter our thresholds become. Specifically, a corollary of what we are about to prove is that increasing $d_\mathcal{G}$ or $d_\mathcal{E}$ (or decreasing $\epsilon$) yields better results.

Recall from \cref{subsec: definition of Rl and Ru} the definitions and properties of a valid $\mathcal{G}$, a valid $\mathcal{E}$, and a valid pair $(\mathcal{G},\mathcal{E})$. The following defines a set $\mathcal{G}'$ obtained by replacing a vertex in $\mathcal{G}$ by its two sons.

\begin{definition}
	\label{def: G'}
    For a valid $\mathcal{G}$ and a vertex $(d',j') \in\mathcal{G}$, let
    \[
    \mathcal{G}'(d',j')  \triangleq \mathcal{G} \cup \left\{(d'+1,2j'),(d'+1,2j'+1) \right\} \setminus \left\{(d',j')\right\} \;.
    \]
\end{definition}%
Note that since we assume that $\mathcal{G}$ is valid, then so is $\mathcal{G}'$.

The following defines the set $\mathcal{E}'$ similarly to the above.%
\begin{definition}
	\label{def: E'}
    For a valid $\mathcal{E}$ and a vertex $(d',j') \in\mathcal{E}$, let
    \[
    \mathcal{E}'(d',j')  \triangleq \mathcal{E} \cup \left\{(d'+1,2j'),(d'+1,2j'+1) \right\} \setminus \left\{(d',j')\right\} \;.
    \]
\end{definition}
As before, since $\mathcal{E}$ is valid, so is $\mathcal{E}'$.

As an example of the above, consider the sets $\mathcal{G}$ and $\mathcal{E}$ depicted in \cref{fig: tree}.
\begin{itemize}
	\item The depiction of $\mathcal{E}'(d'=3, j'=3)$ would be to add two white rectangular sons to $(3,3)$, and to change $(3,3)$ from a white rectangle to a white circle. Note that the pair $(\mathcal{G},\mathcal{E}')$ is valid.
	\item The depiction of $\mathcal{E}'(d'=2, j'=2)$ would be to add two white rectangular sons to $(2,2)$, and to change $(2,2)$ from a red rectangle to a red circle. Note that the pair $(\mathcal{G},\mathcal{E}')$ is valid.
	\item The depiction of $\mathcal{G}'(d'=2, j'=3)$ would be to change the color of $(3,6)$ and $(3,7)$ from white to red, and to change the color of $(2,3)$ from red to white. Note that the pair $(\mathcal{G}',\mathcal{E})$ is valid.
	\item Lastly, note that for $\mathcal{G}'(d'=2,j'=2)$ the pair $(\mathcal{G}',\mathcal{E})$ is not valid. In general, this happens if both $(d',j') \in \mathcal{G}$ and $(d',j') \in \mathcal{E}$. To keep validity for these cases, we enlarge both sets. That is, $\left(\mathcal{G}'(d',j'),\mathcal{E}'(d',j')\right)$ is valid. Thus, the depiction of $\mathcal{G}'(d'=2,j'=2)$ and $\mathcal{E}'(d'=2,j'=2)$ is to add two red rectangular sons to $(2,2)$ and to change $(2,2)$ from a red rectangle to a white circle.
\end{itemize}

The following propositions show that our thresholds become tighter when replacing either $\mathcal{G}$ by $\mathcal{G}'$ or $\mathcal{E}$ by $\mathcal{E}'$. 

\begin{proposition}%
    \label{prop: Ru monotonic in G}
    For a valid $\mathcal{G}$, and a vertex $(d',j') \in\mathcal{G}$,
    \begin{IEEEeqnarray}{rCl}
	    \ourRu(\mathcal{G'}) \leq \ourRu(\mathcal{G}) \; .
	    \label{eq: Ru(G') leq Ru(G)}
    \end{IEEEeqnarray}
\end{proposition}

\begin{proposition}%
    \label{prop: Rl monotonic in E}
    For a valid pair $(\mathcal{G},\mathcal{E})$, and a vertex $(d',j') \in\mathcal{E}$,
    \begin{IEEEeqnarray}{rCl}
	    \ourRl(\mathcal{G}, \mathcal{E}') \geq \ourRl(\mathcal{G}, \mathcal{E}) \; .
	    \label{eq: Rl(G,E') geq Rl(G,E)}
    \end{IEEEeqnarray}
\end{proposition}

\begin{proposition}%
    \label{prop: Rl monotonic in G}
    For a valid pair  $(\mathcal{G},\mathcal{E})$, and a vertex $(d',j')$ such that $(d',j') \in \mathcal{G}$ and $(d',j') \not\in \mathcal{E}$,
    \begin{IEEEeqnarray}{rCl}
	    \ourRl(\mathcal{G}', \mathcal{E}) \geq \ourRl(\mathcal{G}, \mathcal{E}) \; .
	    \label{eq: Rl(G',E) geq Rl(G,E)}
    \end{IEEEeqnarray}
\end{proposition}
We denote $\Gdeep \geq \mathcal{G}$ if $\Gdeep$ is obtained from $\mathcal{G}$ by a finite series of operations as in \cref{def: G'}. Similarly,
we denote $\Edeep \geq \mathcal{E}$ if $\Edeep$ is obtained from $\mathcal{E}$ by a finite series of operations as in \cref{def: E'}.

As we will show, repeated application of the above three propositions yield the following.

\begin{corollary}[Monotonicity of rate thresholds]
	\label{cor: monotonicty of rate thresholds}
	Let the pair $(\mathcal{G},\mathcal{E})$ be valid. Let the pair $(\Gdeep, \Edeep)$ be valid as well, where $\Gdeep \geq \mathcal{G}$ and $\Edeep \geq \mathcal{E}$. Then,
	\begin{IEEEeqnarray}{rCl}
	    \ourRu(\Gdeep) \leq \ourRu(\mathcal{G}) 
	    \label{eq: Ru(G*) leq Ru(G)}
	\end{IEEEeqnarray}
    and
    \begin{IEEEeqnarray}{rCl}
	    \ourRl(\Gdeep, \Edeep) \geq \ourRl(\mathcal{G}, \mathcal{E}) \; .
	    \label{eq: Rl(G*,E*) geq Rl(G,E)}
    \end{IEEEeqnarray}
\end{corollary}

Recall the definition of $\mathcal{G}(V)$ and $\mathcal{E}(M)$, given in \eqref{eq: GV} for $V=2^v$ and $M=2^m$. Note that for $v' \geq v$ and $m' \geq m$, we have $\mathcal{G}(V') \geq \mathcal{G}(V)$ and $\mathcal{E}(M') \geq \mathcal{E}(M)$, where $V' = 2^{v'}$ and $M' = 2^{m'}$. Thus, the above corollary implies that setting the depths $v$ and $m$ larger in \cref{subsec: good labeler} indeed yields tighter thresholds. Namely, $\ourRu(V') \leq \ourRu(V)$ and $\ourRl(V',M') \geq \ourRl(V,M)$. Similarly, increasing $d_\mathcal{G}$ or $d_\mathcal{E}$, or decreasing $\epsilon$ in \cref{subsec: definition of Rl and Ru} also yields tighter thresholds.

\appendices

\section{Non-recursive interpretation of the \MSA}
Recall the definition of $L_N^{(i)}$ in \eqref{eq: LLR defintion}. This definition is explicit (non-recursive).
However, when implementing a decoder the recursive definition given in \eqref{eq: transforms of llrs}--\eqref{eq: LLRs basis condition} is used.
For the \MSA, the corresponding recursive definition of $\tilde{L}_N^{(i)}$ is given in \eqref{eq: transforms of labels}--\eqref{eq: labels basis condition}.
In this appendix we give an explicit definition of $\tilde{L}_N^{(i)}$, under the assumption $\lambda(y) = \mathrm{LLR}(y) = \log_2 \left(W(y;0)/W(y;1)\right)$. This is \eqref{eq: non-recursive tilde-L} below, and is the analog of \eqref{eq: LLR defintion}, which is rephrased below as \eqref{eq: non-recursive L}.
To save space, we use standard shorthand. For example, $\Pr(Y_0^{N-1} = y_0^{N-1}, U_0^{i-1} = u_0^{i-1}, U_i = 0, U_{i+1}^{N-1} = u_{i+1}^{N-1} )$ is shortened to $p(y_0^{N-1}, u_0^{i-1}, u_i = 0, u_{i+1}^{N-1})$.

\begin{proposition}
	\label{prop: non-recursive labels}
	For the non-approximated setting,
\begin{multline}
	L_{N}^{(i)} \left( y_0^{N-1}, u_{0}^{i-1} \right) = \\
\log_2 \left( 
	\frac{\sum \limits_{u_{i+1}^{N-1}} p(y_0^{N-1}, u_0^{i-1}, u_i = 0, u_{i+1}^{N-1})}{\sum \limits_{u_{i+1}^{N-1}} p(y_0^{N-1}, u_0^{i-1}, u_i = 1, u_{i+1}^{N-1})} 
\right) \; .
\label{eq: non-recursive L}
\end{multline}

Under the \MSA with $\lambda(y) = \log_2 \left(W(y;0)/W(y;1)\right)$,
\begin{multline}
	\tilde{L}_{N}^{(i)} \left( y_0^{N-1}, u_{0}^{i-1} \right) = \\
\log_2 \left( 
	\frac{\max \limits_{u_{i+1}^{N-1}} p(y_0^{N-1}, u_0^{i-1}, u_i = 0, u_{i+1}^{N-1})}{\max \limits_{u_{i+1}^{N-1}} p(y_0^{N-1}, u_0^{i-1}, u_i = 1, u_{i+1}^{N-1})} 
\right) \; .
\label{eq: non-recursive tilde-L}
\end{multline}
\end{proposition}%
Notice that the only difference between \eqref{eq: non-recursive L} and \eqref{eq: non-recursive tilde-L} is that in the former we use a ``$\sum$'' while in the latter we use a ``$\max$''.

Although our paper would be self contained without this appendix, we feel that the explicit definition \eqref{eq: non-recursive tilde-L} gives intuition about the \MSA. Specifically, consider stage $i$ of the decoding, in which we have already decided on $\hat{u}_0^{i-1}$, and must now decide the value of $\hat{u}_i$. Define
\begin{IEEEeqnarray*}{rCl}
	\mathcal{C}^{(i)}_{0} & \triangleq &  \big\{ u_0^{N-1} \in \{0,1\}^N : u_i=0, u_0^{i-1} = \hat{u}_0^{i-1}  \big\} \; , \\
	\mathcal{C}^{(i)}_{1} & \triangleq & \big\{ u_0^{N-1} \in \{0,1\}^N : u_i=1, u_0^{i-1} = \hat{u}_0^{i-1}  \big\} \; .
\end{IEEEeqnarray*}
Recall from \cite[Equation 4]{Arikan:09p} the definition of the combined channel
\[
	W_N(y_0^{N-1}|u_0^{N-1}) = \prod_{i=0}^{N-1} W(y_i|x_i) \; ,
\]
where $x_0^{N-1} = u_0^{N-1} B_N F^{\otimes n}$ is the codeword corresponding to $u_0^{N-1}$, $B_N$ is the bit-reversal matrix, and  $F \triangleq \left( \begin{smallmatrix} 1 & 0 \\ 1 & 1 \end{smallmatrix} \right)$ is the \arikan kernel. 
Recall also that in both the non-approximated and approximated settings the decision rule is based on the sign of $L_{N}^{(i)}$ and $\tilde{L}_{N}^{(i)}$, respectively.
Therefore, an immediate corollary of \cref{prop: non-recursive labels} is the following decision rules.
The non-approximated SC decoder sets $\hat{u}_i$ according to
\begin{equation}
	\sum \limits_{{u_0^{N-1}} \in \mathcal{C}^{(i)}_0} W_N(y_0^{N-1}| u_0^{N-1}) \substack{\hat{u}_i=0 \\ \geqslant \\ < \\ \hat{u}_i=1} \sum \limits_{{u_0^{N-1}} \in \mathcal{C}^{(i)}_1} W_N(y_0^{N-1}| u_0^{N-1}) \; ,
	\label{eq: SCinter}
\end{equation}
whereas the min-sum SC decoder sets $\hat{u}_i$ according to
\begin{equation}
	\max \limits_{{u_0^{N-1}} \in \mathcal{C}^{(i)}_0} W_N(y_0^{N-1}| u_0^{N-1}) \substack{\hat{u}_i=0 \\ \geqslant \\ < \\ \hat{u}_i=1} \max \limits_{{u_0^{N-1}} \in \mathcal{C}^{(i)}_1} W_N(y_0^{N-1}| u_0^{N-1}) \; .
\label{eq: MSSCinter}
\end{equation}
Indeed, the above follows by the assumption that the  input distribution is i.i.d.\ symmetric, which implies that a-priori, all $u_0^{N-1}$ are equally likely.

Informally, in both settings we must choose one of two cosets: $\mathcal{C}^{(i)}_0$ or $\mathcal{C}^{(i)}_1$. In the non-approximated setting we base our decision on a weighting of all the words in each coset, whereas in the min-sum setting we base our decision on only the most probable word in each coset.

As we will see, the proof of \eqref{eq: non-recursive L} is straightforward.
To prove \eqref{eq: non-recursive tilde-L}, we take an indirect but simple route.
Namely, we define $L^{\ourStar (i)}_N$ as the RHS of \eqref{eq: non-recursive tilde-L}.
That is,
\begin{multline}
	L_{N}^{\ourStar(i)} \left( y_0^{N-1}, u_{0}^{i-1} \right) \triangleq \\
	\log_2 \left( 
	\frac{\max \limits_{u_{i+1}^{N-1}} p(y_0^{N-1}, u_0^{i-1}, u_i = 0, u_{i+1}^{N-1})}{\max \limits_{u_{i+1}^{N-1}} p(y_0^{N-1}, u_0^{i-1}, u_i = 1, u_{i+1}^{N-1})} 
\right) \; .
\label{eq: non-recursive star-L}
\end{multline}
Our proof will follow by showing that $L^{\ourStar (i)}_N$ satisfies the same recursive relations as $\tilde{L}_N^{(i)}$ in \eqref{eq: transforms of labels}--\eqref{eq: labels basis condition}.
Indeed, by inspection of \eqref{eq: non-recursive star-L}, 
\[
	L_1^{\ourStar(0)}(y)= \mathrm{LLR}(y) = \log_2 \left(W(y;0)/W(y;1)\right) \; .
\]
Recalling \eqref{eq: labels basis condition} and our assumption that $\labeler(y) = \mathrm{LLR}(y)$, we have that the starting condition is the same for $L^{\ourStar (i)}_N$ and $\tilde{L}_N^{(i)}$.
Thus, to prove \eqref{eq: non-recursive tilde-L} all that remains is to show that \eqref{eq: transforms of labels} holds with ``$\ourStar$'' in place of ``$\sim$''.

In aid of the above we introduce the following notation:
\begin{IEEEeqnarray}{l}
   \mu_{N}^{\ourStar(i)} (y_{0}^{N-1},u_{0}^{i-1};u_{i}) \label{eq: mu-star definition} \\
    \qquad\qquad {\triangleq} \max_{u_{i+1}^{N-1}} p(y_{0}^{N-1},u_{0}^{i-1},u_{i}, u_{i+1}^{N-1}) \IEEEnonumber
\end{IEEEeqnarray}
and
\begin{IEEEeqnarray}{rCl}
    \ourell_N^{\ourStar(i)}(y_0^{N-1},u_0^{i-1})
    &{=}& {\left( \frac
    { \mu_{N}^{\ourStar(i)} (y_{0}^{N-1},u_{0}^{i-1};u_{i}=0)}
    { \mu_{N}^{\ourStar(i)} (y_{0}^{N-1},u_{0}^{i-1};u_{i}=1)} \right)} .
    \label{eq: star-likelihood using mu-star}
\end{IEEEeqnarray}
Therefore, $L_N^{\ourStar(i)}(y_0^{N-1},u_0^{i-1}) = \log_2 \ourell_N^{\ourStar(i)}(y_0^{N-1},u_0^{i-1})$.

To derive the required recursive relations of $L_N^{\ourStar(i)}$ we first derive the following recursive relations of $\mu_N^{\ourStar(i)}$.

\begin{lemma}[$\mu^{\ourStar}$ minus and plus transforms]
\label{lemm: mu transforms}
For $N \geq 2$ and $0 \leq j < N/2$,
\begin{multline}
    \mu_{N}^{\ourStar(2j)} (y_{0}^{N-1},u_{0}^{2j-1};u_{2j}) = \label{eq: mu minus transform}
    \\ \max_{u_{2j+1}}
    \Big\{ \mu_{N/2}^{\ourStar(j)} \Bigl(y_0^{N/2-1},u_{0,e}^{2j-1} \oplus u_{0,o}^{2j-1} ; u_{2j} \oplus u_{2j+1}\Bigr) 
    \\ \cdot \mu_{N/2}^{\ourStar(j)} \Bigl(y_{N/2}^{N-1},u_{0,o}^{2j-1} ; u_{2j+1}\Bigr) \Big\}
\end{multline}
and
\begin{multline}
    \mu_{N}^{\ourStar(2j+1)} (y_{0}^{N-1},u_{0}^{2j-1};u_{2j}) = \label{eq: mu plus transform}
    \\ \mu_{N/2}^{\ourStar(j)} \Bigl(y_0^{N/2-1},u_{0,e}^{2j-1} \oplus u_{0,o}^{2j-1} ; u_{2j} \oplus u_{2j+1}\Bigr) 
    \\ \cdot \mu_{N/2}^{\ourStar(j)} \Bigl(y_{N/2}^{N-1},u_{0,o}^{2j-1} ; u_{2j+1}\Bigr) \; .
\end{multline}
\end{lemma}

\begin{IEEEproof}%
For \eqref{eq: mu minus transform} we have
\begin{IEEEeqnarray*}{rCl}
    \IEEEeqnarraymulticol{3}{l}{\mu_{N}^{\ourStar(2j)} (y_{0}^{N-1},u_{0}^{2j-1};u_{2j})} \\
    \quad &\eqann{a}& \max_{u_{2j+1}^{N-1}}
    p \Bigl(y_{0}^{N-1},u_{0}^{2j-1},u_{2j}, u_{2j+1}^{N-1}\Bigr) \\
	  &\eqann{b}& \max_{u_{2j+1}} \;  \max_{u_{2j+2}^{N-1}}
    p \Bigl(y_{0}^{N/2-1},y_{N/2}^{N-1},u_{0,e}^{2j-1} \oplus u_{0,o}^{2j-1}, u_{0,o}^{2j-1}\\ 
    \IEEEeqnarraymulticol{3}{r}{,u_{2j} \oplus u_{2j+1}, u_{2j+1}, u_{2j+2,e}^{N-1} \oplus u_{2j+2,o}^{N-1}, u_{2j+2,o}^{N-1}  \Bigr)} \\
	  &\eqann{c}& \max_{u_{2j+1}} \; \max_{u_{2j+2}^{N-1}}
    p \Bigl(y_{0}^{N/2-1},u_{0,e}^{2j-1} \oplus u_{0,o}^{2j-1} \\
    \IEEEeqnarraymulticol{3}{r}{,u_{2j} \oplus u_{2j+1},u_{2j+2,e}^{N-1} \oplus u_{2j+2,o}^{N-1}\Bigr)} \\
	  &&\! {} \cdot
    p \Bigl( y_{N/2}^{N-1},u_{0,o}^{2j-1}, u_{2j+1}, u_{2j+2,o}^{N-1} | y_{0}^{N/2-1},u_{0,e}^{2j-1} \oplus u_{0,o}^{2j-1} \\ 
     \IEEEeqnarraymulticol{3}{r}{,u_{2j} \oplus ,u_{2j+1}, u_{2j+2,e}^{N-1} \oplus u_{2j+2,o}^{N-1}
    \Bigr)} \\
    &\eqann{d}& \max_{u_{2j+1}} \; \max_{u_{2j+2}^{N-1}}
    p \Bigl(y_{0}^{N/2-1},u_{0,e}^{2j-1} \oplus u_{0,o}^{2j-1} \\
    \IEEEeqnarraymulticol{3}{r}{,u_{2j} \oplus u_{2j+1},u_{2j+2,e}^{N-1} \oplus u_{2j+2,o}^{N-1}\Bigr)} \\
    \IEEEeqnarraymulticol{3}{r}{{} \cdot
    p \Bigl( y_{N/2}^{N-1},u_{0,o}^{2j-1}, u_{2j+1}, u_{2j+2,o}^{N-1} \Bigr)} \\
    &\eqann{e}&
    \max_{u_{2j+1}} \max_{u_{2j+2,o}^{N-1}} \Bigl\{
    p \Bigl( y_{N/2}^{N-1},u_{0,o}^{2j-1}, u_{2j+1}, u_{2j+2,o}^{N-1} \Bigr) \\
    && \quad {} \cdot
    \max_{u_{2j+2,e}^{N-1}}
    p \Bigl(y_{0}^{N/2-1},u_{0,e}^{2j-1} \oplus u_{0,o}^{2j-1}\\
    \IEEEeqnarraymulticol{3}{r}{,u_{2j} \oplus u_{2j+1},u_{2j+2,e}^{N-1} \oplus u_{2j+2,o}^{N-1}\Bigr) \Bigr\}} \\
    &\eqann{f}&
    \max_{u_{2j+1}} \; \max_{u_{2j+2,o}^{N-1}} \Bigl\{
     p \Bigl( y_{N/2}^{N-1},u_{0,o}^{2j-1}, u_{2j+1}, u_{2j+2,o}^{N-1} \Bigr) \\
    \IEEEeqnarraymulticol{3}{r}{\mu_{N/2}^{\ourStar(j)} \Bigl(y_0^{N/2-1},u_{0,e}^{2j-1} \oplus u_{0,o}^{2j-1} ; u_{2j} \oplus u_{2j+1}\Bigr) \Bigr\}} \\
    &\eqann{g}&
    \max_{u_{2j+1}}
    \mu_{N/2}^{\ourStar(j)} \Bigl(y_0^{N/2-1},u_{0,e}^{2j-1} \oplus u_{0,o}^{2j-1} ; u_{2j} \oplus u_{2j+1}\Bigr) \\ 
   \IEEEeqnarraymulticol{3}{r}{\cdot \mu_{N/2}^{\ourStar(j)} \Bigl(y_{N/2}^{N-1},u_{0,o}^{2j-1} ; u_{2j+1}\Bigr) \; ,}
\end{IEEEeqnarray*}
where \eqannref{a} is by \eqref{eq: mu-star definition}, \eqannref{b} follows since there is a one-to-one and onto mapping between the arguments of $p$ on the LHS and the arguments of $p$ on the RHS,
\eqannref{c} is by the definition of conditional probability, \eqannref{d} is since we need not condition on independent random variables,
\eqannref{e} is since $\max_{x,y} \{f(x) \cdot g(x,y) \} = \max_x \{f(x) \cdot \max_y g(x,y) \}$, \eqannref{f} is by \eqref{eq: mu-star definition} since for fixed $u_{2j+2,o}^{N-1}$ the maximization over $u_{2j+2,e}^{N-1}$ ranges over all possible values in $\mathcal{X}^{N/2-1-j}$, and \eqannref{g} is again by \eqref{eq: mu-star definition}.

For \eqref{eq: mu plus transform}, we follow the same steps with maximization over $u_{2j+2}^{N-1}$ instead of $u_{2j+1}^{N-1}$. Therefore, all the above equalities remain the same, apart from not containing the outer $\max_{u_{2j+1}}$. 
\end{IEEEproof}

\begin{IEEEproof}[proof of \cref{prop: non-recursive labels}]
	For \eqref{eq: non-recursive L} we notice that 
	\begin{IEEEeqnarray*}{rCl}
   	 W_N^{(i)}(y_0^{N-1},u_0^{i-1};u_i)
   	 &=&  p(y_0^{N-1},u_0^{i-1},u_i) \\
   	 &=& \sum_{u_{i+1}^{N-1}} p(y_0^{N-1},u_0^{i-1},u_i, u_{i+1}^{N-1}) \; .
	\end{IEEEeqnarray*}
	Thus, substituting the above into \eqref{eq: LLR defintion} with $u_i=0$ for the numerator and $u_i=1$ for the denominator yields \eqref{eq: non-recursive L}.

	For \eqref{eq: non-recursive tilde-L} recall from the above discussion that the proof will be completed once we show that \eqref{eq: transforms of labels} holds with ``$\ourStar$'' in place of ``$\sim$''.
	We begin by proving the minus case \eqref{eq: minus transform of labels}, and then prove the plus case \eqref{eq: plus transform of labels}. 
    We define the following shorthand: 
  \begin{IEEEeqnarray*}{rClCrCl}
	  y_L &\triangleq& y_0^{N/2-1} \; , &\qquad \quad& y_R &\triangleq& y_{N/2-1}^{N-1} \; , \\
	  u_{\oplus} &\triangleq& u_{0,e}^{2j-1} \oplus u_{0,o}^{2j-1} \; , &\quad&  u_o &\triangleq& u_{0,o}^{2j-1} \; , \\
	  u &\triangleq& u_{2j} \; , &\quad&  v &\triangleq& u_{2j+1} \; , \\ 
	  \ourell_a^{\ourStar} &\triangleq& \ourell_{N/2}^{\ourStar(j)} (y_L, u_{\oplus}) \; ,&\quad&  \ourell_b^{\ourStar} &\triangleq& \ourell_{N/2}^{\ourStar(j)} (y_R,u_o) \; , \\
      L_a^{\ourStar} &\triangleq& \log_2 \ourell_a^{\ourStar} \; , &\quad&  L_b^{\ourStar} &\triangleq& \log_2 \ourell_b^{\ourStar} \; .
  \end{IEEEeqnarray*}
    For \eqref{eq: minus transform of labels} we have
    \begin{IEEEeqnarray*}{rCl}
        \IEEEeqnarraymulticol{3}{l}{\ourell_N^{\ourStar(2j)}(y_0^{N-1},u_0^{2j-1})} \\
        \quad\quad &\eqann{a}& {\left( \frac
        { \mu_{N}^{\ourStar(2j)} (y_{0}^{N-1},u_{0}^{2j-1};u_{2j}=0)}
        { \mu_{N}^{\ourStar(2j)} (y_{0}^{N-1},u_{0}^{2j-1};u_{2j}=1)} \right)} \\
        &\eqann{b}&
        {\left( \frac
        { \max_{v}
        \Big\{ \mu_{N/2}^{\ourStar(j)} (y_L,u_{\oplus} ; 0 \oplus v) 
        \cdot \mu_{N/2}^{\ourStar(j)} (y_R,u_{o} ; v) \Big\} }
        { \max_{v}
        \Big\{ \mu_{N/2}^{\ourStar(j)} (y_L,u_{\oplus} ; 1 \oplus v) 
        \cdot \mu_{N/2}^{\ourStar(j)} (y_R,u_{o} ; v) \Big\} } \right)} \\
        &\eqann{c}&
        {\left( \frac
        { \max \{ \alpha, \beta \} }
        { \max\{ \gamma, \delta \} } \right)} \; ,
    \end{IEEEeqnarray*}
    where \eqannref{a} is by \eqref{eq: star-likelihood using mu-star}, \eqannref{b} is by \eqref{eq: mu minus transform}, and \eqannref{c} due to the following notation:
    \begin{IEEEeqnarray*}{rCl}
        \alpha &=&
        \mu_{N/2}^{\ourStar(j)} (y_L,u_{\oplus} ; 0) 
        \cdot \mu_{N/2}^{\ourStar(j)} (y_R,u_{o} ; 0) \; ,\\
        \beta &=&
        \mu_{N/2}^{\ourStar(j)} (y_L,u_{\oplus} ; 1) 
        \cdot \mu_{N/2}^{\ourStar(j)} (y_R,u_{o} ; 1) \; , \\
        \gamma &=&
        \mu_{N/2}^{\ourStar(j)} (y_L,u_{\oplus} ; 1) 
        \cdot \mu_{N/2}^{\ourStar(j)} (y_R,u_{o} ; 0) \; , \\
        \delta &=&
        \mu_{N/2}^{\ourStar(j)} (y_L,u_{\oplus} ; 0) 
        \cdot \mu_{N/2}^{\ourStar(j)} (y_R,u_{o} ; 1) \; .
    \end{IEEEeqnarray*}
    Thus, $\ourell_N^{\ourStar(2j)}(y_0^{N-1},u_0^{2j-1})$ can equal one of four possible values: $\alpha/\gamma$, $\beta/\gamma$, $\alpha/\delta$, $\beta/\delta$. We consider each case separately.
    \begin{itemize}
    \item Consider the case $\alpha/\gamma$. We have
    \begin{IEEEeqnarray*}{rCl}
    \IEEEeqnarraymulticol{3}{l}{\ourell_N^{\ourStar(2j)}(y_0^{N-1},u_0^{2j-1})}\\
    \quad\quad &=& \alpha / \gamma \\
    &=& \frac{\mu_{N/2}^{\ourStar(j)} (y_L,u_{\oplus} ; 0) \cdot \mu_{N/2}^{\ourStar(j)} (y_R,u_{o} ; 0)}{\mu_{N/2}^{\ourStar(j)} (y_L,u_{\oplus} ; 1) 
    \cdot \mu_{N/2}^{\ourStar(j)} (y_R,u_{o} ; 0)}\\
    &=& \frac{\mu_{N/2}^{\ourStar(j)} (y_L,u_{\oplus} ; 0)}{\mu_{N/2}^{\ourStar(j)} (y_L,u_{\oplus} ; 1)}\\
    &=& \ourell_a^{\ourStar}
    \end{IEEEeqnarray*}

    $\alpha \geq \beta$ yields $1/\ourell^{\ourStar}_b \leq \ourell^{\ourStar}_a$. That is $-L^{\ourStar}_b \leq L^{\ourStar}_a$.
    
    $\gamma \geq \delta$ yields $\ourell^{\ourStar}_a \leq \ourell^{\ourStar}_b$. That is $L^{\ourStar}_a \leq L^{\ourStar}_b$.

    Combining both we have $-L^{\ourStar}_b \leq L^{\ourStar}_a \leq L^{\ourStar}_b$. Thus, $\sign(L^{\ourStar}_b) \geq 0$ and $\min \{|L^{\ourStar}_a|,|L^{\ourStar}_b|\}=|L^{\ourStar}_a|$. Therefore,
    \begin{IEEEeqnarray*}{rCl}
	    \IEEEeqnarraymulticol{3}{l}{L_N^{\ourStar(2j)}(y_0^{N-1},u_0^{2j-1})} \\
	   \quad\quad &=& L_a^{\ourStar} \\
       &=& \sign(L^{\ourStar}_a) \cdot |L^{\ourStar}_a| \\
       &=& \sign(L^{\ourStar}_a) \cdot \sign(L^{\ourStar}_b) \cdot \min \{|L^{\ourStar}_a|,|L^{\ourStar}_b|\} \\
       &=& \fminsum(L^{\ourStar}_a, L^{\ourStar}_b)
    \end{IEEEeqnarray*}

\item Consider the case $\alpha/\delta$. We have
    \begin{IEEEeqnarray*}{rCl}
    \IEEEeqnarraymulticol{3}{l}{\ourell_N^{\ourStar(2j)}(y_0^{N-1},u_0^{2j-1})}\\
    \quad\quad &=& \alpha / \delta \\
    &=& \frac{\mu_{N/2}^{\ourStar(j)} (y_L,u_{\oplus} ; 0) \cdot \mu_{N/2}^{\ourStar(j)} (y_R,u_{o} ; 0)}{\mu_{N/2}^{\ourStar(j)} (y_L,u_{\oplus} ; 0) 
    \cdot \mu_{N/2}^{\ourStar(j)} (y_R,u_{o} ; 1)}\\
    &=& \frac{\mu_{N/2}^{\ourStar(j)} (y_R,u_{o} ; 0)}{\mu_{N/2}^{\ourStar(j)} (y_R,u_{o} ; 1)}\\
    &=& \ourell_b^{\ourStar}
    \end{IEEEeqnarray*}

    $\alpha \geq \beta$ yields $1/\ourell^{\ourStar}_a \leq \ourell^{\ourStar}_b$. That is $-L^{\ourStar}_a \leq L^{\ourStar}_b$.
    
    $\delta \geq \gamma$ yields $\ourell^{\ourStar}_b \leq \ourell^{\ourStar}_a$. That is $L^{\ourStar}_b \leq L^{\ourStar}_a$.

    Combining both we have $-L^{\ourStar}_a \leq L^{\ourStar}_b \leq L^{\ourStar}_a$. Thus, $\sign(L^{\ourStar}_a) \geq 0$ and $\min \{|L^{\ourStar}_a|,|L^{\ourStar}_b|\}=|L^{\ourStar}_b|$. Therefore,
    \begin{IEEEeqnarray*}{rCl}
	    \IEEEeqnarraymulticol{3}{l}{L_N^{\ourStar(2j)}(y_0^{N-1},u_0^{2j-1})} \\
	   \quad\quad &=& L_b^{\ourStar} \\
       &=& \sign(L^{\ourStar}_b) \cdot |L^{\ourStar}_b| \\
       &=& \sign(L^{\ourStar}_a) \cdot \sign(L^{\ourStar}_b) \cdot \min \{|L^{\ourStar}_a|,|L^{\ourStar}_b|\} \\
       &=& \fminsum(L^{\ourStar}_a, L^{\ourStar}_b)
    \end{IEEEeqnarray*}

\item Consider the case $\beta/\gamma$. We have
    \begin{IEEEeqnarray*}{rCl}
    \IEEEeqnarraymulticol{3}{l}{\ourell_N^{\ourStar(2j)}(y_0^{N-1},u_0^{2j-1})}\\
    \quad\quad &=& \beta / \gamma \\
    &=& \frac{\mu_{N/2}^{\ourStar(j)} (y_L,u_{\oplus} ; 1) \cdot \mu_{N/2}^{\ourStar(j)} (y_R,u_{o} ; 1)}{\mu_{N/2}^{\ourStar(j)} (y_L,u_{\oplus} ; 1) 
    \cdot \mu_{N/2}^{\ourStar(j)} (y_R,u_{o} ; 0)}\\
    &=& \frac{\mu_{N/2}^{\ourStar(j)} (y_R,u_o ; 1)}{\mu_{N/2}^{\ourStar(j)} (y_R,u_o ; 0)}\\
    &=& 1 / \ourell_b^{\ourStar}
    \end{IEEEeqnarray*}

    $\beta \geq \alpha$ yields $\ourell^{\ourStar}_b \leq 1/\ourell^{\ourStar}_a$. That is $L^{\ourStar}_b \leq -L^{\ourStar}_a$.
    
    $\gamma \geq \delta$ yields $\ourell^{\ourStar}_a \leq \ourell^{\ourStar}_b$. That is $L^{\ourStar}_a \leq L^{\ourStar}_b$.

    Combining both we have $L^{\ourStar}_a \leq L^{\ourStar}_b \leq -L^{\ourStar}_a$. Thus, $\sign(L^{\ourStar}_a) \leq 0$ and $\min \{|L^{\ourStar}_a|,|L^{\ourStar}_b|\}=|L^{\ourStar}_b|$. Therefore,
    \begin{IEEEeqnarray*}{rCl}
	    \IEEEeqnarraymulticol{3}{l}{L_N^{\ourStar(2j)}(y_0^{N-1},u_0^{2j-1})} \\
	  \quad\quad &=& -L_b^{\ourStar} \\
       &=& - \sign(L^{\ourStar}_b) \cdot |L^{\ourStar}_b| \\
       &=& \sign(L^{\ourStar}_a) \cdot \sign(L^{\ourStar}_b) \cdot \min \{|L^{\ourStar}_a|,|L^{\ourStar}_b|\} \\
       &=& \fminsum(L^{\ourStar}_a, L^{\ourStar}_b)
    \end{IEEEeqnarray*}

\item Consider the case $\beta/\delta$. We have
    \begin{IEEEeqnarray*}{rCl}
    \IEEEeqnarraymulticol{3}{l}{\ourell_N^{\ourStar(2j)}(y_0^{N-1},u_0^{2j-1})}\\
    \quad\quad &=& \beta / \delta \\
    &=& \frac{\mu_{N/2}^{\ourStar(j)} (y_L,u_{\oplus} ; 1) \cdot \mu_{N/2}^{\ourStar(j)} (y_R,u_{o} ; 1)}{\mu_{N/2}^{\ourStar(j)} (y_L,u_{\oplus} ; 0) 
    \cdot \mu_{N/2}^{\ourStar(j)} (y_R,u_{o} ; 1)}\\
    &=& \frac{\mu_{N/2}^{\ourStar(j)} (y_L,u_{\oplus} ; 1)}{\mu_{N/2}^{\ourStar(j)} (y_L,u_{\oplus} ; 0)}\\
    &=& 1 / \ourell_a^{\ourStar}
    \end{IEEEeqnarray*}

    $\beta \geq \alpha$ yields $\ourell^{\ourStar}_a \leq 1/\ourell^{\ourStar}_b$. That is $L^{\ourStar}_a \leq -L^{\ourStar}_b$.
    
    $\delta \geq \gamma$ yields $\ourell^{\ourStar}_b \leq \ourell^{\ourStar}_a$. That is $L^{\ourStar}_b \leq L^{\ourStar}_a$.

    Combining both we have $L^{\ourStar}_b \leq L^{\ourStar}_a \leq -L^{\ourStar}_b$. Thus, $\sign(L^{\ourStar}_b) \leq 0$ and $\min \{|L^{\ourStar}_a|,|L^{\ourStar}_b|\}=|L^{\ourStar}_a|$. Therefore,
    \begin{IEEEeqnarray*}{rCl}
	    \IEEEeqnarraymulticol{3}{l}{L_N^{\ourStar(2j)}(y_0^{N-1},u_0^{2j-1})} \\
	  \quad\quad &=& -L_a^{\ourStar} \\
       &=& - \sign(L^{\ourStar}_a) \cdot |L^{\ourStar}_a| \\
       &=& \sign(L^{\ourStar}_a) \cdot \sign(L^{\ourStar}_b) \cdot \min \{|L^{\ourStar}_a|,|L^{\ourStar}_b|\} \\
       &=& \fminsum(L^{\ourStar}_a, L^{\ourStar}_b)
    \end{IEEEeqnarray*}
    \end{itemize}
    To summarize, in all four cases \eqref{eq: minus transform of labels} holds with ``$\ourStar$'' in place of ``$\sim$''.

    For \eqref{eq: plus transform of labels} we have
    \begin{IEEEeqnarray*}{rCl}
	    \IEEEeqnarraymulticol{3}{l}{\ourell_N^{\ourStar(2j+1)} (y_{0}^{N-1},u_{0}^{2j})} \\
	\quad\quad &\eqann{a}& {\left( \frac
        { \mu_{N}^{\ourStar(2j+1)} (y_{0}^{N-1},u_{0}^{2j};u_{2j+1}=0)}
        { \mu_{N}^{\ourStar(2j+1)} (y_{0}^{N-1},u_{0}^{2j};u_{2j+1}=1)} \right)} \\
	      &\eqann{b}& \frac{\mu_{N/2}^{\ourStar(j)} (y_R,u_o ; 0)}{\mu_{N/2}^{\ourStar(j)} (y_R,u_o ; 1)} \cdot \frac{\mu_{N/2}^{\ourStar(j)} (y_L,u_\oplus ; u \oplus 0)}{\mu_{N/2}^{\ourStar(j)} (y_L, u_\oplus ; u \oplus 1)}\\
	      &\eqann{c}& \begin{cases}
            \ourell_{N/2}^{\ourStar(j)} (y_R,u_o) \cdot \ourell_{N/2}^{\ourStar(j)} (y_L,u_\oplus) & \;\mbox{if}\; u=0 \; ,\\ %
            \ourell_{N/2}^{\ourStar(j)} (y_{N/2}^{N-1},u_{0,o}^{2j-1}) / \ourell_{N/2}^{\ourStar(j)} (y_L,u_\oplus) & \;\mbox{if}\; u=1  %
        \end{cases} \\
	      &=& \begin{cases}
            \ourell_{b}^{\ourStar} \cdot \ourell_{a}^{\ourStar} & \;\mbox{if}\; u=0 \; ,\\ %
            \ourell_{b}^{\ourStar} / \ourell_{a}^{\ourStar} & \;\mbox{if}\; u=1 \; ,  %
        \end{cases}
    \end{IEEEeqnarray*}
    where \eqannref{a} is by \eqref{eq: star-likelihood using mu-star}, \eqannref{b} is by \eqref{eq: mu plus transform}, and \eqannref{c} is again by \eqref{eq: star-likelihood using mu-star}. In log-domain the above is simply
    \begin{IEEEeqnarray*}{rCl}
       L_N^{\ourStar(2j+1)} (y_{0}^{N-1},u_{0}^{2j})
        &=& \begin{cases}
            L_{b}^{\ourStar} + L_{a}^{\ourStar} & \;\mbox{if}\; u=0\\ %
            L_{b}^{\ourStar} - L_{a}^{\ourStar} & \;\mbox{if}\; u=1  %
        \end{cases} \\
        &=&
        g_u(L_{a}^{\ourStar},L_{b}^{\ourStar})
    \end{IEEEeqnarray*}
Therefore, \eqref{eq: plus transform of labels} also holds with ``$\ourStar$'' in place of ``$\sim$''.
\end{IEEEproof}

\section{Proofs}
\subsection{Proofs for \cref{sec: posynomial representation}}

\begin{IEEEproof}[proof of \cref{lemm: recursive relation on Q-tilde joint-distribution}]
    We prove \eqref{eq: minus transform of Q-tilde joint-distribution} and \eqref{eq: plus transform of Q-tilde joint-distribution}.
    We begin with \eqref{eq: minus transform of Q-tilde joint-distribution}.
    Define $\theta \triangleq (y_0^{N-1},u_0^{2j-1})$, then by the definition of the synthetic min-sum joint distribution in \eqref{eq: definition of Q-tilde joint-distribution} we have
    \begin{IEEEeqnarray}{rcl}
        \label{eq: proof minus transform on jd}
        \tilde{Q}_N^{(2j)}(t;u_{2j}) \;\; = \sum_{\theta: \tilde{L}_N^{(2j)}\!(\theta) = t} W_N^{(2j)}(\theta;u_{2j}) \; .
    \end{IEEEeqnarray}
    Further define the following: $\alpha \triangleq (y_0^{N/2-1}, u_{0,e}^{2j-1} \oplus u_{0,o}^{2j-1} )$, $\beta \triangleq (y_{N/2}^{N-1}, u_{0,o}^{2j-1})$, $\tau_a \triangleq \tilde{L}_{N/2}^{(j)}(\alpha)$, and $\tau_b \triangleq \tilde{L}_{N/2}^{(j)}(\beta)$. 
    By \eqref{eq: minus transform of W joint-distribution} we have
    \[
        W_N^{(2j)}(\theta ; u_{2j}) = \!\! \sum_{u_{2j+1}} W_{N/2}^{(j)}(\alpha; u_{2j} \oplus u_{2j+1}) \cdot W_{N/2}^{(j)}(\beta; u_{2j+1}), 
    \]
    and by \eqref{eq: minus transform of labels} we have
    \[
        \tilde{L}_N^{(2j)}\!(\theta) = \fminsum \left( \tilde{L}_{N/2}^{(j)}(\alpha),\tilde{L}_{N/2}^{(j)}(\beta) \right) = \fminsum(\tau_a,\tau_b) \; .
    \]
    Therefore, by applying a change of variables from $\theta$ to $(\alpha, \beta)$, which by inspection iterate over the same set of possible values, we can rewrite the sum in \eqref{eq: proof minus transform on jd} as follows:
    \begin{IEEEeqnarray*}{c}
        \sum_{u_{2j+1}} \sum_{\substack{\alpha, \beta: \\ \fminsum(\tau_a,\tau_b) = t}} W_{N/2}^{(j)}(\alpha; u_{2j} \oplus u_{2j+1}) \cdot W_{N/2}^{(j)}(\beta; u_{2j+1}) \; .
    \end{IEEEeqnarray*}
    The sum over $\{{\alpha, \beta: \\ \fminsum(\tau_a,\tau_b) = t}\}$ can be modified into two sums: an outer sum over $\{{t_a, t_b: \\ \fminsum(t_a,t_b) = t}\}$ and an inner sum over $\{{\alpha, \beta: \\ \tilde{L}_{N/2}^{(j)}(\alpha) = t_a, \tilde{L}_{N/2}^{(j)}(\beta) = t_b}\}$.
    By doing that, the innermost sum becomes
    \begin{IEEEeqnarray}{rcl}
        \label{eq: proof minus transform on jd inner sum}
        \sum_{\substack{\alpha, \beta: \\ \tilde{L}_{N/2}^{(j)}(\alpha) = t_a \\ \tilde{L}_{N/2}^{(j)}(\beta) = t_b}} W_{N/2}^{(j)}(\alpha; u_{2j} \oplus u_{2j+1}) \cdot W_{N/2}^{(j)}(\beta; u_{2j+1}) \IEEEnonumber \\
         = \tilde{Q}_{N/2}^{(j)}(t_a; u_{2j} \oplus u_{2j+1}) \cdot \tilde{Q}_{N/2}^{(j)}(t_b; u_{2j+1}) \; .
    \end{IEEEeqnarray}
    where the equality follows by the definition in \eqref{eq: definition of Q-tilde joint-distribution}.
    Now, the result in \eqref{eq: proof minus transform on jd inner sum} is summed over $\{{u_{2j+1},t_a, t_b: \\ \fminsum(t_a,t_b) = t}\}$, which is \eqref{eq: minus transform of Q-tilde joint-distribution}.

    Similarly, \eqref{eq: plus transform of Q-tilde joint-distribution} can be obtained by following the same steps, using \eqref{eq: plus transform of W joint-distribution} instead of \eqref{eq: minus transform of W joint-distribution} and \eqref{eq: plus transform of labels} instead of \eqref{eq: minus transform of labels}.
    That is, define $\theta' \triangleq (\theta, u_{2j})$. Then by \eqref{eq: definition of Q-tilde joint-distribution} we have
    \begin{IEEEeqnarray}{rcl}
        \label{eq: proof plus transform on jd}
        \tilde{Q}_N^{(2j+1)}(t;u_{2j+1}) = \!\!\!\! \sum_{\theta': \tilde{L}_N^{(2j+1)}\!(\theta') = t} \!\!\!\! W_N^{(2j+1)}(\theta';u_{2j+1}) \; .
    \end{IEEEeqnarray}
    By \eqref{eq: plus transform of W joint-distribution} we have
    \[
        W_N^{(2j+1)}(\theta' ; u_{2j+1}) \! = \! W_{N/2}^{(j)}(\alpha; u_{2j} \oplus u_{2j+1}) \cdot W_{N/2}^{(j)}(\beta; u_{2j+1}), 
    \]
    and by \eqref{eq: plus transform of labels} we have
    \[
        \tilde{L}_N^{(2j)}\!(\theta') = g_{u_{2j}} \left( \tilde{L}_{N/2}^{(j)}(\alpha),\tilde{L}_{N/2}^{(j)}(\beta) \right) = g_{u_{2j}}(\tau_a,\tau_b) \; .
    \]
    Therefore, by applying a change of variables from $\theta'$ to $(\alpha, \beta, u_{2j})$, we can rewrite the sum in \eqref{eq: proof plus transform on jd} as follows:
    \begin{IEEEeqnarray*}{c}
        \sum_{u_{2j}} \sum_{\substack{\alpha, \beta: \\ g_{u_{2j}}(\tau_a,\tau_b) = t}} W_{N/2}^{(j)}(\alpha; u_{2j} \oplus u_{2j+1}) \cdot W_{N/2}^{(j)}(\beta; u_{2j+1})
    \end{IEEEeqnarray*}
    The rest of the proof is proceeds as before, by modifying the inner sum and using the definition in \eqref{eq: definition of Q-tilde joint-distribution}.
\end{IEEEproof}

\begin{IEEEproof}[proof of \cref{lemm: symmetry of Q-tilde joint-distribution}]
    We prove \eqref{eq: symmetry of Q-tilde joint-distribution} by induction.
    For the base case $(N=1, i=0)$, we have $W_1^{(0)}(y;x)=W(y;x)$ and $\tilde{L}_1^{(0)}(y) = \labeler(y)$, see \eqref{eq: labels basis condition}.
    By the first item of \cref{def: good labeler} we have $\labeler(\pi(y))=-\labeler(y)$.
    Therefore, by \eqref{eq: definition of Q-tilde joint-distribution} we have %
    \begin{IEEEeqnarray*}{rcl}
     \tilde{Q}_1^{(0)}(-t;x_i \oplus 1) \; &=& \sum_{y : \labeler(y)= -t} W(y;x_i \oplus 1) \\
     &=& \!\!\! \sum_{y : \labeler(\pi(y))=t} W(\pi(y);x_i) = \tilde{Q}_1^{(0)}(t;x_i) \; .
    \end{IEEEeqnarray*}
    We now assume that \eqref{eq: symmetry of Q-tilde joint-distribution} holds for $(\frac{N}{2},j)$ and show that it also holds for $(N,2j)$ and $(N,2j+1)$.
    We define the shorthands: $u_{2j} \triangleq u$, and $u_{2j+1} \triangleq v$.
    Then, for $(N,2j)$,
    \begin{IEEEeqnarray*}{rcl}
        \tilde{Q}_N^{(2j)}(-t,u \oplus 1) &\eqann{a}&
        \!\!\! \sum_{\substack{t_a,t_b,v:\\ \fminsum(t_a,t_b) = -t}} \!\!\! \tilde{Q}_{N/2}^{(j)}(t_a; u \oplus 1 \oplus v) \cdot \tilde{Q}_{N/2}^{(j)}(t_b;v) \\
        &\eqann{b}&
        \!\!\! \sum_{\substack{t_a,t_b,v:\\ \fminsum(t_a,t_b) = -t}} \!\!\! \tilde{Q}_{N/2}^{(j)}(-t_a; u \oplus v) \cdot \tilde{Q}_{N/2}^{(j)}(t_b;v) \\
        &\eqann{c}&
        \!\!\! \sum_{\substack{t_A,t_b,v:\\ \fminsum(-t_A,t_b) = -t}} \!\!\! \tilde{Q}_{N/2}^{(j)}(t_A; u \oplus v) \cdot \tilde{Q}_{N/2}^{(j)}(t_b;v) \\
        &\eqann{d}&
        \!\!\! \sum_{\substack{t_A,t_b,v:\\ \fminsum(t_A,t_b) = t}} \!\!\! \tilde{Q}_{N/2}^{(j)}(t_A; u \oplus v) \cdot \tilde{Q}_{N/2}^{(j)}(t_b;v) \\
        &\eqann{e}&
        \; \tilde{Q}_N^{(2j)}(t,u) \; ,
    \end{IEEEeqnarray*}
where \eqannref{a} and \eqannref{e} are by \eqref{eq: minus transform of Q-tilde joint-distribution}, \eqannref{b} is by the induction hypothesis, \eqannref{c} is by changing variables $t_A=-t_a$, and \eqannref{d} is by inspection of \eqref{eq: definition of min-sum f function} which reveals that $\fminsum(-t_A,t_b)=-t$ is the same condition as $\fminsum(t_A,t_b)=t$.

Similarly for $(N,2j+1)$,
    \begin{IEEEeqnarray*}{l}
        \tilde{Q}_N^{(2j+1)}(-t,v \oplus 1) \\
        \quad\quad \eqann{a}
        \!\!\! \sum_{\substack{t_a,t_b,u:\\ g_{u}(t_a,t_b) = -t}} \!\!\! \tilde{Q}_{N/2}^{(j)}(t_a; u \oplus v \oplus 1) \cdot \tilde{Q}_{N/2}^{(j)}(t_b;v \oplus 1) \\
        \quad\quad \eqann{b}
        \!\!\! \sum_{\substack{t_a,t_b,u:\\ g_{u}(t_a,t_b) = -t}} \!\!\! \tilde{Q}_{N/2}^{(j)}(-t_a; u \oplus v) \cdot \tilde{Q}_{N/2}^{(j)}(-t_b;v) \\
        \quad\quad \eqann{c}
        \!\!\! \sum_{\substack{t_A,t_B,u:\\ g_{u}(-t_A,-t_B) = -t}} \!\!\! \tilde{Q}_{N/2}^{(j)}(t_A; u \oplus v) \cdot \tilde{Q}_{N/2}^{(j)}(t_B;v) \\
        \quad\quad \eqann{d}
        \!\!\! \sum_{\substack{t_A,t_B,u:\\ g_{u}(t_A,t_B) = t}} \!\!\! \tilde{Q}_{N/2}^{(j)}(t_A; u \oplus v) \cdot \tilde{Q}_{N/2}^{(j)}(t_B;v) \\
        \quad\quad \eqann{e}
        \tilde{Q}_N^{(2j)}(t,v) \; ,
    \end{IEEEeqnarray*}
    where now \eqannref{a} and \eqannref{e} are by \eqref{eq: plus transform of Q-tilde joint-distribution}, \eqannref{b} is by the induction hypothesis, \eqannref{c} is by changing variables $t_A=-t_a$ and $t_B=-t_b$, and \eqannref{d} is by inspection of \eqref{eq: definition of g fucntion} which reveals that $g_v(-t_A,-t_b)=-t$ is the same condition as $g_v(t_A,t_B)=t$.

We now prove the claim in \cref{lemm: symmetry of Q-tilde joint-distribution} with the tildes removed. For the base case, the symmetry of $W$ implies that $L_1^{(0)}(\pi(y)) = -L_1^{(0)}(y)$, see \eqref{eq: LLRs basis condition}. Thus, by \eqref{eq: definition of Q joint-distribution}, we have 
    \begin{IEEEeqnarray*}{rcl}
	    Q_1^{(0)}(-t;x_i \oplus 1) \; &=& \sum_{y : L_1^{(0)}(y) = -t} W(y;x_i \oplus 1) \\
	                               \; &=& \sum_{y : L_1^{(0)}(\pi(y)) = t} W(\pi(y);x_i) = Q_1^{(0)}(t;x_i) \; .
    \end{IEEEeqnarray*}
    The induction is unchanged by removing the tildes. All that must be verified is that $f(-t_A,t_b) = -t$ is the same condition as $f(t_A,t_b) = t$. This follows by \eqref{eq: definition of f fucntion}, and recalling that $\tanh$ is an odd function.
\end{IEEEproof}

\begin{IEEEproof}[Proof of \cref{lemm: pe using Q-tilde joint-distribution}]
We prove \eqref{eq: pe using Q-tilde joint-distribution}. Denote by $\ind{A}$ an indicator of the event $A$.
\begin{IEEEeqnarray*}{rCl}
    \ourPe \left(\tilde{Q}_N^{(i)} \right) &=& \Pr(\hat{u}_i \neq u_i) = \sum_{t, u_i} \tilde{Q}_N^{(i)}(t;u_i) \cdot \ind{\hat{u}_i \neq u_i}  \\
    &=& \sum_{t} \tilde{Q}_N^{(i)}(t;0) {\cdot} \ind{\hat{u}_i \neq 0} + \sum_{t} \tilde{Q}_N^{(i)}(t;1) {\cdot} \ind{\hat{u}_i \neq 1}  \\
    &\eqann{a}& \sum_{t<0} \tilde{Q}_N^{(i)}(t;0) + \sum_{t \geq 0} \tilde{Q}_N^{(i)}(t;1)\\
    &\eqann{b}& \sum_{t<0} \tilde{Q}_N^{(i)}(t;0) + \sum_{t \geq 0} \tilde{Q}_N^{(i)}(-t;0)\\
    &=& \sum_{t<0} \tilde{Q}_N^{(i)}(t;0) + \sum_{t \leq 0 } \tilde{Q}_N^{(i)}(t;0)\\
    &=& \tilde{Q}_N^{(i)}(0;0) + 2 \cdot \sum_{t<0} \tilde{Q}_N^{(i)}(t;0) \; ,
\end{IEEEeqnarray*}
where \eqannref{a} is by the decision rule of the decoder as described in the $\algMakeDecision$ function given in \cref{alg: makeDecision}, and \eqannref{b} is by the symmetry property in \eqref{eq: symmetry of Q-tilde joint-distribution}.
\end{IEEEproof}
\begin{IEEEproof}[proof of \cref{cor: symmetry of posynomial}]
    The coefficient of $\xi^t$ in $\tilde{Q}_N^{(i)}(1/\xi)$ is the coefficient of $\xi^{-t}$ in $\tilde{Q}_N^{(i)}(\xi)$, which equals $\tilde{Q}_N^{(i)}(-t;0)$. By \eqref{eq: symmetry of Q-tilde joint-distribution} this is $\tilde{Q}_N^{(i)}(t;1)$.
\end{IEEEproof}

\begin{IEEEproof}[proof of \cref{lemm: Bhattacharyya-like bound}]
We recall \eqref{eq: pe using Q-tilde joint-distribution}--\eqref{eq: z-ish definition} and prove \eqref{eq: Bhattacharyya-like bound}.
 We have
 \begin{IEEEeqnarray*}{rCl}
	 \ourZ\left(\tilde{Q}_N^{(i)},\xi_0\right) & \eqann{a} & 2 \cdot  \sum_{t} \tilde{Q}_N^{(i)}(t;0) \cdot \xi_0^t \\ 
						   & \eqann[\geq]{b}& \tilde{Q}_N^{(i)}(0;0) + 2 \cdot \sum_{t < 0} \tilde{Q}_N^{(i)}(t;0) \cdot \xi_0^t \\\
						   &\eqann[\geq]{c}& \tilde{Q}_N^{(i)}(0;0) + 2 \cdot \sum_{t < 0} \tilde{Q}_N^{(i)}(t;0) \\
						   &\eqann{d}& \ourPe\left(\tilde{Q}_N^{(i)}\right) \; ,
 \end{IEEEeqnarray*}
 where \eqannref{a} is by \eqref{eq: posynomial definition} and \eqref{eq: z-ish definition}, \eqannref{b} is since we have thrown away non-negative terms, \eqannref{c} is since $\xi_0^t \geq 1$, for $t < 0$ and $0 < \xi_0 \leq 1$, and \eqannref{d} is by \eqref{eq: pe using Q-tilde joint-distribution}.
\end{IEEEproof}

We now prove \cref{lemm: posynomial minus transform bound,lemm: posynomial plus transform}, which directly leads to the proof of \cref{lemm: Bhattacharyya-like evolutions}.
\begin{IEEEproof}[proof of \cref{lemm: posynomial minus transform bound}]
    We prove \eqref{eq: posynomial minus transform bound}.  Using the shorthand $u_{2j+1} \triangleq v$ we have
    \begin{IEEEeqnarray*}{rCl}
         \tilde{Q}_N^{(2j)}(\xi) &\eqann{a}& \sum_t \tilde{Q}_N^{(2j)}(t;0) \xi^t 
         \IEEEnonumber \\
         &\eqann{b}& \sum_t \Bigg( \sum_{\substack{t_a,t_b,v:\\ \fminsum(t_a,t_b) = t}} \!\!\! \tilde{Q}_{N/2}^{(j)}(t_a; v) \cdot \tilde{Q}_{N/2}^{(j)}(t_b; v) \Bigg) \cdot \xi^t
    \end{IEEEeqnarray*}
    where \eqannref{a} is by \eqref{eq: posynomial definition} and \eqannref{b} is by \eqref{eq: minus transform of Q-tilde joint-distribution}.
    Evaluating the inner sum for the case $v=1$ yields
    \begin{IEEEeqnarray*}{l}
         \sum_{\substack{t_a,t_b:\\ \fminsum(t_a,t_b) = t}} \!\!\! \tilde{Q}_{N/2}^{(j)}(t_a; 1) \cdot \tilde{Q}_{N/2}^{(j)}(t_b; 1)  \\
         \quad \eqann{a} \sum_{\substack{t_a,t_b:\\ \fminsum(t_a,t_b) = t}} \!\!\! \tilde{Q}_{N/2}^{(j)}(-t_a; 0) \cdot \tilde{Q}_{N/2}^{(j)}(-t_b; 0) \\
         \quad \eqann{b} \sum_{\substack{t_A,t_B:\\ \fminsum(-t_A,-t_B) = t}} \!\!\! \tilde{Q}_{N/2}^{(j)}(t_A; 0) \cdot \tilde{Q}_{N/2}^{(j)}(t_B; 0) \\
         \quad \eqann{c} \sum_{\substack{t_A,t_B:\\ \fminsum(t_A,t_B) = t}} \!\!\! \tilde{Q}_{N/2}^{(j)}(t_A; 0) \cdot \tilde{Q}_{N/2}^{(j)}(t_B; 0) \; ,
    \end{IEEEeqnarray*}
    where \eqannref{a} is by \eqref{eq: symmetry of Q-tilde joint-distribution}, \eqannref{b} is by changing variables $t_A=-t_a$ and $t_B=-t_b$, and \eqannref{c} is by inspection of \eqref{eq: definition of min-sum f function} which reveals that $\fminsum(-t_A,-t_B)=\fminsum(t_A,t_B)$.
    Therefore the inner sum is the same for $v=1$ and for $v=0$.
    Therefore we have
    \begin{IEEEeqnarray}{rCl}
         \tilde{Q}_N^{(2j)}(\xi) &=& 2\cdot \sum_t \Bigg( \sum_{\substack{t_a,t_b:\\ \fminsum(t_a,t_b) = t}} \!\!\! \tilde{Q}_{N/2}^{(j)}(t_a; 0) \cdot \tilde{Q}_{N/2}^{(j)}(t_b; 0) \Bigg) \cdot \xi^t \IEEEnonumber \\
         &=& 2\cdot \sum_{t_a,t_b} \tilde{Q}_{N/2}^{(j)}(t_a; 0) \cdot \tilde{Q}_{N/2}^{(j)}(t_b; 0) \cdot \xi^{\fminsum(t_a,t_b)}
	 \label{eq: QN2j with tilde-f}
    \end{IEEEeqnarray}
     To upper bound the above expression for $0 < \xi_0 \leq 1$ in place of $\xi$, we divide all pairs $(t_a,t_b) \in \tilde{\mathcal{T}}_{N/2}^{(j)} \times \tilde{\mathcal{T}}_{N/2}^{(j)}$ into eight disjoint sets denoted $\{S_k\}_{k=1}^8$.
     For each set we evaluate $\xi_0^{\fminsum(t_a,t_b)}$ and upper bound this expression by either $\xi_0^{t_a}$ or $\xi_0^{t_b}$ as described in \cref{tbl: S18}.
     All the upper bounds are justified since for $0<\xi_0 \leq1$ the function $\xi_0^t$ is non-increasing in $t$.
     
     \begin{table}[ht!]
	\renewcommand{\arraystretch}{1.8}
        \centering
        \begin{tabular}{|c|c|c|c|c|} 
            \hline
	    & {\bm{$\!\sign(t_a)$}\!} & {\bm{$\!\sign(t_b)\!$}} & \bm{{$\!\!\min\{|t_a|,|t_b|\}\!\!$}} & \bm{$\!\xi_{0}^{\fminsum(t_a,t_b)}\!$} \\
            \noalign{\hrule height 2pt}
            \rowcolor{colorSetA}
            $S_1$ & $+/0$ & $+/0$ & $|t_a|$ & $\xi_0^{|t_a|}=\xi_0^{t_a}$ \\
            \hline
            \rowcolor{colorSetB}
            $S_2$ & $+/0$ & $-$ & $|t_a|$ & $\xi_0^{-|t_a|}=\xi_0^{-t_a} \leq \xi_0^{t_b}$ \\
            \hline
            \rowcolor{colorSetA}
            $S_3$ & $-$ & $+/0$ & $|t_a|$ & $\xi_0^{-|t_a|}=\xi_0^{t_a} $ \\
            \hline
            \rowcolor{colorSetB}
            $S_4$ & $-$ & $-$ & $|t_a|$ & $\xi_0^{|t_a|}=\xi_0^{-t_a} \leq \xi_0^{t_b}$ \\
            \hline
            \rowcolor{colorSetB}
            $S_5$ & $+/0$ & $+/0$ & $|t_b|$ & $\xi_0^{|t_b|}=\xi_0^{t_b}$ \\
            \hline
            \rowcolor{colorSetB}
            $S_6$ & $+/0$ & $-$ & $|t_b|$ & $\xi_0^{-|t_b|}=\xi_0^{t_b}$ \\
            \hline
            \rowcolor{colorSetA}
            $S_7$ & $-$ & $+/0$ & $|t_b|$ & $\xi_0^{-|t_b|}=\xi_0^{-t_b} \leq \xi_0^{t_a}$ \\
            \hline
            \rowcolor{colorSetA}
            $S_8$ & $-$ & $-$ & $|t_b|$ & $\xi_0^{|t_b|}=\xi_0^{-t_b} \leq \xi_0^{t_a}$ \\
            \hline
        \end{tabular}
        \vspace{5pt}
        \caption{The sets $S_1$ to $S_8$.} \label{tbl: S18}
    \end{table}%
We now define two disjoint sets regarding the two possible values for upper bounds described in the table. That is,
\begin{IEEEeqnarray*}{rCl}
    \highlight{colorSetA}{\mathcal{A}} & \triangleq & \highlight{colorSetA}{S_1 \sqcup S_3 \sqcup S_7 \sqcup S_8} \\
		& \subseteq & \left\{(t_a,t_b) \in \tilde{\mathcal{T}}_{N/2}^{(j)} \times \tilde{\mathcal{T}}_{N/2}^{(j)} : \xi_0^{\fminsum(t_a,t_b)} \leq \xi_0^{t_a} \right\} \; , \\
    \highlight{colorSetB}{\mathcal{B}} & \triangleq & \highlight{colorSetB}{S_2 \sqcup S_4 \sqcup S_5 \sqcup S_6} \\ 
		& \subseteq & \left\{(t_a,t_b) \in \tilde{\mathcal{T}}_{N/2}^{(j)} \times \tilde{\mathcal{T}}_{N/2}^{(j)} : \xi_0^{\fminsum(t_a,t_b)} \leq \xi_0^{t_b} \right\} \; ,
\end{IEEEeqnarray*}
where ``$\sqcup$'' denotes disjoint union, and the ``$\subseteq$'' relations follow from the last column of \cref{tbl: S18}.
Note that by definition
\begin{equation}
	\tilde{\mathcal{T}}_{N/2}^{(j)} \times \tilde{\mathcal{T}}_{N/2}^{(j)} = \mathcal{A} \sqcup \mathcal{B} \; .
	\label{eq: AcupB}
\end{equation}
Denote the shorthands $q \triangleq \tilde{Q}_{N/2}^{(j)}(t_a; 0) \cdot \tilde{Q}_{N/2}^{(j)}(t_b; 0)$ and $\tilde{\mathcal{T}} \triangleq\tilde{\mathcal{T}}_{N/2}^{(j)} \times \tilde{\mathcal{T}}_{N/2}^{(j)}$. By \eqref{eq: QN2j with tilde-f} we have
\begin{IEEEeqnarray*}{rCl}
	\tilde{Q}_N^{(2j)}(\xi_0) &=& 2\cdot \sum_{t_a,t_b \in \tilde{\mathcal{T}}} q \cdot \xi_0^{\fminsum(t_a,t_b)} \\
	 &\eqann{a}& 2\cdot \sum_{t_a,t_b \in \mathcal{A}} q \cdot \xi_0^{\fminsum(t_a,t_b)} + 2\cdot \sum_{t_a,t_b \in \mathcal{B}} q \cdot \xi_0^{\fminsum(t_a,t_b)} \\
	 &\eqann[\leq]{b}& 2\cdot \sum_{t_a,t_b \in \mathcal{A}} q \cdot \xi_0^{t_a} + 2\cdot \sum_{t_a,t_b \in \mathcal{B}} q \cdot \xi_0^{t_b} \\
	 &\eqann[\leq]{c}& 2\cdot \sum_{t_a,t_b \in \tilde{\mathcal{T}}} q \cdot \xi_0^{t_a} + 2\cdot \sum_{t_a,t_b \in \tilde{\mathcal{T}}} q \cdot \xi_0^{t_b} \\
	&=& 2\cdot \sum_{t_a} \tilde{Q}_{N/2}^{(j)}(t_a; 0) \cdot \xi_0^{t_a} \cdot \sum_{t_b} \tilde{Q}_{N/2}^{(j)}(t_b; 0) \\
	\IEEEeqnarraymulticol{3}{r}{+  2\cdot \sum_{t_a} \tilde{Q}_{N/2}^{(j)}(t_a; 0) \cdot \sum_{t_b} \tilde{Q}_{N/2}^{(j)}(t_b; 0) \cdot \xi_0^{t_b}} \\
	&\eqann{d}& 2 \cdot \frac{1}{2} \cdot \sum_{t_a} \tilde{Q}_{N/2}^{(j)}(t_a; 0) \cdot \xi_0^{t_a} \\
	\IEEEeqnarraymulticol{3}{r}{+ 2 \cdot \frac{1}{2} \cdot \sum_{t_b} \tilde{Q}_{N/2}^{(j)}(t_b; 0) \cdot \xi_0^{t_b}} \\
	&\eqann{e}& 2 \cdot \tilde{Q}_{N/2}^{(j)}(\xi_0) \; ,
\end{IEEEeqnarray*}
where \eqannref{a} is by \eqref{eq: AcupB}, \eqannref{b} is by the last column in \cref{tbl: S18}, \eqannref{c} is since we are adding non-negative terms, \eqannref{d} is since $\tilde{Q}_{N/2}^{(j)}(t;v)$ is a joint distribution and summing over all $t$ yields $\Pr(v=0)=1/2$, and \eqannref{e} is by \eqref{eq: posynomial definition}.
\end{IEEEproof}

\begin{IEEEproof}[proof of \cref{lemm: posynomial plus transform}]
    We prove \eqref{eq: posynomial plus transform}. Using the shorthand $u_{2j} \triangleq u$ we have
    \begin{IEEEeqnarray*}{rCl}
        \tilde{Q}_N^{(2j+1)}(\xi) &\eqann{a}& \sum_t \tilde{Q}_N^{(2j+1)}(t;0) \xi^t \\
        &\eqann{b}& \sum_t \Bigg( \sum_{\substack{t_a,t_b,u:\\ g_u(t_a,t_b) = t}} \!\!\! \tilde{Q}_{N/2}^{(j)}(t_a; u) \cdot \tilde{Q}_{N/2}^{(j)}(t_b; 0) \Bigg) \cdot \xi^t \\
        &=& \sum_{\substack{t_a,t_b,u}} \tilde{Q}_{N/2}^{(j)}(t_a; u) \cdot \tilde{Q}_{N/2}^{(j)}(t_b; 0) \cdot \xi^{g_u(t_a,t_b)} \\
        &=& \sum_{\substack{t_a,t_b}} \tilde{Q}_{N/2}^{(j)}(t_a; 0) \cdot \tilde{Q}_{N/2}^{(j)}(t_b; 0) \cdot \xi^{g_0(t_a,t_b)} \\
	\IEEEeqnarraymulticol{3}{r}{ + \sum_{\substack{t_a,t_b}} \tilde{Q}_{N/2}^{(j)}(t_a; 1) \cdot \tilde{Q}_{N/2}^{(j)}(t_b; 0) \cdot \xi^{g_1(t_a,t_b)}} \\
	&\eqann{c}& \sum_{\substack{t_a,t_b}} \tilde{Q}_{N/2}^{(j)}(t_a; 0) \cdot \tilde{Q}_{N/2}^{(j)}(t_b; 0) \cdot \xi^{t_a+t_b} \\
	\IEEEeqnarraymulticol{3}{r}{ + \sum_{\substack{t_a,t_b}} \tilde{Q}_{N/2}^{(j)}(t_a; 1) \cdot \tilde{Q}_{N/2}^{(j)}(t_b; 0) \cdot \xi^{t_b-t_a}} \\
        &=& \sum_{t_a} \tilde{Q}_{N/2}^{(j)}(t_a; 0) \cdot \xi^{t_a} \cdot \sum_{t_b} \tilde{Q}_{N/2}^{(j)}(t_b; 0) \cdot \xi^{t_b} \\
	\IEEEeqnarraymulticol{3}{r}{ + \sum_{t_a} \tilde{Q}_{N/2}^{(j)}(t_a; 1) \cdot \xi^{-t_a} \cdot \sum_{t_b} \tilde{Q}_{N/2}^{(j)}(t_b; 0) \cdot \xi^{t_b}} \\
	&\eqann{d}& 2 \cdot \left( \sum_{t_a} \tilde{Q}_{N/2}^{(j)}(t_a; 0) \cdot \xi^{t_a} \right)^2 = 2 \cdot \left( \tilde{Q}_{N/2}^{(j)}(\xi) \right)^2 \; ,
    \end{IEEEeqnarray*}
    where \eqannref{a} is by \eqref{eq: posynomial definition}, \eqannref{b} is by \eqref{eq: plus transform of Q-tilde joint-distribution}, \eqannref{c} is by \eqref{eq: definition of g fucntion}, and \eqannref{d} is by using change of variables $t_A=-t_a$ and the symmetry property in \eqref{eq: symmetry of Q-tilde joint-distribution}.
    That is, \eqannref{d} follows since
    \begin{IEEEeqnarray*}{rCl}
    \sum_{t_a} \tilde{Q}_{N/2}^{(j)}(t_a; 1) \cdot \xi^{-t_a} &=& \sum_{t_A} \tilde{Q}_{N/2}^{(j)}(-t_A; 1) \cdot \xi^{t_A} \\
    &=& \sum_{t_A} \tilde{Q}_{N/2}^{(j)}(t_A; 0) \cdot \xi^{t_A} \; .
    \end{IEEEeqnarray*}
\end{IEEEproof}

\begin{IEEEproof}[proof of \cref{lemm: Bhattacharyya-like evolutions}]
	We first consider the ``-'' case, and prove \eqref{eq: Z-ish minus evolution} and \eqref{eq: Z-ish-opt minus evolution}. To prove \eqref{eq: Z-ish minus evolution}, we have by \eqref{eq: posynomial minus transform bound} that for all $0<\xi_0\leq1$ that $\tilde{Q}_N^{(2j)}(\xi_0) \leq 2 \cdot \tilde{Q}_{N/2}^{(j)}(\xi_0)$. Plugging the definition of $\ourZ$ from \eqref{eq: z-ish definition}, which is $\tilde{Q}_N^{(2j)}(\xi_0) = \frac{1}{2} \cdot \ourZ\left(\tilde{Q}_N^{(2j)},\xi_0\right)$, yields \eqref{eq: Z-ish minus evolution}.
    To obtain \eqref{eq: Z-ish-opt minus evolution} we optimize both sides of \eqref{eq: Z-ish minus evolution} separately, such that $\ourZopt\left(\tilde{Q}_{N/2}^{(j)}\right)=\ourZ\left(\tilde{Q}_{N/2}^{(j)},\xi_{opt1}\right)$, and $\ourZopt\left(\tilde{Q}_{N}^{(2j)}\right)=\ourZ\left(\tilde{Q}_{N}^{(2j)},\xi_{opt2}\right)$. Therefore,
    \begin{multline*}
    \ourZopt\left(\tilde{Q}_{N}^{(2j)}\right) = \ourZ\left(\tilde{Q}_{N}^{(2j)},\xi_{opt2}\right) 
    \leq \ourZ\left(\tilde{Q}_{N}^{(2j)},\xi_{opt1 }\right) \\
    \leq 2 \cdot \ourZ\left(\tilde{Q}_{N/2}^{(j)},\xi_{opt1}\right) = 2 \cdot \ourZopt\left(\tilde{Q}_{N/2}^{(j)}\right) \; ,
    \end{multline*}
    where the first inequality is by the optimization and the second inequality is by \eqref{eq: Z-ish minus evolution}.
    
    We now consider the ``$+$'' case, and prove \eqref{eq: Z-ish plus evolution} and \eqref{eq: Z-ish-opt plus evolution}. To prove \eqref{eq: Z-ish plus evolution} we have by \eqref{eq: posynomial plus transform} (for all $\xi$ and specifically for $0<\xi_0\leq1$) that $\tilde{Q}_N^{(2j+1)}(\xi_0) = 2 \cdot \tilde{Q}_{N/2}^{(j)}(\xi_0)$. Again, plugging $\tilde{Q}_N^{(2j+1)}(\xi_0) = \frac{1}{2} \cdot \ourZ\left(\tilde{Q}_N^{(2j+1)},\xi_0\right)$, yields \eqref{eq: Z-ish plus evolution}.
    To obtain \eqref{eq: Z-ish-opt plus evolution} we use the same optimization argument as before. This concludes the proof. We remark that the equality in \eqref{eq: Z-ish plus evolution} implies that now $\xi_{opt1}=\xi_{opt2}$. This has a practical computational advantage: it implies that roughly half of the optimizations in \cref{sec: improved thresholds} need not be carried out.
\end{IEEEproof}

\begin{IEEEproof}[proof of \cref{lemm: posynomial minus transform}]
We prove \eqref{eq: posynomial minus transform} by considering each coefficient of $\tilde{Q}_N^{(2j)}(\xi)$ separately.
\begin{IEEEeqnarray*}{rCl}
	[\xi^t] \; \tilde{Q}_N^{(2j)}(\xi) &\eqann{a}& \tilde{Q}_N^{(2j)}(t;0) \\
					   &\eqann{b}& \sum_{\substack{t_a,t_b,u:\\ \fminsum(t_a,t_b) = t}} \!\!\! \tilde{Q}_{N/2}^{(j)}(t_a;u) \cdot \tilde{Q}_{N/2}^{(j)}(t_b;u) \\
				       &=& \sum_{\substack{t_a,t_b:\\ \fminsum(t_a,t_b) = t}} \!\!\! \tilde{Q}_{N/2}^{(j)}(t_a;0) \cdot \tilde{Q}_{N/2}^{(j)}(t_b;0) \\
    && \quad + \sum_{\substack{t_a,t_b:\\ \fminsum(t_a,t_b) = t}} \!\!\! \tilde{Q}_{N/2}^{(j)}(t_a;1) \cdot \tilde{Q}_{N/2}^{(j)}(t_b;1) \\
      &\eqann{c}& 2 \cdot \!\!\!\!\!\! \sum_{\substack{t_a,t_b:\\ \fminsum(t_a,t_b) = t}} \!\!\! \tilde{Q}_{N/2}^{(j)}(t_a;0) \cdot \tilde{Q}_{N/2}^{(j)}(t_b;0) \; . \IEEEyesnumber \label{eq: minus posynomial coefficient}
\end{IEEEeqnarray*}
where \eqannref{a} is by \eqref{eq: posynomial definition}, \eqannref{b} is by \eqref{eq: minus transform of Q-tilde joint-distribution}, and \eqannref{c} is since by
\eqref{eq: symmetry of Q-tilde joint-distribution} we have
\[\tilde{Q}_{N/2}^{(j)}(t_a;1) \cdot \tilde{Q}_{N/2}^{(j)}(t_b;1) = \tilde{Q}_{N/2}^{(j)}(-t_a;0) \cdot \tilde{Q}_{N/2}^{(j)}(-t_b;0) \; ,\]
and by \eqref{eq: definition of min-sum f function} we have $\fminsum(-t_a,-t_b)=f(t_a,t_b)$.

We define the set $\mathcal{C}_t$ as the set of pairs $(t_a,t_b)$ which contribute to the sum in \eqref{eq: minus posynomial coefficient},
\begin{IEEEeqnarray}{rCl}
\mathcal{C}_t = \left\{ (t_a,t_b) \in \tilde{\mathcal{T}}_{N/2}^{(j)} \times \tilde{\mathcal{T}}_{N/2}^{(j)} : \fminsum(t_a,t_b) = t \right\} \; .
\label{eq: Ct set}
\end{IEEEeqnarray}
We first consider the case $t>0$.
By inspection of $\fminsum$ in \eqref{eq: definition of min-sum f function}, the set $\mathcal{C}_t$ can be partitioned into six disjoint sets denoted $S_{\mathrm{I}}, S_\mathrm{{I}I}, \ldots S_\mathrm{{VI}}$ and defined as follows:
\begin{IEEEeqnarray}{lClCrClCr}
    \subnumberinglabel{eq: Roman sets}
	S_{\mathrm{I}} &\triangleq& \{ (t_a,t_b) \in \mathcal{C}_t : t_a &=& t &,& t_b &>& t \} \; ,  
    \label{eq: Roman set I} \\
	S_{\mathrm{II}} &\triangleq& \{ (t_a,t_b) \in \mathcal{C}_t : t_a &>& t &,& t_b &=& t \} \; , 
    \label{eq: Roman set II} \\
	S_{\mathrm{III}} &\triangleq& \{ (t_a,t_b) \in \mathcal{C}_t : t_a &=& -t &,& t_b &<& -t \} \; ,
    \label{eq: Roman set III} \\
	S_{\mathrm{IV}} &\triangleq& \{ (t_a,t_b) \in \mathcal{C}_t : t_a &<& -t &,& t_b &=& -t \} \; ,
    \label{eq: Roman set IV} \\
	S_{\mathrm{V}} &\triangleq& \{ (t_a,t_b) \in \mathcal{C}_t : t_a &=& t &,& t_b &=& t \} \; ,
    \label{eq: Roman set V} \\
	S_{\mathrm{VI}} &\triangleq& \{ (t_a,t_b) \in \mathcal{C}_t : t_a &=& -t &,& t_b &=& -t \} \; .
    \label{eq: Roman set VI}
\end{IEEEeqnarray}
Thus, the sum in \eqref{eq: minus posynomial coefficient} can be broken into six sums.
By the symmetry between $t_a$ and $t_b$ in \eqref{eq: minus posynomial coefficient}, both \eqref{eq: Roman set I} and \eqref{eq: Roman set II} have the same contribution, as well as both \eqref{eq: Roman set III} and \eqref{eq: Roman set IV}.
We now consider the contribution of each sum.

For $S_{\mathrm{I}}$ (and also for $S_{\mathrm{II}}$, as explained above)
we have
\begin{IEEEeqnarray*}{rCl}
	\IEEEeqnarraymulticol{3}{l}{\sum_{(t_a,t_b) \in S_{\mathrm{I}}} \tilde{Q}_{N/2}^{(j)}(t_a;0) \cdot \tilde{Q}_{N/2}^{(j)}(t_b;0)} \\
	\quad &\eqann{a}& \sum_{t_a=t} \tilde{Q}_{N/2}^{(j)}(t_a;0) \cdot \sum_{t_b>t}  \tilde{Q}_{N/2}^{(j)}(t_b;0) \\
	\quad &\eqann{b}&  [\xi^t] \; \tilde{Q}_{N/2}^{(j)}(\xi) \cdot [\xi^t] \; \tilde{A}_{N/2}^{(j)}(\xi)  \\
    \quad &\eqann{c}&  [\xi^t] \; \tilde{Q}_{N/2}^{(j)}(\xi) \cdot [\xi^t] \; \ourpos{\tilde{A}_{N/2}^{(j)}(\xi)} \; ,
\end{IEEEeqnarray*}
where \eqannref{a} is by \eqref{eq: Roman set I}, \eqannref{b} is by \eqref{eq: posynomial definition} and \eqref{eq: definition of A posynomial}, and \eqannref{c} is by \eqref{eq: definition of pos operator} since $t>0$.

For $S_{\mathrm{III}}$ (and also for $S_{\mathrm{IV}}$, as explained above) we have
\begin{IEEEeqnarray*}{rCl}
	\IEEEeqnarraymulticol{3}{l}{\sum_{(t_a,t_b) \in S_{\mathrm{III}}} \tilde{Q}_{N/2}^{(j)}(t_a;0) \cdot \tilde{Q}_{N/2}^{(j)}(t_b;0)} \\
	\quad &\eqann{a}& \sum_{t_a=-t} \tilde{Q}_{N/2}^{(j)}(t_a;0) \cdot \sum_{t_b<-t}  \tilde{Q}_{N/2}^{(j)}(t_b;0) \\
	\quad &\eqann{b}&  [\xi^{-t}] \; \tilde{Q}_{N/2}^{(j)}(\xi) \cdot [\xi^{-t}] \; \tilde{B}_{N/2}^{(j)}(\xi)  \\
    \quad &\eqann{c}&  [\xi^t] \; \tilde{Q}_{N/2}^{(j)}(1/\xi) \cdot [\xi^t] \; \ourneg{\tilde{B}_{N/2}^{(j)}(\xi)} \; ,
\end{IEEEeqnarray*}
where \eqannref{a} is by \eqref{eq: Roman set III}, \eqannref{b} is by \eqref{eq: posynomial definition} and \eqref{eq: definition of B posynomial}, and \eqannref{c} is by \eqref{eq: definition of neg operator} since $t>0$.

For $S_{\mathrm{V}}$ we have
\begin{IEEEeqnarray*}{rCl}
	\IEEEeqnarraymulticol{3}{l}{\sum_{(t_a,t_b) \in S_{\mathrm{V}}} \tilde{Q}_{N/2}^{(j)}(t_a;0) \cdot \tilde{Q}_{N/2}^{(j)}(t_b;0)} \\
	\quad &\eqann{a}& \sum_{t_a=t} \tilde{Q}_{N/2}^{(j)}(t_a;0) \cdot \sum_{t_b=t}  \tilde{Q}_{N/2}^{(j)}(t_b;0) \\
	\quad &\eqann{b}&  [\xi^{t}] \; \tilde{Q}_{N/2}^{(j)}(\xi) \cdot [\xi^{t}] \; \tilde{Q}_{N/2}^{(j)}(\xi) \; , \IEEEyesnumber \label{eq: SV sum} 
\end{IEEEeqnarray*}
where \eqannref{a} is by \eqref{eq: Roman set V} and \eqannref{b} is by \eqref{eq: posynomial definition}.%

For $S_{\mathrm{VI}}$ we have
\begin{IEEEeqnarray*}{rCl}
	\IEEEeqnarraymulticol{3}{l}{\sum_{(t_a,t_b) \in S_{\mathrm{VI}}} \tilde{Q}_{N/2}^{(j)}(t_a;0) \cdot \tilde{Q}_{N/2}^{(j)}(t_b;0)} \\
	\quad &\eqann{a}& \sum_{t_a=-t} \tilde{Q}_{N/2}^{(j)}(t_a;0) \cdot \sum_{t_b=-t}  \tilde{Q}_{N/2}^{(j)}(t_b;0) \\
	\quad &\eqann{b}&  [\xi^{-t}] \; \tilde{Q}_{N/2}^{(j)}(\xi) \cdot [\xi^{-t}] \; \tilde{Q}_{N/2}^{(j)}(\xi) \\ 
	\quad &=&  [\xi^{t}] \; \tilde{Q}_{N/2}^{(j)}(1/\xi) \cdot [\xi^{t}] \; \tilde{Q}_{N/2}^{(j)}(1/\xi)  \; , \IEEEyesnumber \label{eq: SVI sum}
\end{IEEEeqnarray*}
where \eqannref{a} is by \eqref{eq: Roman set VI} and \eqannref{b} is by \eqref{eq: posynomial definition}.%

Plugging all six sums into \eqref{eq: minus posynomial coefficient} yields (for $t>0$)
\begin{IEEEeqnarray}{rCl}
    \IEEEeqnarraymulticol{3}{l}{[\xi^t] \; \tilde{Q}_N^{(2j)}(\xi) = } 
    \label{eq: minus posynomial coefficient t>0} \\
    \quad &&\! \phantom{{} + {} }  [\xi^t] \; 2 \left(\tilde{Q}_{N/2}^{(j)}(\xi) \odot \left( 2 \cdot \ourpos{\tilde{A}_{N/2}^{(j)}(\xi)} + \tilde{Q}_{N/2}^{(j)}(\xi)\right) \right) \IEEEnonumber \\
	  && \! {} + [\xi^t] \; 2 \left(\tilde{Q}_{N/2}^{(j)}(1/\xi) \odot \left( 2 \cdot \ourneg{\tilde{B}_{N/2}^{(j)}(\xi)} + \tilde{Q}_{N/2}^{(j)}(1/\xi) \right) \right) \; , \IEEEnonumber
    \end{IEEEeqnarray}
where we used the definition of ``$\odot$'' in \eqref{eq: definition of Hadamard product}.

We now consider the case $t<0$.
Similarly to what we did before for the case $t>0$, we partition the set $\mathcal{C}_t$ defined in \eqref{eq: Ct set} into six disjoined sets denoted $\tilde{S}_{\mathrm{I}}, \tilde{S}_\mathrm{{I}I}, \ldots \tilde{S}_\mathrm{{VI}}$ as follows:
\begin{IEEEeqnarray}{lClCrClCr}
    \subnumberinglabel{eq: Roman sets tilde}
	\tilde{S}_{\mathrm{I}} &\triangleq& \{ (t_a,t_b) \in \mathcal{C}_t : t_a &=& t &,& t_b &>& -t \} \; ,  
    \label{eq: Roman set I tilde} \\
	\tilde{S}_{\mathrm{II}} &\triangleq& \{ (t_a,t_b) \in \mathcal{C}_t : t_a &>& -t &,& t_b &=& t \} \; , 
    \label{eq: Roman set II tilde} \\
	\tilde{S}_{\mathrm{III}} &\triangleq& \{ (t_a,t_b) \in \mathcal{C}_t : t_a &=& -t &,& t_b &<& t \} \; ,
    \label{eq: Roman set III tilde} \\
	\tilde{S}_{\mathrm{IV}} &\triangleq& \{ (t_a,t_b) \in \mathcal{C}_t : t_a &<& t &,& t_b &=& -t \} \; ,
    \label{eq: Roman set IV tilde} \\
	\tilde{S}_{\mathrm{V}} &\triangleq& \{ (t_a,t_b) \in \mathcal{C}_t : t_a &=& t &,& t_b &=& t \} \; ,
    \label{eq: Roman set V tilde} \\
	\tilde{S}_{\mathrm{VI}} &\triangleq& \{ (t_a,t_b) \in \mathcal{C}_t : t_a &=& -t &,& t_b &=& -t \} \; .
    \label{eq: Roman set VI tilde}
\end{IEEEeqnarray}
We now consider the contribution of each sum.

For $\tilde{S}_{\mathrm{I}}$ (and also for $\tilde{S}_{\mathrm{II}}$) we have
\begin{IEEEeqnarray*}{rCl}
	\IEEEeqnarraymulticol{3}{l}{\sum_{(t_a,t_b) \in \tilde{S}_{\mathrm{I}}} \tilde{Q}_{N/2}^{(j)}(t_a;0) \cdot \tilde{Q}_{N/2}^{(j)}(t_b;0)} \\
	\quad &\eqann{a}& \sum_{t_a=t} \tilde{Q}_{N/2}^{(j)}(t_a;0) \cdot \sum_{t_b>-t}  \tilde{Q}_{N/2}^{(j)}(t_b;0) \\
	\quad &\eqann{b}&  [\xi^t] \; \tilde{Q}_{N/2}^{(j)}(\xi) \cdot [\xi^{-t}] \; \tilde{A}_{N/2}^{(j)}(\xi)  \\
    \quad &\eqann{c}&  [\xi^t] \; \tilde{Q}_{N/2}^{(j)}(\xi) \cdot [\xi^t] \; \ourpos{\tilde{A}_{N/2}^{(j)}(\xi)} \; ,
\end{IEEEeqnarray*}
where \eqannref{a} is by \eqref{eq: Roman set I tilde}, \eqannref{b} is by \eqref{eq: posynomial definition} and \eqref{eq: definition of A posynomial}, and \eqannref{c} is by \eqref{eq: definition of pos operator} since $t<0$.

For $\tilde{S}_{\mathrm{III}}$ (and also for $\tilde{S}_{\mathrm{IV}}$) we have
\begin{IEEEeqnarray*}{rCl}
	\IEEEeqnarraymulticol{3}{l}{\sum_{(t_a,t_b) \in \tilde{S}_{\mathrm{III}}} \tilde{Q}_{N/2}^{(j)}(t_a;0) \cdot \tilde{Q}_{N/2}^{(j)}(t_b;0)} \\
	\quad &\eqann{a}& \sum_{t_a=-t} \tilde{Q}_{N/2}^{(j)}(t_a;0) \cdot \sum_{t_b<t}  \tilde{Q}_{N/2}^{(j)}(t_b;0) \\
	\quad &\eqann{b}&  [\xi^{-t}] \; \tilde{Q}_{N/2}^{(j)}(\xi) \cdot [\xi^{t}] \; \tilde{B}_{N/2}^{(j)}(\xi)  \\
    \quad &\eqann{c}&  [\xi^t] \; \tilde{Q}_{N/2}^{(j)}(1/\xi) \cdot [\xi^t] \; \ourneg{\tilde{B}_{N/2}^{(j)}(\xi)} \; ,
\end{IEEEeqnarray*}
where \eqannref{a} is by \eqref{eq: Roman set III tilde}, \eqannref{b} is by \eqref{eq: posynomial definition} and \eqref{eq: definition of B posynomial}, and \eqannref{c} is by \eqref{eq: definition of neg operator} since $t<0$.

For $\tilde{S}_{\mathrm{V}}$ and for $\tilde{S}_{\mathrm{VI}}$ we have exactly the same expressions as for $S_{\mathrm{V}}$ and $S_{\mathrm{VI}}$ given in \eqref{eq: SV sum} and \eqref{eq: SVI sum}, respectively. Indeed, this is because in deriving these we did not use the assumption that $t>0$, and by comparing \eqref{eq: Roman set V} and \eqref{eq: Roman set VI} to \eqref{eq: Roman set V tilde} and \eqref{eq: Roman set VI tilde}, respectively.  
Plugging all six sums into \eqref{eq: minus posynomial coefficient}, reveals that \eqref{eq: minus posynomial coefficient t>0} also holds for $t<0$.

For the case $t=0$ we follow the same steps as those for the case $t>0$, but note that now $S_{\mathrm{V}}=S_{\mathrm{VI}}$. That is, the contribution of $S_{\mathrm{V}} \cup S_{\mathrm{VI}}$ is now
\[
	[\xi^{0}] \; \tilde{Q}_{N/2}^{(j)}(\xi) \cdot [\xi^{0}] \; \tilde{Q}_{N/2}^{(j)}(\xi) = [\xi^{0}] \; \tilde{Q}_{N/2}^{(j)}(1/\xi) \cdot [\xi^{0}] \; \tilde{Q}_{N/2}^{(j)}(1/ \xi) \; ,
\]
as opposed to the contribution for $t>0$: 
\[
	[\xi^{t}] \; \tilde{Q}_{N/2}^{(j)}(\xi) \cdot [\xi^{t}] \; \tilde{Q}_{N/2}^{(j)}(\xi) +  
	[\xi^{t}] \; \tilde{Q}_{N/2}^{(j)}(1/\xi) \cdot [\xi^{t}] \; \tilde{Q}_{N/2}^{(j)}(1/ \xi) \; .
\]
Thus, \eqref{eq: minus posynomial coefficient t>0} will hold also for $t = 0$, once we subtract $2 \left([\xi^0] \; \tilde{Q}_{N/2}^{(j)} \right)^2$, which yields \eqref{eq: posynomial minus transform}.
\end{IEEEproof}

\subsection{Proofs for \cref{sec: finite length case}}
This subsection is devoted to proving the results in \cref{sec: finite length case} regarding  the complexity of calculations for the finite-length case. That is, we show how the posynomials $\tilde{Q}_N^{(i)}(\xi)$ are efficiently calculated. Recall that $\tilde{Q}_N^{(i)}(\xi)$ is defined in \eqref{eq: posynomial definition}, where the summation index $t$ ranges over $\mathcal{\tilde{T}}_N^{(i)}$ given in \eqref{eq: tilde TNi}.

Recall that $\mathcal{\tilde{T}}_N^{(i)}$ contains all integers from $-\gamma \cdot 2^{\mathrm{wt}(i)}$ to $\gamma \cdot 2^{\mathrm{wt}(i)}$. Thus, we represent $\tilde{Q}_N^{(i)}(\xi)$ by an array indexed over this integer range, where entry $t$ contains the coefficient of $\xi^t$.
Namely, if we denote this array as $q[\cdot]$, then $q[t]=[\xi^t] \; \tilde{Q}_N^{(i)}(\xi)$.
The same representation is used for all posynomials arising from intermediate steps in the calculation.

\begin{IEEEproof}[proof of \cref{lemm: complexity of calc minus posynomial}]
	We first note by inspection that all intermediate posynomials taking part in the calculation have the same range of indices. That is, all these posynomials have indices ranging over $\mathcal{\tilde{T}}_{N/2}^{(j)} = \mathcal{\tilde{T}}_{N}^{(2j)}$, where the equality is also in accordance with \eqref{eq: tilde TNi} since $\mathrm{wt}(2j) = \mathrm{wt}(j)$.
	The result will follow by showing that all the intermediate calculations in \eqref{eq: posynomial minus transform} can be carried out in linear time. That is, in time $\bigO\left(|\mathcal{\tilde{T}}_{N/2}^{(j)}|\right)$.

Denote the array of $\tilde{Q}_{N/2}^{(j)}(\xi)$ as $q[\cdot]$.
First consider the calculation of $\tilde{A}_{N/2}^{(j)}(\xi)$, defined in \eqref{eq: definition of A posynomial}.
This is done by allocating an array $a[\cdot]$, indexed over $\mathcal{\tilde{T}}_{N/2}^{(j)}$, and populating its entries from highest to lowest. That is, we set $a[\gamma \cdot 2^{\mathrm{wt}(j)}]=0$ and for all smaller $t$ we set $a[t]=a[t+1] + q[t+1]$. Clearly, this calculation is linear in the size of $a[\cdot]$. Similarly, $\tilde{B}_{N/2}^{(j)}(\xi)$, defined in \eqref{eq: definition of B posynomial}, is calculated by allocating an array $b[\cdot]$, and populating its entries from lowest to highest such that $b[-\gamma \cdot 2^{\mathrm{wt}(j)}]=0$ and for all larger $t$, $b[t]=b[t-1] + q[t-1]$. This operation is linear as well. By inspection, all the other operations involved in \eqref{eq: posynomial minus transform} are also linear. Note that $\tilde{Q}_{N/2}^{(j)}(1/\xi)$ is simply the posynomial $\tilde{Q}_{N/2}^{(j)}(\xi)$, reversed. That is, $[\xi^t] \; \tilde{Q}_{N/2}^{(j)}(1/\xi) = [\xi^{-t}] \; \tilde{Q}_{N/2}^{(j)}(\xi)$.
\end{IEEEproof}

\begin{IEEEproof}[proof of \cref{lemm: complexity of calc plus posynomial}]
By \eqref{eq: tilde TNi}, the largest and smallest powers of $\tilde{Q}_{N/2}^{(j)}(\xi)$ are $\gamma' \triangleq 2^{\mathrm{wt}(j)} \gamma$ and $-\gamma'$, respectively. We recast \eqref{eq: posynomial plus transform} as follows:
\[
\tilde{Q}_N^{(2j+1)}(\xi)  = 2 \cdot \left(\tilde{Q}_{N/2}^{(j)}(\xi) \right)^2 = 2 \cdot \xi^{-2 \gamma'} \cdot \left(\xi^{\gamma'} \cdot \tilde{Q}_{N/2}^{(j)}(\xi) \right)^2 \; .
\]
Notice that $\xi^{\gamma'} \cdot \tilde{Q}_{N/2}^{(j)}(\xi)$ is a polynomial.
Therefore, the complexity of calculating $\tilde{Q}_N^{(2j+1)}(\xi)$ is that of squaring a polynomial of degree  $2 \gamma'$.
By \cite[Chapter 30]{CLRS:22b}, this can be done using fast Fourier transform in time $\bigO(2\gamma' \cdot \log( 2 \gamma')) = \bigO\left(|\mathcal{\tilde{T}}_{N/2}^{(j)}|\cdot \log(|\mathcal{\tilde{T}}_{N/2}^{(j)}|)\right)$, where the equality follows by \eqref{eq: tilde TNi}. 
\end{IEEEproof}

\begin{IEEEproof}[proof of \cref{thm: complexity of finite length case}]
	We first note that both $|\tilde{T}_{N}^{(2j)}|$ and $|\tilde{T}_{N}^{(2j+1)}|$ are at least $|\tilde{T}_{N/2}^{(j)}|$, by \eqref{eq: tilde TNi}. Hence, by
	\cref{lemm: complexity of calc minus posynomial,lemm: complexity of calc plus posynomial} we may bound the computational complexity of calculating $\tilde{Q}^{(i)}(\xi)$ from $\tilde{Q}^{(\lfloor i/2\rfloor)}(\xi)$ by $\bigO(|\tilde{T}_{N}^{(i)}| \log|\tilde{T}_{N}^{(i)}|)$. That is, using \eqref{eq: tilde TNi}, by $\bigO(2^{\mathrm{wt}(i)} \cdot \gamma \cdot   \log(2^{\mathrm{wt}(i)} \cdot \gamma))$.

	For $1 \leq m \leq n$, consider the complexity of the last step of calculating all $\tilde{Q}_{M}^{(j)}(\xi)$, where $0 \leq j < M = 2^m$. That is, calculating the $\tilde{Q}_{M}^{(j)}(\xi)$, when we have already calculated all $\tilde{Q}_{M/2}^{(k)}(\xi)$, $0 \leq k < M/2 = 2^{m-1}$. Since the number of indices $j$ of weight $w$ is $\binom{m}{w}$, the complexity is of order
	\[
\sum_{w=0}^m \binom{m}{w} 2^w \gamma \log_2(2^w \cdot \gamma) \; .
	\]
	Thus, the overall complexity is of order
	\begin{equation}
		\label{eq: overall complexity double sum}
		\sum_{m=1}^n \sum_{w=0}^m \binom{m}{w} 2^w \gamma \log_2(2^w \cdot \gamma) \; .
	\end{equation}	
	We start by bounding the inner sum in \eqref{eq: overall complexity double sum}. We have
	\begin{multline}
		\label{eq: two sums}
\sum_{w=0}^m \binom{m}{w} 2^w \gamma \log_2(2^w \cdot \gamma) \\
= \gamma \sum_{w=0}^m \binom{m}{w} 2^w  w +  \gamma \log_2 \gamma \sum_{w=0}^m \binom{m}{w} 2^w  \; .
\end{multline}
The first sum on the RHS of \eqref{eq: two sums} is bounded by	
\begin{IEEEeqnarray*}{rCl}
	\sum_{w=0}^m \binom{m}{w} 2^w w & = & \sum_{w=1}^m \binom{m}{w} 2^w w \\
	      & = & \sum_{w=1}^m \frac{m!}{w! (m-w)!} 2^w w \\
	      & = & \sum_{w=1}^m \frac{m \cdot (m-1)!}{(w-1)! (m-w)!} 2^w \\
	      & = & \sum_{w=1}^m m \binom{m-1}{w-1} 2^w \\
	      & = & 2 m \sum_{w=1}^m  \binom{m-1}{w-1} 2^{w-1} 1^{(m-1)-(w-1)} \\
	      & = & 2 m \sum_{k=0}^{m-1}  \binom{m-1}{k} 2^{k} 1^{(m-1)-k} \\
	      & \eqann{a} & 2 m (1 + 2)^{m-1} \\
	      & = & 2 m \cdot 3^{m-1} \; ,
\end{IEEEeqnarray*}
where \eqannref{a} follows by the binomial theorem.

The second sum on the RHS of \eqref{eq: two sums} is bounded by	
\[
	\sum_{w=0}^m \binom{m}{w} 2^w  = \sum_{w=0}^m \binom{m}{w} 2^w 1^{m-w} = 3^m \; .
\]

Plugging the above simplifications into \eqref{eq: overall complexity double sum} yields an overall complexity of order
\begin{IEEEeqnarray*}{rCl}
	\IEEEeqnarraymulticol{3}{l}{\sum_{m=1}^n \gamma \cdot 2m \cdot 3^{m-1} + \gamma \log_2 \gamma \cdot  3^m} \\
	\quad &=& \gamma \sum_{m=1}^n \left(\frac{2}{3} \cdot m + \log_2 \gamma \right) \cdot 3^m \\
	      & \eqann{a} & \gamma \left( \frac{2}{3} \cdot \frac{3}{4} \left( 1 + (2n-1) \cdot 3^n \right) + \log_2 \gamma \cdot \frac{3^{n+1} - 3}{3-1} \right) \\
	      & \leq & \gamma \left( \frac{1}{2} \left( 2n \cdot 3^n \right) + \log_2 \gamma \cdot 3^{n+1} \right) \\
	      & = & \gamma \left( n \cdot 3^n  + \log_2 \gamma \cdot 3^{n+1} \right) \\
	      & = & \gamma \left( n \cdot 2^{n \log_2 3}  + \log_2 \gamma \cdot 3 \cdot 2^{n \log_2 3} \right) \\
	      & = & \gamma \left( \log_2 N \cdot N^{\log_2 3}  + \log_2 \gamma \cdot 3 \cdot N^{\log_2 3} \right) \\
	      & = & \bigO( \gamma \cdot N^{\log_2 3} \log N  + \gamma \log \gamma \cdot N^{\log_2 3} ) \; ,
\end{IEEEeqnarray*}
where \eqannref{a} follows by the methods in \cite[Section 2.6]{GKP:94b}.
The above can be further bounded as $\bigO( \gamma \log \gamma \cdot N^{\log_2 3} \log N)$. Since $1.585 > \log_2 3$, we can also bound this as $\bigO( \gamma \log \gamma \cdot N^{1.585})$. 
\end{IEEEproof}

\subsection{Proofs for \cref{sec: asymptotic case}}
\begin{IEEEproof}[proof of \cref{corollary: of universal}]
    The corollary is obtained by a reduction to \cref{prop: fromUniversal} with appropriate parameters.
    Since the same symbols $n'$, $\eta$, $\epsilon'$, and $\delta'$ are used in both \cref{corollary: of universal} and \cref{prop: fromUniversal}, we apply hats to all symbols in \cref{corollary: of universal} to avoid ambiguity.

    Thus, our setting is that we are given $\hat{\epsilon}'$, $\hat{\eta}$, and must find $\hat{n}'(\hat{\epsilon}', \hat{\eta})$ such that if $S_0 \leq \hat{\eta}$, then
	\begin{equation*}
		\Pr \left( S_n \leq \hat{\epsilon}' \textup{ for all } n \geq \hat{n}' \right) \geq 1 - \hat{\delta}'(\eta) \; ,
\end{equation*}
where 
\begin{equation}
	\hat{\delta}'(\hat{\eta}) = 2 \cdot (8 \hat{\eta})^{\log_2 \varphi} \; .
\label{eq: delta' definition hat}
\end{equation}

We now consider \cref{prop: fromUniversal} with $\epsilon' = \hat{\epsilon}'$, $\delta' = \hat{\delta}'(\hat{\eta})$, and $\kappa = 2$. Consider first the corresponding $\eta$. By \eqref{eq: eta of delta} and \eqref{eq: delta' definition hat}, this is
\begin{IEEEeqnarray*}{rCl}
    \eta = \eta(\epsilon', \delta')
&=& \frac{1}{8} \left( \frac{\delta'}{2} \right)^{1/\log_2{\varphi}}
= \frac{1}{8} \left( \frac{\hat{\delta}'(\hat{\eta})}{2} \right)^{1/\log_2{\varphi}} \\
&=& \frac{1}{8} \left( \frac{1}{2} \cdot 2 (8\cdot \hat{\eta})^{\log_2{\varphi}} \right)^{1/\log_2{\varphi}}
= \hat{\eta}
\; .
\end{IEEEeqnarray*}
Thus, if $S_0 \leq \hat{\eta}$, we have for $n'(\epsilon',\delta',\kappa = 2)$ that \eqref{eq: fromUniversal} holds. Comparing \eqref{eq: fromUniversal} and \eqref{eq: of universal}, we deduce that we may take $\hat{n}'(\hat{\epsilon}', \hat{\eta}) \triangleq n'(\hat{\epsilon}', \hat{\delta}'(\hat{\eta}), \kappa = 2)$, where $n'$ is the function promised in \cref{prop: fromUniversal}.
\end{IEEEproof}

\begin{IEEEproof}[proof of \cref{prop: augmented proposition}]
    We prove \eqref{eq: augmented proposition} by following similar steps as in \cite{Tal:17.2p}.
    The proof is given for completeness.
    Let $\varepsilon_a, \varepsilon_b >0$ and $n_a<n_b$ be parameters.
    Define the following events:
    \begin{IEEEeqnarray}{rl}
        A: &\quad S_n \leq \varepsilon_a \;\mbox{for all}\; n \geq n_a \; ,\\
        B: &\quad \left| \frac{|\{n_a<i<n : T_i = t\}|}{n-n_a} - \frac{1}{2} \right| \leq \varepsilon_b \; , \IEEEnonumber \\
        & \;\mbox{for all }\; n \geq n_b \;\mbox{and all}\; t \in \{0,1\} \; .
    \end{IEEEeqnarray}

    We will use the following three observations shortly:
    \begin{enumerate}
	    \item \label{it:na} By \cref{corollary: of universal}, for given $\varepsilon_a > 0$ and $\eta > 0$ there exists an $n_a$ such that
    \begin{IEEEeqnarray}{c}
        \Pr(A) \geq 1-\delta'(\eta) \; .
        \label{eq: event A geq}
    \end{IEEEeqnarray}
\item \label{it:nb} By the strong law of large numbers, for given $\varepsilon_b$, $n_a$, and $\delta - \delta'(\eta) > 0$ there exists $n_b>n_a$ such that
    \begin{IEEEeqnarray}{c}
        \Pr(B) \geq 1-(\delta-\delta'(\eta)) \; .
         \label{eq: event B geq}
    \end{IEEEeqnarray}
\item  \label{it:nanb} If the inequalities \eqref{eq: event A geq} and \eqref{eq: event B geq} hold, then
    \begin{IEEEeqnarray*}{rCl}
        \Pr(A \cap B) &=& \Pr(A)+\Pr(B)-\Pr(A \cup B) \\
        &=& \Pr(A)+\Pr(B)-\left( 1- \Pr(\Bar{A} \cap \Bar{B}) \right) \\
        &\eqann{a}& \Pr(A)+\Pr(B) - 1 + \Pr(\Bar{A} \cap \Bar{B}) \\
	&\eqann[\geq]{b}& \Pr(A) + \Pr(B) -1 \\
	&\eqann[\geq]{c}& \Big(1-\delta'(\eta)\Big) + \Big(1-(\delta-\delta'(\eta))\Big) - 1 \\
        &=& 1 - \delta \IEEEyesnumber \; ,
        \label{eq: event A and B geq}
    \end{IEEEeqnarray*}
    where \eqannref{a} is by De Morgan's laws, \eqannref{b} is since we throw away a non-negative term, and \eqannref{c} is by \eqref{eq: event A geq} and \eqref{eq: event B geq}.
\end{enumerate}

    Define the shorthand 
    \begin{IEEEeqnarray*}{c}
        \theta \triangleq - \log_{\varepsilon_a}(\kappa) %
	= \log_{1/\varepsilon_a}(2) \; .
    \end{IEEEeqnarray*}
    Note that for $\varepsilon_a<1$ we have $\theta>0$ and $\lim_{\varepsilon_a\rightarrow0}\theta=0$.
    Moreover define the shorthands $d_0 = 1$ and $d_1 = 2$.

    Following the same steps in \cite{Tal:17.2p}, we require that $\varepsilon_a$ is small enough such that $d_t-\theta > 0$ for $t \in \{0,1\}$, and also require $\varepsilon_a, \varepsilon_b<1/2$. For $\varepsilon_a$, $\varepsilon_b$ satisfying the above and $n_b > n_a$ that are yet to be fixed, we have under $A \cap B$ that for all $n > n_b$,
    \begin{IEEEeqnarray*}{c}
    S_n \leq 2^{-2^{\left(\frac{1}{2}-\Delta\right)n}} \; ,
    \end{IEEEeqnarray*}
    where
    \begin{multline}
	    \Delta = \sum_{t\in\{0,1\}}\frac{1}{2}\log_2 \left(\frac{d_t}{d_t-\theta}\right)-\sum_{t\in \{0,1\}}\pm \varepsilon_b\log_2(d_t-\theta)\\
	    +\sum_{t\in\{0,1\}}\frac{n_a}{n}\left(\frac{1}{2}\pm \varepsilon_b\right)\log_2(d_t-\theta) \; ,
         \label{eq: Delta}
    \end{multline}
    and
    \begin{IEEEeqnarray*}{c}
        \pm \triangleq
        \begin{cases}
            + & \mbox{if} \;\; d_t-\theta \leq 1, \\
            - & \mbox{otherwise}.
        \end{cases}
    \end{IEEEeqnarray*}
    Now, for a given $0<\beta< 1/2$, $\eta>0$ and $\delta>\delta'(\eta)$, our aim is to show that there exists a choice of parameters $\varepsilon_a, \varepsilon_b>0$ and $n_a<n_b$ such that \eqref{eq: event A and B geq} holds and $\Delta<1/2-\beta$.
    We choose $\varepsilon_a$ small enough such that the first sum in \eqref{eq: Delta} is less than $\frac{1/2-\beta}{3}$, $d_t-\theta > 0$ for $t \in \{0,1\}$, and $\varepsilon_a<1/2$.
    We choose $\varepsilon_b$ small enough such that the second sum in \eqref{eq: Delta} is less than $\frac{1/2-\beta}{3}$ and that $\varepsilon_b<1/2$.
    Note that $\varepsilon_a$ has been set and $\eta$ is given. As justified by Observation \ref{it:na}, we choose $n_a$ large enough such that \eqref{eq: event A geq} holds. 
    Then, we choose $n_b$ large enough such that both \eqref{eq: event B geq} holds (as justified by Observation \ref{it:nb}) and the third sum in \eqref{eq: Delta} is less than $\frac{1/2-\beta}{3}$.
    The above choices indeed satisfy our aim: by Observation \ref{it:nanb}, \eqref{eq: event A and B geq} holds since both \eqref{eq: event A geq} and \eqref{eq: event B geq} hold, and $\Delta < 1/2 -\beta$  since each one of the three sums in \eqref{eq: Delta} is smaller than $\frac{1}{3}(1/2-\beta)$. Therefore, setting $n_0=n_b$ ensures \eqref{eq: augmented proposition} holds.
\end{IEEEproof}

\subsection{Additional proofs for \cref{sec: improved thresholds}}
\begin{IEEEproof}[proof of \cref{prop: Ru monotonic in G}]
    By \eqref{eq: general Ru} we have 
    \begin{multline*}
    \ourRu(\mathcal{G'}) = \ourRu(\mathcal{G}) 
    - \frac{1}{2^{d'}} \cdot I\left(\tilde{Q}_{2^{d'}}^{(j')}\right) \\
    + \frac{1}{2^{d'+1}} \cdot I\left(\tilde{Q}_{2^{d'+1}}^{(2j')}\right)
    + \frac{1}{2^{d'+1}} \cdot I\left(\tilde{Q}_{2^{d'+1}}^{(2j'+1)}\right)    
    \end{multline*}
    Therefore, the claim will follow by showing that 
    \[
    I\left(\tilde{Q}_{2^{d'+1}}^{(2j')}\right) + I\left(\tilde{Q}_{2^{d'+1}}^{(2j'+1)}\right)
    \leq 2\cdot I\left(\tilde{Q}_{2^{d'}}^{(j')}\right) \; .
    \]
    For brevity denote 
    \begin{IEEEeqnarray*}{lCl}
	    \Gamma &\triangleq& \tilde{Q}_{2^{d'}}^{(j')} \; , \\
	    \Lambda^- &\triangleq& \tilde{Q}_{2^{d'+1}}^{(2j')} \; , \\
	    \Lambda^+ &\triangleq& \tilde{Q}_{2^{d'+1}}^{(2j'+1)} \; ,
    \end{IEEEeqnarray*}
    and
    \[
	    u \triangleq u_{2j'} \; , \quad v \triangleq u_{2j'+1} \; .
    \]
    Hence, our goal is to show that
    \begin{IEEEeqnarray*}{rCl}
   I(\Lambda^-) + I(\Lambda^+) \leq 2 \cdot I(\Gamma) \; . 
    \end{IEEEeqnarray*}
    For this, we further denote by $\Gamma^-$ and $\Gamma^+$ the minus and plus transforms of $\Gamma$, respectively. That is,
    \begin{IEEEeqnarray}{l}
        \subnumberinglabel{eq: Gamma transforms}
	    \Gamma^-(t_a,t_b;u) = 
	    \sum_v \Gamma(t_a; u \oplus v) \cdot \Gamma(t_b; v) \; , \label{eq: Gamma minus} \\
        \Gamma^+(t_a,t_b,u;v) = 
        \Gamma(t_a; u \oplus v) \cdot \Gamma(t_b; v) \; . \label{eq: Gamma plus}
    \end{IEEEeqnarray}
    By the chain rule,
    \[
I(\Gamma^-) + I(\Gamma^+) = 2 \cdot I(\Gamma) \; .
    \]
    Since (stochastic) degradation reduces mutual information, we will be done once we prove that $\Lambda^-$ is degraded with respect to $\Gamma^-$, and $\Lambda^+$ is degraded with respect to $\Gamma^+$. By inspection of \eqref{eq: Transforms of Q-tilde joint-distributions} versus \eqref{eq: Gamma transforms}, this is indeed the case. Namely, $\Gamma^{-}$ is degraded to $\Lambda^{-}$ by deterministically mapping $(t_a,t_b)$ to $\fminsum(t_a,t_b)$ while $\Gamma^{+}$ is degraded to $\Lambda^{+}$ by deterministically mapping $(t_a,t_b,u)$ to $g_u(t_a,t_b)$.
\end{IEEEproof}

\begin{IEEEproof}[proof of \cref{prop: Rl monotonic in E}]
	By \eqref{eq: general Rl} we have 
    \begin{multline*}
	    \ourRl(\mathcal{G},\mathcal{E'}) = \ourRl(\mathcal{G},\mathcal{E})
    - \frac{1}{2^{d'}} \cdot \max \left\{1 - \delta'(\ourzeta_{2^{d'}}^{(j')}), 0\right\} \\
    + \frac{1}{2^{d'+1}} \cdot \max \left\{1 - \delta'(\ourzeta_{2^{d'+1}}^{(2j')}), 0\right\} \\
    + \frac{1}{2^{d'+1}} \cdot \max \left\{1 - \delta'(\ourzeta_{2^{d'+1}}^{(2j'+1)}), 0\right\} \; .
    \end{multline*}
    Therefore, the claim will follow by showing that 
    \begin{multline*}
        2\cdot \max \left\{1 - \delta'(\ourzeta_{2^{d'}}^{(j')}), 0\right\} \leq 
        \max \left\{1 - \delta'(\ourzeta_{2^{d'+1}}^{(2j')}), 0\right\} \\
        + \max \left\{1 - \delta'(\ourzeta_{2^{d'+1}}^{(2j'+1)}), 0\right\} \; .
    \end{multline*}
    Recall \eqref{eq: zeta recursive} and denote for brevity 
    \begin{IEEEeqnarray*}{lClCl}
	    \ourzeta &\triangleq& (\ourzeta_{2^{d'}}^{(j')}) \; , \\
	    \ourzeta^- &\triangleq& (\ourzeta_{2^{d'+1}}^{(2j')}) &=& 2 \cdot \ourzeta \; ,  \\
	    \ourzeta^+ &\triangleq& (\ourzeta_{2^{d'+1}}^{(2j'+1)}) &=& \ourzeta^2 \; . 
    \end{IEEEeqnarray*}
Hence, our goal is to show that
\begin{multline}
	2 \cdot \max \{1-\delta'(\ourzeta),0\} \leq  \max \{1-\delta'(\ourzeta^-),0\} \\ +  \max \{1-\delta'(\ourzeta^+),0\}  \; .
	\label{eq: zeta condision for monotonicity in E}
\end{multline}
We assume that the LHS is positive, otherwise the claim is trivial.    
Under this assumption, we show that
\begin{IEEEeqnarray*}{C}
	2 \cdot \left(1-\delta'(\ourzeta)\right) \leq  \left(1-\delta'(\ourzeta^-)\right) + \left(1-\delta'(\ourzeta^+)\right) \; ,
	\label{eq: zeta condision for monotonicity in E - simpilied}
\end{IEEEeqnarray*}
which implies \eqref{eq: zeta condision for monotonicity in E}.
Using the definition of $\delta'(\cdot)$ in \eqref{eq: delta' definition} and plugging $\ourzeta^- = 2 \cdot \ourzeta$ and $\ourzeta^+ = \ourzeta^2$, the above simplifies to
\begin{IEEEeqnarray}{C}
	x \cdot \left(x-(2-2^{\log_2 \varphi})\right) \leq 0 \; ,
\end{IEEEeqnarray}
where we use the shorthand $x \triangleq \ourzeta^{\log_2 \varphi}$.
Therefore, the inequality holds for $x \in [0, 2-2^{\log_2 \varphi}]$, which is  $\ourzeta \in [0,1/4]$. That is, \eqref{eq: zeta condision for monotonicity in E} holds if $\ourzeta \in [0,1/4]$. Since $\ourzeta$ is non-negative, it remains to show that $\ourzeta \leq 1/4$. Indeed, by our assumption that the LHS in \eqref{eq: zeta condision for monotonicity in E} is positive and by \eqref{eq: delta' definition},
\begin{IEEEeqnarray*}{C}
	\ourzeta < \frac{1}{8} \cdot \left(\frac{1}{2}\right)^{1/\log_2 \varphi} \approx 0.046 \; . 
	\label{eq: zeta leq 0.046}
\end{IEEEeqnarray*}
\end{IEEEproof}

\begin{IEEEproof}[proof of \cref{prop: Rl monotonic in G}]
	Recall \eqref{eq: general Rl} and consider the calculation of $\ourRl(\mathcal{G},\mathcal{E})$ versus $\ourRl(\mathcal{G}',\mathcal{E})$.
	To distinguish and compare between the terms of these two sums,  we use the notation $\hat{\ourzeta}$ for $\mathcal{G}'$.
	That is,
\begin{IEEEeqnarray*}{rCl}
	\ourRl(\mathcal{G}, \mathcal{E}) &=& \sum_{(d,j) \in \mathcal{E}} \frac{1}{2^d} \cdot \max \left\{1 - \delta'(\ourzeta_{2^d}^{(j)}), 0\right\} \; , \\
	\ourRl(\mathcal{G}', \mathcal{E}) &=& \sum_{(d,j) \in \mathcal{E}} \frac{1}{2^d} \cdot \max \left\{1 - \delta'(\hat{\ourzeta}_{2^d}^{(j)}), 0\right\} \; .
\end{IEEEeqnarray*}
By \cref{def: G'} of $\mathcal{G}'(d',j')$, only nodes $(d,j) \in \mathcal{E}$ which are descendants of $(d',j')$ contribute differently to the sum.
The proof will follow by showing that for each such term, the contribution does not decrease when changing $\mathcal{G}$ to $\mathcal{G}'$. Thus, our goal is to show that for all $(d,j) \in \mathcal{E}$ which is a descendant of $(d',j')$ it holds that
\begin{IEEEeqnarray*}{rCl}
	\delta'(\hat{\ourzeta}_{2^d}^{(j)}) \leq \delta'(\ourzeta_{2^d}^{(j)}) \; .
\end{IEEEeqnarray*}
By the monotonicity of $\delta'(\cdot)$, defined in \eqref{eq: delta' definition}, it is sufficient to show that
\begin{IEEEeqnarray}{rCl}
	\hat{\ourzeta}_{2^d}^{(j)} \leq \ourzeta_{2^d}^{(j)} \; .
	\label{eq: zetahat leq zeta}
\end{IEEEeqnarray}

Recall how $\ourzeta_{2^d}^{(j)}$ is calculated: we use \eqref{eq: zeta recursive}, where the base of the recursion is node $(d',j')$ whose corresponding $\ourzeta_{2^{d'}}^{(j')}$ equals $\ourZopt(\tilde{Q}_{2^{d'}}^{(j')})$. In contrast, for $\hat{\ourzeta}_{2^d}^{(j)}$, we use the same recursive relations in \eqref{eq: zeta recursive}, but the base case is one level deeper: either $\hat{\ourzeta}_{2^{d'+1}}^{(2j')} = \ourZopt(\tilde{Q}_{2^{d'+1}}^{(2j')})$ or $\hat{\ourzeta}_{2^{d'+1}}^{(2j'+1)}  = \ourZopt(\tilde{Q}_{2^{d'+1}}^{(2j' + 1)})$, depending on the value of $j$. By inspection of \eqref{eq: zeta recursive} versus \eqref{eq: Z-ish-opt evolutions}, we have that $\hat{\ourzeta}_{2^{d'+1}}^{(2j')} \leq \ourzeta_{2^{d'+1}}^{(2j')}$ and $\hat{\ourzeta}_{2^{d'+1}}^{(2j'+1)} \leq \ourzeta_{2^{d'+1}}^{(2j'+1)}$. By the monotonicity of the two operations in \eqref{eq: zeta recursive}, doubling and squaring, we deduce that this inequality persists throughout the path to $(d,j)$, and hence \eqref{eq: zetahat leq zeta} indeed holds.
\end{IEEEproof}

\begin{IEEEproof}[proof of \cref{cor: monotonicty of rate thresholds}]
	To prove \eqref{eq: Ru(G*) leq Ru(G)}, recall the definition of $\Gdeep$. That is, for some finite integer $T$, there exists a sequence
	\begin{equation}
		\mathcal{G} = \mathcal{G}_0, \mathcal{G}_1, \ldots, \mathcal{G}_T = \Gdeep \; ,
		\label{eq: G series}
	\end{equation}
	where for each $0 \leq t < T$ we have $\mathcal{G}_{t+1} = \mathcal{G}_{t}'(d,j)$, for some $(d,j) \in \mathcal{G}_t$. Thus,
	by \eqref{eq: Ru(G') leq Ru(G)}, for each $0 \leq t < T$ we have $\ourRu(\mathcal{G}_{t+1}) \leq \ourRu(\mathcal{G}_{t})$. Hence, \eqref{eq: Ru(G*) leq Ru(G)} follows.

	To prove \eqref{eq: Rl(G*,E*) geq Rl(G,E)} we show that
	\begin{equation}
	 \ourRl(\mathcal{G}, \mathcal{E}) \leq \ourRl(\mathcal{G}, \Edeep) \leq	\ourRl(\Gdeep, \Edeep)\; .
	 \label{eq: Rl intermediate inequality}
 \end{equation}

 Consider the first inequality. For this, note that as before, we have for some finite integer $S$ the sequence
	\begin{equation}
		\mathcal{E} = \mathcal{E}_0, \mathcal{E}_1, \ldots, \mathcal{E}_S = \Edeep \; ,
		\label{eq: E series}
	\end{equation}
	where for each $0 \leq s < S$ we have $\mathcal{E}_{s+1} = \mathcal{E}_{s}'(d,j)$, for some $(d,j) \in \mathcal{E}_s$. Thus, by \eqref{eq: Rl(G,E') geq Rl(G,E)}, for each $0 \leq s < S$ we have $\ourRl(\mathcal{G}, \mathcal{E}_{s+1}) \geq \ourRl(\mathcal{G}, \mathcal{E}_s)$. Hence, the first inequality in \eqref{eq: Rl intermediate inequality} follows.

Consider now the second inequality in \eqref{eq: Rl intermediate inequality}. 
Recall that $(\Gdeep, \Edeep)$ is a valid pair. Next, note that if $(\mathcal{G}_{t+1},\Edeep)$ is a valid pair, then so is $(\mathcal{G}_{t},\Edeep)$.
Hence, since $\Gdeep = \mathcal{G}_T$, all pairs $(\mathcal{G}_{t},\Edeep)$ are valid for $0 \leq t \leq T$.
We wish to apply \eqref{eq: Rl(G',E) geq Rl(G,E)} repeatedly to show that $\ourRl(\mathcal{G}_{t+1},\Edeep) \geq \ourRl(\mathcal{G}_{t},\Edeep)$, from which the second inequality in \eqref{eq: Rl intermediate inequality} follows.
Recalling the conditions in \cref{prop: Rl monotonic in G}, we must show that $\mathcal{G}_{t+1}$ is not constructed from $\mathcal{G}_{t}$ by a node $(d',j') \in \Edeep$.
Indeed, if this were the case, then $(\mathcal{G}_{t+1},\Edeep)$ would not be valid, contradicting what we have already established.
\end{IEEEproof}

\begin{IEEEproof}[proof of \cref{prop: Ru leq C}]
	Note that $I(\tilde{Q}_1^{(0)}) \leq I(W) \triangleq C$, since $\tilde{Q}_1^{(0)}$ is obtained by stochastically degrading $W$ using the labeling function $\labeler(\cdot)$. Next, we define the set $\mathcal{G}_{(0,0)} \triangleq \{ (0,0) \}$. By inspection of \eqref{eq: general Ru}, $\ourRu(\mathcal{G}_{(0,0)}) = I(\tilde{Q}_1^{(0)})$. Hence, $\ourRu(\mathcal{G}_{(0,0)}) \leq C$. The claim follows by noting that each valid $\mathcal{G}$ satisfies $\mathcal{G} \geq \mathcal{G}_{(0,0)}$. Therefore by applying \eqref{eq: Ru(G*) leq Ru(G)} we have $\ourRu(\mathcal{G}) \leq \ourRu(\mathcal{G}_{(0,0)}) \leq C$.
\end{IEEEproof}

\twobibs{
\bibliographystyle{IEEEtran}
\bibliography{mybib.bib}
}
{
\ifdefined\bibstar\else\newcommand{\bibstar}[1]{}\fi

}
\end{document}